\documentclass[preprint,review,12pt,authoryear]{elsarticle}

\usepackage{amssymb}
\usepackage{amsthm}
\usepackage{amsmath}
\usepackage{siunitx}
\usepackage{adjustbox}
\usepackage{booktabs}
\usepackage{multirow}

\usepackage{lineno}

\usepackage{xcolor}
\usepackage[hidelinks,breaklinks]{hyperref}
\usepackage{xurl}
\usepackage{makecell}
\usepackage{pdfpages}

\DeclareSIUnit\year{yr}

\begin{document}

\begin{frontmatter}

\title{Downscaling GRACE-derived ocean bottom pressure anomalies using self-supervised data fusion}

\author[ETH]{Junyang Gou}\ead{jungou@ethz.ch}
\author[Bonn]{Lara Börger}
\author[Bonn]{Michael Schindelegger}
\author[ETH]{Benedikt Soja}

\affiliation[ETH]{organization={Institute of Geodesy and Photogrammetry, ETH Zurich}, city={Zurich}, country={Switzerland}}

\affiliation[Bonn]{organization={Institute of Geodesy and Geoinformation, University of Bonn}, city={Bonn}, country={Germany}}

\begin{abstract}
The gravimetry measurements from the Gravity Recovery and Climate Experiment (GRACE) and its follow-on (GRACE-FO) satellite mission provide an essential way to monitor changes in ocean bottom pressure ($p_b$), which is a critical variable in understanding ocean circulation. However, the coarse spatial resolution of the GRACE(-FO) fields blurs important spatial details, such as $p_b$ gradients. In this study, we employ a self-supervised deep learning algorithm to downscale global monthly $p_b$ anomalies derived from GRACE(-FO) observations to an equal-angle \SI{0.25}{\degree} grid in the absence of high-resolution ground truth. The optimization process is realized by constraining the outputs to follow the large-scale mass conservation contained in the gravity field estimates while learning the spatial details from two ocean reanalysis products. The downscaled product agrees with GRACE(-FO) solutions over large ocean basins at the millimeter level in terms of equivalent water height and shows signs of outperforming them when evaluating short spatial scale variability. In particular, the downscaled $p_b$ product has more realistic signal content near the coast and exhibits better agreement with tide gauge measurements at around \SI{80}{\percent} of 465 globally distributed stations. Our method presents a novel way of combining the advantages of satellite measurements and ocean models at the product level, with potential downstream applications for studies of the large-scale ocean circulation, coastal sea level variability, and changes in global geodetic parameters.

\end{abstract}


\end{frontmatter}


\section{Introduction}
\label{sec:Introduction}
Ocean bottom pressure ($p_b$) fluctuations indicate variations in the amount and spatio-temporal distribution of ocean mass. These variations are primarily caused by atmospheric forces and the transfer of continental freshwater into the ocean~\citep{church2013SeaLevelChange}, but they can also emerge internally from the unstable ocean circulation~\citep{zhao2021IntrinsicVariability}. The information contained in $p_b$ fields is valuable to understanding ocean dynamics~\citep{olbers2012oceandynamics}, including monitoring large-scale ocean circulation~\citep{hughes2018window2deepocean,mccarthy2020ReviewAMOC} and mesoscale turbulence ~\citep{beech2022longtermEddy}. Moreover, changes in $p_b$ fields are also closely linked to essential geodetic parameters, such as Earth orientation parameters~\citep{boergerOceanReanalysesEarthRotation2023} and non-tidal ocean loading effects~\citep{williams2011non-tidalOceanLoading}. Accurate $p_b$ estimations with high spatio-temporal resolution are required for the success of the aforementioned applications.

Variations in $p_b$ can be directly measured by bottom pressure recorders (BPR), which are also employed to observe components of the Atlantic meridional overturning circulation (AMOC), see, e.g.,~\cite{elipot2013coherence} or~\cite{worthington2019AMOCfromBPR}. However, in-situ recorders are costly to maintain and require apt post-processing approaches~\citep{watts1990BPR,Macrander2010BPR,poropat2018BPRpostcessing}, whilst installation of a global BPR network calls for more resources than presently available. For decades, a significant amount of effort has been put into modeling the ocean state by considering the equations for ocean motions~\citep{olbers2012oceandynamics}, and indeed, ocean models have contributed substantially to our understanding of $p_b$ variability and its implications~\citep[e.g.,][]{ponteGlobalOBP1999,weijer2010almostfree,piecuch2015vertical,hughes2018window2deepocean,rohith2019basin}. However, the fidelity of modeled $p_b$ variations may be compromised by model errors, including dynamical simplifications, errors in forcing fields and boundary conditions (e.g., bathymetry), or uncertainties associated with choices of viscosity, momentum schemes, and physical parameterizations in general \citep{forget2015ecco4,fox2019challenges}. Therefore, numerical models need to be validated against, and ideally also constrained to observations. 

Since 2002, the Gravity Recovery and Climate Experiment (GRACE) and its follow-on (GRACE-FO) satellite mission have been providing an opportunity to monitor monthly gravity variations of the Earth with unprecedented accuracy~\citep{tapley2004GracePrinciple,wahr2004GracePrinciple,landerer2020GRACE-FO}. Gravity variations over the ocean can be converted into $p_b$ anomalies to infer ocean mass changes~\citep{chambers2010GRACEOcean,chen2019GRACEOcean} and the broad characteristics of the seasonal cycle in manometric sea level \citep{johnson13seasonal}. In addition, large-scale volume transports can be, in principle, recovered from gravimetry-based $p_b$ gradients \citep[see, e.g.,][]{peralta2014arctic}, but the coarse effective resolution of the GRACE(-FO) fields of around \SI{3}{\degree} still limits the range of possible applications~\citep{chen2022GRACEreview}. For example, the potential of GRACE(-FO) products for monitoring deep AMOC variability has been demonstrated at 26.5$^{\circ}$N~\citep{landerer2015GRACE4AMOC,bentel2015monitoring}, but the method remains unproven for wider sections of the North Atlantic. Progress to this end would require sufficiently dense $p_b$ information along the continental slope~\citep{roussenov2008BoundaryWaveCommunication,hughes2018window2deepocean}. Similarly, knowledge of short-scale $p_b$ variability is essential to understand the drivers for coastal sea level changes, which are related to, e.g., upwelling, coastal trapped circulations, and boundary waves~\citep{woodworth2019forcingCoastSeaLevel}.

In general, syntheses of GRACE(-FO) derived mass changes with ocean models may leverage the advantages of either component, thus potentially resulting in more accurate estimates of $p_b$. Such syntheses have been realized by assimilating GRACE(-FO) measurements into ocean models~\citep{kohl2012impactOBPassimilation,menemenlis2008ecco2,forget2015ecco4} or by using ocean model outputs to guide the downscaling of GRACE(-FO) products~\citep{delman2022downscalingOBP}. Recent progress in deep learning methods called attention to the potential of these techniques for enhancing climate and general circulation model outputs~\citep{schneider2017EarthSystemModeling,fox2019challenges,irrgang2021towards}. Particular opportunities arise in downscaling Earth observations~\citep{reichstein2019deeplearning4geoscience} and in utilizing machine learning approaches to assimilate Earth observations into climate models~\citep{schneider2023AI4ClimateModel}. \cite{schneider2023AI4ClimateModel} also argue that climate modeling should use cell spacings of about 10 to \SI{50}{\kilo\metre}, which is feasible to resolve mesoscale turbulence~\citep{oldenburg2022resolution}. However, efforts to downscale $p_b$ variations to such level of detail with deep learning methods are far behind similar considerations for GRACE(-FO) based terrestrial water storage changes~\citep{miro2018downscaling,seyoum2019downscaling,yin2022Downscaling,irrgang2020self,gou2024Global}.

This study contributes to the above dimensions by applying a convolution-based neural network to fuse the GRACE(-FO) products and two eddy-permitting ocean reanalysis products with the help of additional features (e.g., bathymetry and wind stress). High-resolution $p_b$ measurements are not available with sufficient global coverage to serve as ground truth and provide supervision signals. Therefore, the classical supervised learning approaches are not applicable. We employ a self-supervised deep learning algorithm that receives supervision signals based on parts of input features~\citep{wang2022SSLinRS}. The pipeline was originally designed by \cite{gou2024Global} and adapted to fit the requirements of downscaling $p_b$. Specifically, two ocean reanalysis products are considered together as high-resolution guidance, and their weights are dynamically determined, while the GRACE(-FO) fields are used to constrain mass conservation. The model has promising generalizability and provides, within one network, global results covering both GRACE and GRACE-FO eras (April 2002 to December 2020), where the end time is defined by the availability of ocean reanalysis data. The downscaled product is evaluated for its general signal content, including global and basin-averaged mass changes, spatially distributed trends, and seasonal oscillations. The added value of the high-resolution signals is assessed with comparisons against in-situ BPR measurements and coastal tide gauge measurements.

The rest of this paper is structured as follows: We first clarify the definition of $p_b$ and introduce all relevant datasets in Section~\ref{sec:Data}. The main methods, with a focus on the architecture of the self-supervised data fusion neural network, are reported in Section~\ref{sec:Method}. Selected characteristics of the downscaled product are shown in Section~\ref{sec:Results}, complemented by evaluations with in-situ bottom pressure measurements and coastal sea level observations from tide gauges. The potential benefits of our approach and possible directions for further analysis are discussed in Section~\ref{sec:Conclusions and outlook}.

\section{Definitions and Data}
\label{sec:Data}
\subsection{Ocean bottom pressure}
Given the integration of the hydrostatic equation over the full water column, $p_b$ can be written as follows \citep[e.g.,][]{ponteGlobalOBP1999}:
\begin{equation}
\label{eq::obp_vert_int}
p_b = p_a + \int_{-H}^{0} \rho g \mathrm{d}z + \rho_0 g \eta,
\end{equation}
where $p_a$ is the atmospheric surface pressure (or its fluctuation), $H$ is the local water depth, $g$ is the gravitational acceleration, $\rho=\rho(z)$ is the total density of seawater, $\rho_0$ represents a constant reference density, and $z$ is the vertical coordinate pointing upwards. The sea level anomaly relative to $z=0$ is denoted by $\eta$. Given the ocean's tendency for an inverted barometer (IB) response to changes in $p_a$, one typically writes the sea level anomaly as $\eta = \eta^\mathrm{IB} + \eta'$, comprising the IB term $\eta^\mathrm{IB} = \left(\overline{p}_a - p_a \right)/\left(\rho_o g\right)$ \citep{ponte1994IB} and the remaining dynamic component $\eta'$. Thus, the first and the last term in Eq.~(\ref{eq::obp_vert_int}) yield $\overline{p}_a + g \rho_0 \eta'$, where $\overline{p}_a$ is the spatially averaged $p_a$ over the global ocean \citep{ponteGlobalOBP1999}. Throughout this study, we reckon $p_b$ values into the form of equivalent water height (EWH) in millimeters. Given the area of a grid cell, changes in EWH describe the changes in water heights needed to cause the observed mass changes.

\subsection{GRACE(-FO) mascon solution}
Three analysis centers of GRACE(-FO) missions, Center for Space Research (CSR), Goddard Space Flight Center (GSFC), and Jet Propulsion Laboratory (JPL), are operationally providing mass concentration (mascon) solutions, which are gridded mass changes derived from the inter-satellite range-rate measurements~\citep{save2016CSRMascon,loomis2019GSFCMascon,watkins2015JPLMascon}. Some common processing steps are performed for generating the mascon data, including the replacement of poorly constrained low-degree spherical harmonic coefficients, a glacial isostatic adjustment (GIA) correction~\citep{richard2018GIAcorrection}, and removal of the non-tidal atmosphere and ocean mass variability product (AOD1B) for de-aliasing~\citep{dobslaw2017AOD1BRL06}. Note that the temporal averages of $p_b$ as simulated by the AOD1B ocean model (GAD products) are restored. Therefore, the GRACE(-FO) mascons over the ocean represent the full bottom pressure caused by ocean dynamics and changes in $\overline{p}_a$.

The mascon solutions have been proven to reflect $p_b$ variations near coastlines better than spherical harmonic solutions~\citep{piecuch2018tide,mu2020investigation}. The three different solutions agree well over large scales, such as ocean basins, but show some disparities on a regional scale, especially close to landmasses, owing to different processing strategies and approaches for suppressing leakage from terrestrial signals~\citep{sakumura2014ensemble}. In this study, we consider all three mascon products individually for deriving our downscaled products and select the product obtained from CSR mascon (CSRM) as an example. Similar analyses on the other two downscaled products can be found in the supplementary material.

\subsection{Ocean model and reanalysis products}
We use two out of four members of the eddy-permitting ocean reanalysis ensemble provided by CMEMS (Copernicus Marine Environment Monitoring Service, \citealt{desportesQID2019}), which are the Global Ocean Reanalysis and Simulation 2 version 4 (GLORYS2v4, for short: GLORYS, \citealt{lellouche2013GLORYS}) and the Ocean Reanalysis System 5 (ORAS5, \citealt{zuo2017ORAS5}). Our choice of GLORYS and ORAS5 is largely motivated by the skill displayed by these products in an analysis of Earth rotation variations \citep{boergerOceanReanalysesEarthRotation2023}. Both reanalyses were run on an eddy-permitting horizontal 0.25$^\circ$ tri-polar grid, consisting of 75 vertical layers. The Nucleus for European models of the Ocean version 3 (NEMO3) was adopted as a common hydrodynamic core and forced with six-hourly buoyancy and momentum fluxes from ERA-Interim~\citep{deeERAInterimConfiguration2011}, without considering atmospheric pressure loading. Filter approaches were used to constrain the model to observations of sea surface temperature, daily sea level anomalies, sea ice concentration, and hydrographic profiles. Despite the commonalities, the reanalyses' states are not identical due to differences in the chosen data assimilation scheme, analysis window, and surface nudging, and in the treatment of uncertainties, see Table 2 in \cite{boergerOceanReanalysesEarthRotation2023} for a synopsis. Both reanalysis products have global coverage, excluding parts of the Southern Ocean in latitudes higher than \SI{70}{\degree}S.

For our analysis period from 2002 to 2020, we use monthly potential temperature and salinity fields from the selected reanalyses provided by CMEMS \citep{cmemsGlobalOceanEnsemble2019} to derive density $\rho$ using the TEOS-10 subroutines contained in the Gibbs-Seawater (GSW) Oceanographic Toolbox \citep{mcdougall2011GSW}. In the vertical integration, we account for the actual sea surface height as in increment to the uppermost layer (Eq.~(\ref{eq::obp_vert_int})) but neglect contributions from $p_a$, just as in the reanalyses themselves. To be consistent with the actual model bathymetry, we utilize the 1-arcminute ETOPO1 dataset \citep{amanteETOPO1ArcminuteGlobal2009}, average it to the 0.25$^\circ$ grids of the reanalyses, and employ the so-derived field as lower bound in the vertical integration for calculating $p_b$~\eqref{eq::obp_vert_int}. Water bodies underneath ice-shelf cavities are not included in the integration, as they are treated as land in the reanalyses.

\subsection{Auxiliary features}
To enhance the performance of the deep learning method, we include four additional features that are relevant for describing ocean dynamics globally and regionally. First, we include steric sea level anomalies because their changes may indirectly reflect mass changes, especially in the regions where halosteric signals correlate with mass variations~\citep{jorda2013StericvsMass}. Moreover, steric changes in the deeper parts of the ocean can induce mass changes on the shelf to balance horizontal pressure gradients~\citep{bingham2012local,woodworth2019forcingCoastSeaLevel}. Monthly steric heights were computed from EN4 subsurface temperature and salinity~\citep{good2013EN4}, again using the GSW Oceanographic Toolbox. The EN4 profiles are based on ocean floater observations and provide information independent of other features. Second, we consider the fact that the impacts of atmospheric winds on ocean dynamics vary with the ocean depths, i.e., the response in shallow regions is generally more energetic~\citep{bingham2008relationshipSLAandOBP_025}. Therefore, additional features describing bathymetry and wind stress vectors are meaningful. We include the 1-arcminute ETOPO1 dataset for seafloor depths and monthly east/west instantaneous turbulent surface stresses from ERA5~\citep{hersbach2020ERA5}.

\subsection{In-situ bottom pressure recorders}
We use the BPR data initially collected by~\cite{Macrander2010BPR} and further processed by~\cite{schindelegger2021DailyGRACE}. Relevant steps include the concatenation of different deployments at the same BPR site, visual inspection, and manual removal of drifts and occasional spikes. Tidal variability associated with 14 harmonics, including monthly and fortnightly constituents, was subtracted using a tidal atlas \citep[updated version of][]{Egbert2002}. To obtain the monthly BPR measurements, we computed the arithmetic mean of daily measurements within the measuring interval of individual GRACE(-FO) months. To ensure the quality of the time series, we only considered the monthly values containing more than 15 valid daily observations and the stations with more than 12 valid months. With these criteria, 119 out of 132 stations remain valid. We aligned $p_b$ stations to the nearest sampling grids of $p_b$ anomalies to formulate the time series pairs. A temporal average of the common valid months was removed from each pair to generate the anomalies.

\subsection{Tide gauge measurements}
The tide gauge measurements used in this study are monthly revised local reference values distributed by the Permanent Service for Mean Sea Level \citep[PSMSL,][]{holgate2013PSMSL,PSMSLData}. We corrected each record for the IB effect~\citep{ponte1994IB} using ERA5 monthly surface pressures data~\citep{hersbach2020ERA5}. Then, we chose all stations located in the coastal ocean by considering the GRACE(-FO) ocean mask. The tide gauge measured time series were aligned to the GRACE(-FO) measuring epochs (192 months from April 2002 to December 2020) by linear interpolation, and gaps larger than one month were omitted. All tide gauge stations with less than 96 valid monthly samples (\SI{50}{\percent} completeness) were excluded, resulting in a remaining total of 465 stations. The stations were then aligned to their nearest $p_b$ sampling grids to formulate time series pairs and a temporal average over the valid months was removed from each pair to generate anomalies. Moreover, steric heights obtained from ORAS5 were removed from the sea level anomalies to deduce the manometric sea level anomalies. This steric correction was not performed in previous studies~\citep[e.g.][]{piecuch2018tide}, but we found it to yield improved agreements between the tide gauge measurements and all $p_b$ products considered in this study.

\section{Method}
\label{sec:Method}
\subsection{Self-supervised data fusion algorithm}
Given the absence of high-resolution global $p_b$ measurements, we cannot directly formulate a loss function based on some references that can provide supervision signals. To overcome this difficulty, we formulate this problem as a self-supervised task in which the trainable parameters can be optimized by the supervision signals generated from a part of the input data~\citep{wang2022SSLinRS}. The loss function in our study is designed to fulfill two requirements: (1) keeping the large-scale accuracy as in the GRACE(-FO) fields and (2) learning and balancing high-resolution information from reanalysis products. To achieve these two goals, we design the loss function with two terms as follows:
\begin{equation}
    \mathcal{L}\left(\hat{\mathbf{P}}, \mathbf{P}_\mathrm{M}, \mathbf{P}_\mathrm{G}, \mathbf{P}_\mathrm{O}\right) = \frac{1}{B}\sum_{b=1}^B\left\{\mathcal{L}_\mathrm{GRACE}\left(\hat{\mathbf{P}}, \mathbf{P}_\mathrm{M}\right) +  \mathcal{L}_\mathrm{Reanalysis}\left(\hat{\mathbf{P}}, \mathbf{P}_\mathrm{G}, \mathbf{P}_\mathrm{O}\right)\right\},\label{eq:Customized_Loss}
\end{equation}
where $\hat{\mathbf{P}}, \mathbf{P}_\mathrm{M}, \mathbf{P}_\mathrm{G}, \mathbf{P}_\mathrm{O}$ indicate the patches of downscaled products, GRACE(-FO) mascons, GLORYS, and ORAS5. The two terms of the loss function are denoted by $\mathcal{L}$, and batch size is denoted by $B$. The supervision signal for controlling large-scale accuracy comes from minimizing the absolute errors (AE) between the predictions and GRACE(-FO) patches, as defined as:
\begin{equation}
    \mathcal{L}_\mathrm{GRACE}\left(\hat{\mathbf{P}}, \mathbf{P}_\mathrm{M}\right) = \mathrm{AE_{M}}\left(\hat{\mathbf{P}}, \mathbf{P}_\mathrm{M}\right) = \left|\frac{1}{N}\sum_{n=1}^Np_{\mathrm{M}, n} - \frac{1}{N}\sum_{n=1}^N\hat{p}_{n}\right|,\label{eq:Loss_GRACE}
\end{equation}
in which the patch-wise average values of $N$ pixels ($p$) within individual patches are first computed. Then, the absolute differences between the average values are obtained. By minimizing $\mathcal{L}_\mathrm{GRACE}$, the average values of the predictions are forced to be close to GRACE(-FO) solutions.

The second loss term aims to provide supervision on learning high-resolution information from the reanalysis products. Usually, it can be done by maximizing a similarity measure, such as the 2D Pearson correlation ($\mathrm{R}$), or minimizing the pixel-wise mean absolute error ($\mathrm{MAE}$). We combine these two metrics to generate the loss terms based on GLORYS or ORAS5 fields ($\mathcal{L}_\mathrm{G/O}$) as shown in Eq.~\eqref{eq:Loss_Reanalysis}. In this case, the impacts of the potential outliers contained in the reanalysis products are reduced to a certain degree since a large MAE will gain relatively small weight due to the high similarity. Moreover, we introduce additional weights $w_{G/O}$ to the GLORYS and ORAS5 terms, which allows the model to rely on individual high-resolution products in different scenarios dynamically. The weights are computed from $\mathrm{AE}$ between reanalysis products and GRACE measurements to reduce the impact disparities of large-scale deterministic signals. We have found that considering two reanalysis products clearly improves the quality of the downscaled product (see supplementary material). The final formulation of $\mathcal{L}_\mathrm{Reanalysis}$ is:
\begin{multline} \mathcal{L}_\mathrm{Reanalysis}\left(\hat{\mathbf{P}}, \mathbf{P}_\mathrm{G}, \mathbf{P}_\mathrm{O}\right) = \underbrace{\frac{\mathrm{AE_{O}}}{\mathrm{AE_{G}}+\mathrm{AE_{O}}}}_\text{$w_\mathrm{G}$}\cdot\underbrace{\left[1 - \mathrm{R}\left(\mathbf{P}_\mathrm{G}, \hat{\mathbf{P}}\right)\right]\cdot\mathrm{MAE_{G}}\left(\mathbf{P}_\mathrm{G}, \hat{\mathbf{P}}\right)}_\text{$\mathcal{L}_\mathrm{G}$} \\+ \underbrace{\frac{\mathrm{AE_\mathrm{G}}}{\mathrm{AE_{G}}+\mathrm{AE_{O}}}}_\text{$w_\mathrm{O}$}\cdot\underbrace{\left[1 - \mathrm{R}\left(\mathbf{P}_\mathrm{O}, \hat{\mathbf{P}}\right)\right]\cdot\mathrm{MAE_{O}}\left(\mathbf{P}_\mathrm{O}, \hat{\mathbf{P}}\right)}_\text{$\mathcal{L}_\mathrm{O}$},\label{eq:Loss_Reanalysis}
\end{multline}
where the subscripts $\mathrm{G}$ and $\mathrm{O}$ indicate GLORYS and ORAS5 reanalysis, respectively. By minimizing the final loss function, the network tries to learn the high-resolution information contained in ocean reanalysis products while constraining the outputs to force mass conservation. In practice, the network was optimized using Adam~\citep{kingma2014Adam}.

\subsection{Network architecture}
The pipeline used in this study is modified from the self-supervised data assimilation model proposed by~\cite{gou2024Global} to fit our purpose for downscaling $p_b$ anomalies. The model is convolution-based~\citep{lecun1998CNN} with an encoder-decoder architecture, which is beneficial for retaining valuable information while reducing noise levels~\citep{ronneberger2015U-net}. The principle of residual learning is also employed~\citep{He2016Resnet}. The network explicitly approximates a residual function $\mathcal{F}(x) = \mathcal{H}(x) - x$ instead of the full target function $\mathcal{H}(x)$ since it can relieve the complexity of training a deep neural network. The batch normalization is also included to stabilize the training process~\citep{ioffe2015BatchNorm}.

\begin{figure}[!ht]
    \centering
    \begin{adjustbox}{width=18.3cm,center}
    \includegraphics[width=18.3cm]{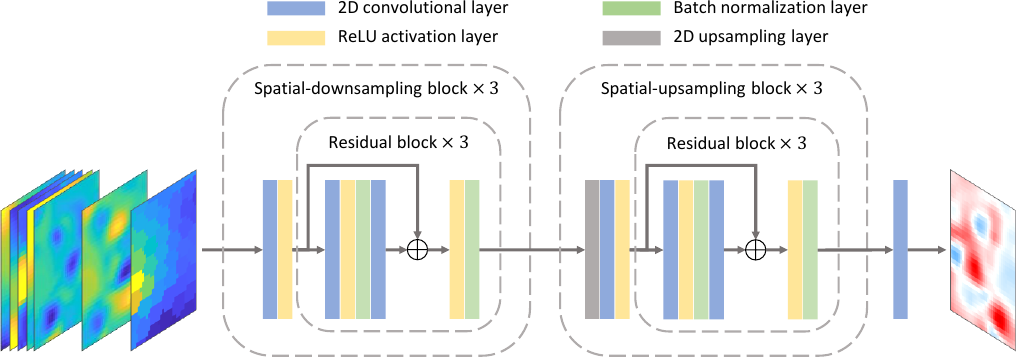}
    \end{adjustbox}
    \caption{The model architecture. The encoder consists of three spatial downsampling blocks, each having one 2D convolutional layer and three residual blocks. The decoder consists of three spatial upsampling blocks, each having one 2D upsampling layer followed by one 2D convolutional layer and three residual blocks. The final 2D convolutional layer serves as the output layer.}
    \label{fig:Model-architecture}
\end{figure}

A schematic diagram of the designed architecture is shown in Fig.~\ref{fig:Model-architecture}, with an input size of $32\times32\times7$.
The encoder has three spatial-downsampling blocks constituted by a 2D convolutional layer followed by a ReLU activation function $\left(\mathrm{ReLU}(x)=\max(0, x)\right)$ and three residual blocks. The first convolutional layers of each spatial-downsampling block have a stride of 2 to reduce the spatial size but increase the receptive field of their output features. The increasing numbers of kernels enable the growth of latent dimension. The final output latent features of the encoder have a dimension of $4\times4\times64$, which compresses the spatial information into the latent space. In the decoder, the latent features are first upsampled using a bilinear upsampling layer, and then the values are refined through the convolutional layers and residual blocks. The important spatial information is reconstructed through the decoder while the noise level is reduced~\citep{bourlard1988autoencoder1,Hinton2006autoencoder2}. The network outputs are then compared to the original GRACE(-FO) and reanalysis $p_b$ anomalies to formulate the different compartments of the loss function, as shown in Eq.~\eqref{eq:Customized_Loss}.

\subsection{Data preprocessing and model optimization}
The global data are split into patches with a size of $32\times32$ grids, equal to $\SI{7.5}{\degree}\times\SI{7.5}{\degree}$, in which the average values of GRACE(-FO) fields are representative. In total, we have about 127.3 million patches covering the global ocean area with a sampling resolution of \SI{0.25}{\degree} from April 2002 to December 2020. The seven features are considered as seven channels for the convolution-based network. The features are normalized to enhance the optimizing stability~\citep{goodfellow2016deeplearning}. To prevent outliers from distorting the data distribution, we employed robustness normalization by considering the 0.01$^\mathrm{th}$ and 99.99$^\mathrm{th}$ percentiles instead of minimum and maximum. All the models and optimization processes were realized using TensorFlow V2.6.0~\citep{tensorflow2015whitepaper}.

\section{Results and Discussion}
\label{sec:Results}
\subsection{Global downscaled $p_b$ with eddy-permitting resolution}
\label{sec:Global downscaled OBP with eddy-permitting resolution}
\begin{figure}[!ht]
    \centering
    \begin{adjustbox}{width=18.3cm,center}
    \includegraphics[width=18.3cm]{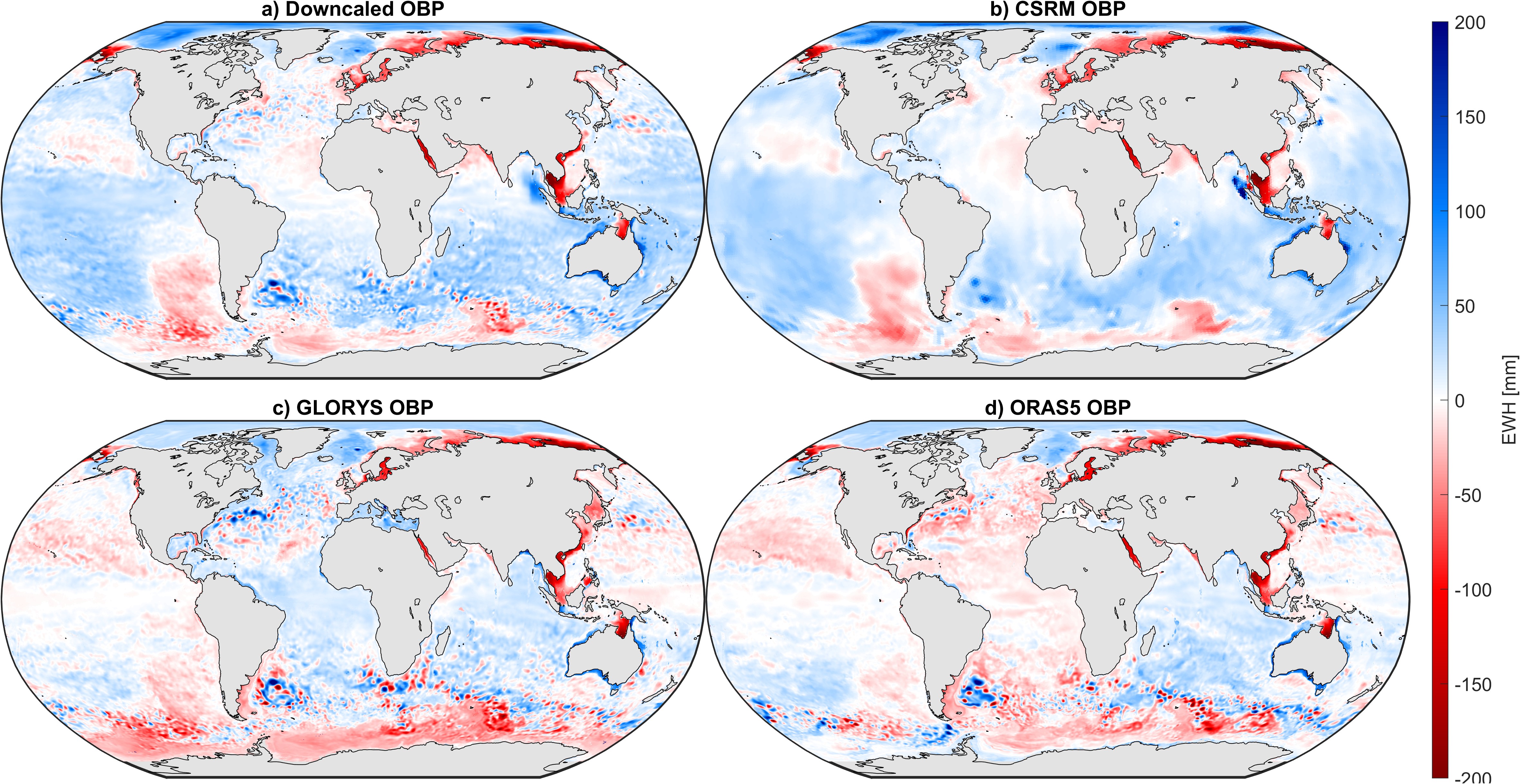}
    \end{adjustbox}
    \caption{An example of a downscaled monthly $p_b$ field along with the three $p_b$ products considered as input features in June 2010.}
    \label{fig:OBP_Compare}
\end{figure}

We generated the global monthly downscaled $p_b$ fields with a spatial resolution of \SI{0.25}{\degree} from April 2002 to December 2020 with the proposed model. As an example, Fig.~\ref{fig:OBP_Compare} illustrates the downscaled $p_b$ for June 2010, along with the other three $p_b$ products that serve as inputs. The downscaled $p_b$ anomalies have clearly higher spatial resolution than the CSRM data, enabling us to observe small-scale $p_b$ gradients. The small-scale information is learned from the two reanalysis products. At the same time, the large-scale signals are forced to agree with CSRM anomalies by considering mass conservation. This is a common drawback of the reanalysis products since their large-scale fidelity compared to the GRACE(-FO) measurements tends to be degraded. For example, in Fig.~\ref{fig:OBP_Compare}, GLORYS has a suspicious basin-wide $p_b$ anomaly in the Pacific Ocean, while ORAS5 disagrees with the average GRACE(-FO) $p_b$ signal in the Atlantic Ocean. We note that atmospheric pressures do not cause these differences since removing the GAD product from CSRM $p_b$ fields did not improve the agreement. The disparities shown here more likely reflect on limitations in the reanalyses and particularly differences in the simulated mesoscale field that can project onto mass fluctuations between entire basins~\citep{zhao2021IntrinsicVariability,zhao2023IntrinsicVariability}. A contribution of our method is that the large-scale signals in the resulting downscaled product are in line with the ones observed by GRACE(-FO). We note that the signals caused by large earthquakes persist in our downscaled product since they exist in the GRACE(-FO) fields~\citep{ghobadi2020grace4earthquakes}. We should not interpret them as ocean mass changes.

\begin{figure}[!ht]
    \centering
    \begin{adjustbox}{width=18.3cm,center}
    \includegraphics[width=18.3cm]{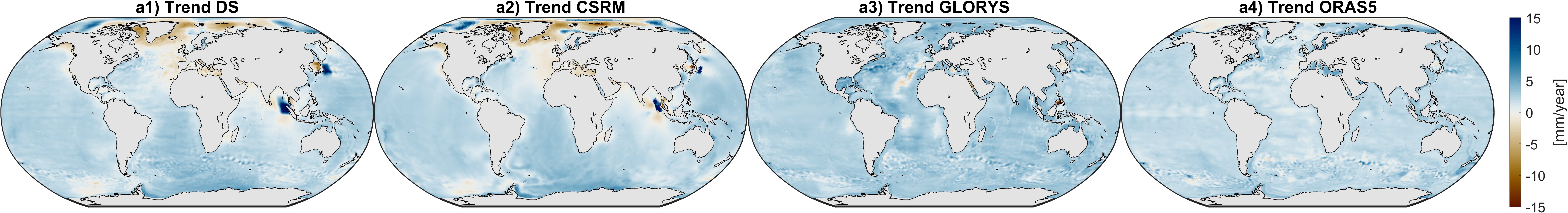}
    \end{adjustbox}
    \begin{adjustbox}{width=18.3cm,center}
    \includegraphics[width=18.3cm]{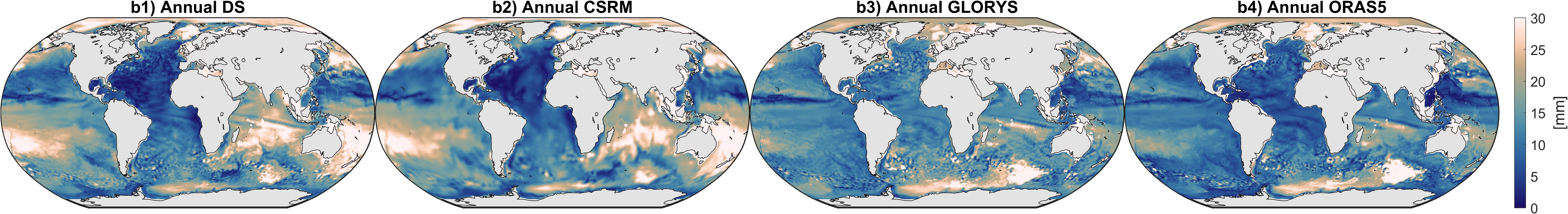}
    \end{adjustbox}
    \begin{adjustbox}{width=18.3cm,center}
    \includegraphics[width=18.3cm]{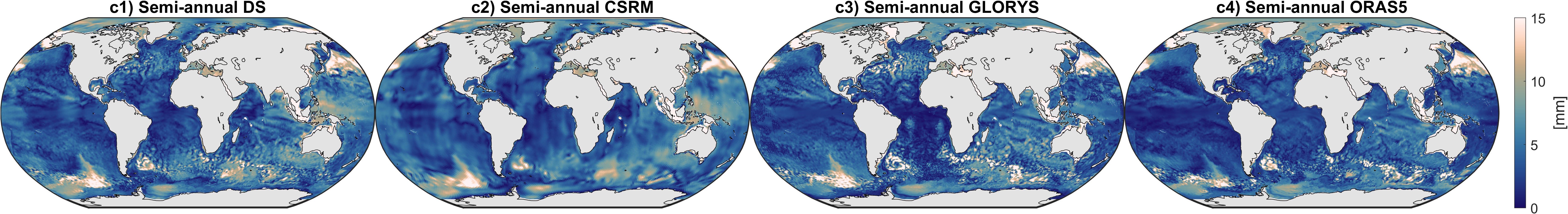}
    \end{adjustbox}
    \begin{adjustbox}{width=18.3cm,center}
    \includegraphics[width=18.3cm]{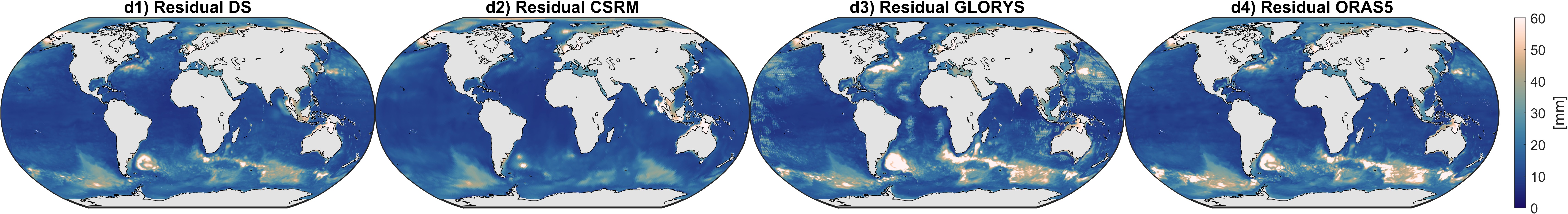}
    \end{adjustbox}
    \caption{Temporal decomposition of four $p_b$ products from April 2002 to December 2020. Shown are a) long-term trends [mm/year], b) annual amplitudes [mm], c) semi-annual amplitudes [mm], and d) standard deviations of residuals [mm]. The four columns represent the respective estimates from the downscaled (DS), CSRM, GLORYS, and ORAS5 $p_b$.}
    \label{fig:OBP_Decompose}
\end{figure}

To examine the signals in different temporal bands, we decomposed the $p_b$ products of the entire study period into linear trends, annual, semi-annual, and post-fit residuals by applying sinusoidal regressions (see Fig.~\ref{fig:OBP_Decompose} for the results). The downscaled trends agree well with the CSRM trends, whereas the reanalysis products only indicate an overall rise in ocean mass without showing long-wavelength patterns. The decreasing trends surrounding Greenland might be associated with the sea-level fingerprints induced by ice sheet melt~\citep{hsu2017IceFingerprints,coulson2022IceFingerprints}. Since the ocean models do not consider gravitational attraction and loading effects, the reanalysis $p_b$ fields cannot show such sea-level response. However, we refrain from interpreting the decreasing trends surrounding Greenland as genuine physically driven static seal level changes, as leakage errors from ice sheets and Arctic glaciers might also be playing a role~\citep{wahr2006accuracyGRACE,chen2022GRACEreview}. On another note, large earthquakes such as the Sumatra–Andaman earthquake~\citep{chen2007grace4earthquake} and Tohoku-Oki earthquake~\citep{wang2012grace4earthquake} affect the downscaled and CSRM trends, resulting in disparities with the reanalysis products in terms of the estimated trends. In the annual band, which is an attractive target for climate analysis~\citep{niu2022mechanisms,qin2022mechanism}, the downscaled amplitudes also conform with CSRM. On the contrary, both reanalyses show widespread low amplitudes in the Indian and South Pacific oceans, pointing to difficulties in numerical models to simulate the large-scale ocean response to annual changes in wind stress and buoyancy fluxes.

The small-scale $p_b$ gradients reveal local ocean dynamics, which feature prominently in the semi-annual $p_b$ component and the post-fit results. For example, the post-fit residuals of monthly $p_b$ products are far from noise but also include plausible and expected signals, such as geostrophic bottom currents~\citep{olbers2012oceandynamics}, topographically constrained barotropic ciculations~\citep{weijer2010almostfree}, or $p_b$ variability caused by the interaction of eddies with seafloor topograhy~\citep{yu2018VariationsArgentineGyre}. Fig.~\ref{fig:OBP_Decompose}d shows that the signal levels of the downscaled $p_b$ post-fit residuals are indeed distinctly higher than the CSRM $p_b$ in regions with active ocean dynamics. Both reanalysis products reveal strong $p_b$ variability in regions of the Gulf Stream, the Kuroshio Current, the Argentine Gyre, the Agulhas Current, and the Antarctic Circumpolar Current, all of which are considered as eddy-rich regions~\citep{hughes2018window2deepocean,beech2022longtermEddy}. These signals are hardly observable in the CSRM fields due to the limited resolution of the GRACE(-FO) solutions, whereas similar patterns are reflected in our downscaled product. Nevertheless, the magnitudes in the post-fit signals of our downscaled product are comparatively smaller than the reanalyses, which is inevitable when balancing the GRACE(-FO) and reanalysis inputs in our deep-learning algorithm. We consider this issue as a trade-off while providing long-term and large-scale fidelity.

\subsection{Global and large-scale ocean mass variability}
\begin{figure}[!ht]
    \centering
    \includegraphics[width=89mm]{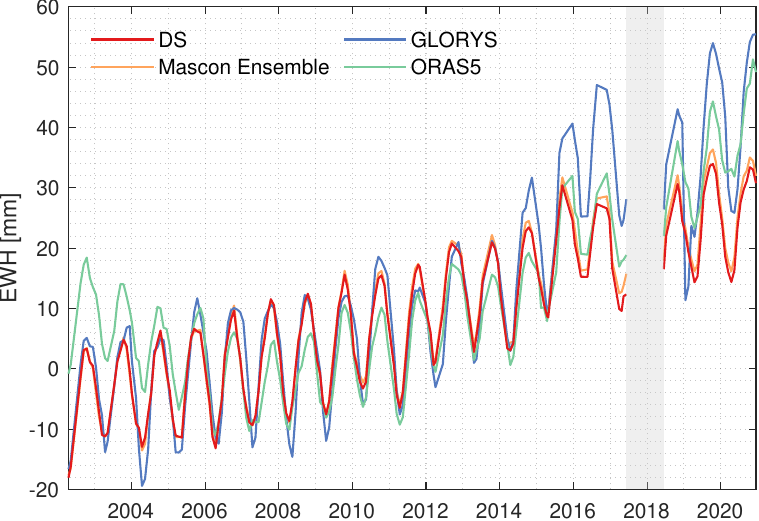}\\
    \caption{The global mean $p_b$ time series [mm] of the four products from April 2002 to December 2020. Static atmospheric effects in $p_b$ have been accounted for by removing the GAD product from the downscaled (DS) and Mascon ensemble $p_b$ variations. The gap between GRACE and GRACE-FO missions is shaded.}
    \label{fig:BigBasinMassConservation_Global}
\end{figure}

The intra-annual $p_b$ variability contained in the reanalysis products can be afflicted with errors from various sources, such as imperfections in atmospheric forcing fields or sensitivities to parameterizations and eddy dynamics~\citep{androsov2020OBPmodel,zhao2023IntrinsicVariability}. It is, therefore, crucial to constrain the high-resolution $p_b$ estimations using the GRACE(-FO) fields at their effective scale. To evaluate the large-scale accuracy, we consider the ensemble mean of the three mascon products as ground truth since the simple arithmetic mean is effective in reducing the noise level at this scale~\citep{sakumura2014ensemble,JPL2021GRACEL3handbook}. Fig.~\ref{fig:BigBasinMassConservation_Global} depicts the global average $p_b$ variations with the numerical metrics reported in Table~\ref{table:BigBasinMassConservation}. To account for the static atmospheric contribution to $p_b$, the GAD product~\citep{dobslaw2017AOD1BRL06} has been removed from downscaled and mascon ensemble $p_b$ fields to obtain the mean ocean mass variations. As a result, they indicate the same quantity as the reanalysis products and can be directly compared to these products.

On the global scale, downscaled $p_b$ agrees well with the mascon ensemble with a correlation of 0.999 and RMSE lower than \SI{1}{\milli\meter}, whereas GLORYS and ORAS5 tend to over-/underestimate the annual and semi-annual amplitudes, respectively~\citep[cf.][]{boergerOceanReanalysesEarthRotation2023}. The long-term trend reflected in the downscaled $p_b$ anomalies (\SI{1.87}{\milli\meter\per\year}) is slightly lower than the trend in the mascon ensemble (\SI{1.99}{\milli\meter\per\year}) because the CSRM $p_b$ anomalies indicate smaller trends compared to the other two mascon solutions (see supplementary material), primarily due to the patterns in the Arctic Ocean, as discussed in Section~\ref{sec:Global downscaled OBP with eddy-permitting resolution}. However, the trends reflected in the downscaled product are still more realistic than the GLORYS product (\SI{2.72}{\milli\meter\per\year}), which overestimates the trend after 2015. The trend from ORAS5 provides the closest match with the mascon ensemble but is the result of a questionable temporal evolution in the reanalyzed ocean mass (Fig.~\ref{fig:BigBasinMassConservation_Global}). In particular, the global $p_b$ average decreased from 2002 to 2008 and increased after this period. The behavior could indicate an inherent model drift, errors in the ERA-Interim mass fluxes, or spurious mass changes incurred by the sequential data assimilation~\citep{boergerOceanReanalysesEarthRotation2023}.

\begin{figure}[!ht]
    \centering
    \begin{adjustbox}{width=18.3cm,center}
    \includegraphics[width=18.3cm]{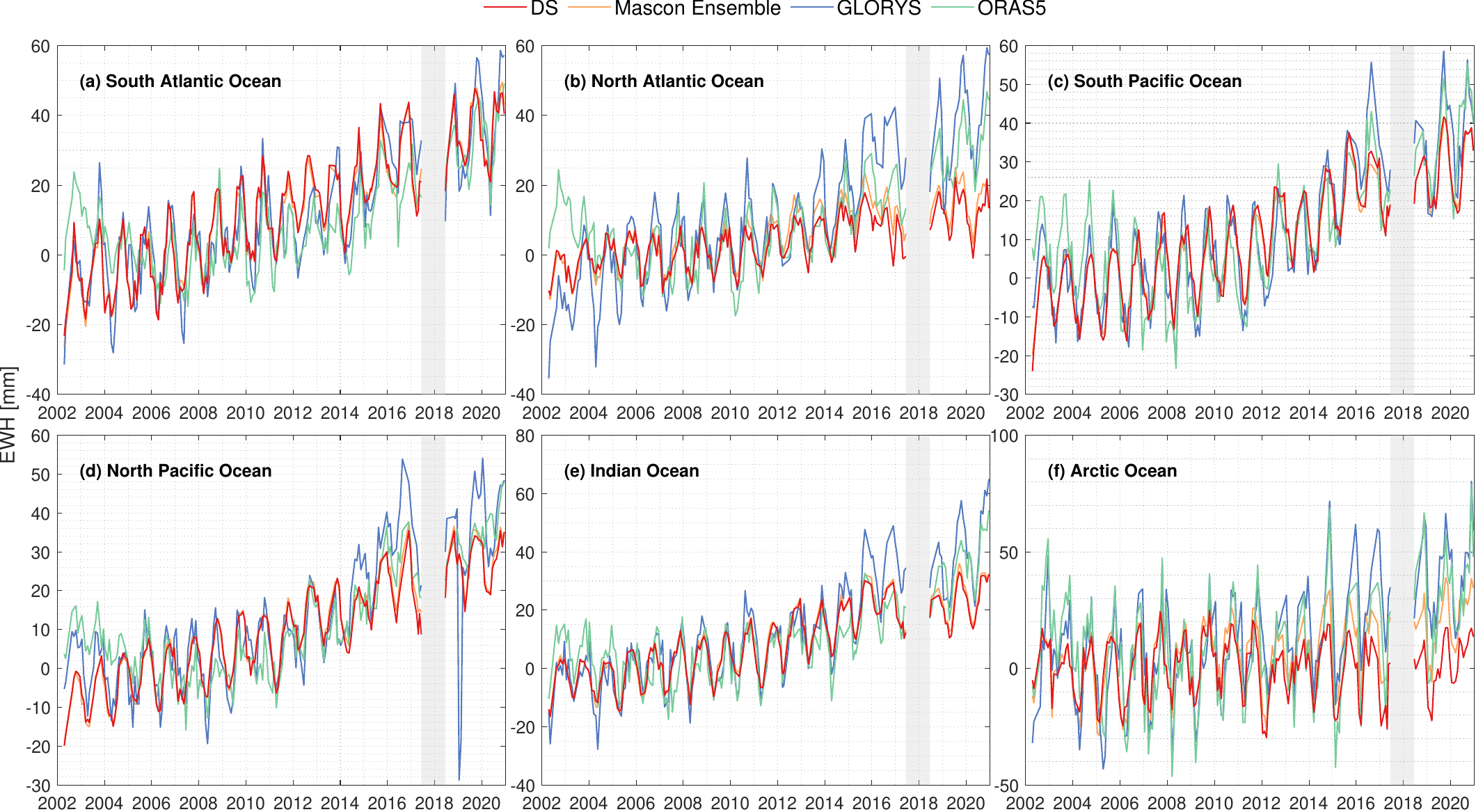}
    \end{adjustbox}
    \caption{Time series of the spatial mean $p_b$ [mm] from the four products in the six large ocean basins from April 2002 to December 2020. Static atmospheric effects in $p_b$ have been accounted for by removing the GAD product from the downscaled (DS) and Mascon ensemble $p_b$ variations. The gap between GRACE and GRACE-FO missions is shaded. Note the different EWH ranges in different regions.}
    \label{fig:BigBasinMassConservation}
\end{figure}

We further separate the whole globe into six major ocean basins based on ocean basin boundaries provided by~\cite{GOAS2021} to study the quality of our product in different regions (Fig.~\ref{fig:BigBasinMassConservation}). The downscaled $p_b$ product tightly agrees with the mascon ensemble solution in the South Atlantic, South and North Pacific, and the Indian Ocean with root-mean-square errors (RMSE) less than \SI{2}{\milli\meter} and correlation higher than 0.99. The increasing trends reflected in the mascon ensemble solution are also preserved in our downscaled product. Conversely, the two reanalysis products have notably larger RMSE of about \SI{10}{\milli\meter} with correlation varying from 0.74 (ORAS5 in the South Atlantic Ocean) to 0.91 (GLORYS in the South Pacific Ocean). The disparities between reanalysis products and the mascon ensemble solution are mainly caused by disagreements in monthly to seasonal variations. Moreover, GLORYS tends to overestimate the trend in the Pacific Ocean, whereas ORAS5 shows the above-noted non-linear long-term variability. Some exceptional outliers, such as the sudden drop in the North Pacific Ocean suggested by GLORYS $p_b$ anomalies in 2019, are also evident.

In the North Atlantic and Arctic Ocean, the performance of our model is slightly degraded, with RMSE of \SI{3.10}{\milli\meter} and \SI{9.71}{\milli\meter}, respectively. The correlation remains higher than 0.96 for the North Atlantic but slightly drops to 0.88 for the Arctic Ocean. However, our product still agrees better with the mascon ensemble solution than the other two products. Due to the existence of island glaciers and strong signals caused by Greenland ice melting, clear separation between cryospheric and oceanic signals becomes challenging. Therefore, the slight degradation of our method in these regions is understandable, given the fact that the internal variations among the three mason solutions are relatively high~\citep[and supplementary materials]{imbie2020MassLossGreenland,velicogna2020MassLossIceSheet}. The ocean mass changes in the Atlantic Ocean are characterized by large seasonal signals. The high correlation seen for the downscaled product implies good phase agreement and confirms that the relatively large RMSE are rather due to a slight excess in amplitudes. On the contrary, the two reanalysis products show various distortions of trends and phases, along with overestimated seasonal amplitudes, resulting in large RMSE of $\gtrsim$~\SI{10}{\milli\meter} in both the North and South Atlantic oceans.

\begin{table}[!ht]
    \caption{Comparison of the three high-resolution $p_b$ products with the mascon ensemble (ME) solution in the six large ocean basins regarding long-term trends, RMSE, and Pearson correlations. The best agreements with ME are highlighted. SA: South Atlantic, NA: North Atlantic, SP: South Pacific, NP: North Pacific, IND: Indian, ARC: Arctic.}
    \centering
    \begin{adjustbox}{width=119mm,center}
    \begin{tabular}{cc|ccccccc}
        \toprule
        & & Global& SA& NA& SP& NP& IND& ARC\\
        \midrule
        \multirow{4}{*}{Trends [\SI{}{\milli\meter\per\year}]}& ME& 1.99& 2.72& 1.22& 2.06& 2.23& 1.90& 1.33\\
                                    & DS& 1.87& \textbf{2.67}& \textbf{0.88}& 2.14& \textbf{2.10}& 1.80& 0.05\\
                                    & GLORYS& 2.72& 2.66& 2.91& 2.52& 2.43& 2.88& 2.55\\
                                    & ORAS5& \textbf{1.97}& 1.72& 1.58& \textbf{2.02}& 1.99& \textbf{1.95}& \textbf{1.35}\\
                                    \midrule
        \multirow{4}{*}{\makecell{Annual \\Amplitudes [mm]}}& ME& 9.27& 10.53& 6.33& 10.08& 7.93& 8.54& 16.45\\
                                                            & DS& \textbf{9.26}& \textbf{11.08}& \textbf{5.76}& \textbf{10.60}& \textbf{8.06}& 8.22& \textbf{14.42}\\
                                                            & GLORYS& 11.3& 12.06& 11.80& 11.70& 8.44& 10.56& 21.15\\
                                                            & ORAS5& 8.10& 7.85& 8.09& 11.24& 4.50& \textbf{8.39}& 22.12\\
                                                            \midrule
        \multirow{4}{*}{\makecell{Semi-annual \\Amplitudes [mm]}}& ME& 0.76& 1.13& 1.49& 0.67& 1.02& 1.70& 3.62\\
                                                                 & DS& \textbf{0.82}& \textbf{1.24}&\textbf{ 1.34}& \textbf{1.07}& \textbf{1.00}&\textbf{ 1.66}& \textbf{3.91}\\
                                                                 & GLORYS& 1.73& 0.14& 1.00& 2.87& 2.00& 2.89& 5.43\\
                                                                 & ORAS5& 0.18& 1.96& 1.76& 1.55& 1.39& 2.40& 4.31\\
                                                                 \midrule
        \multirow{3}{*}{RMSE [mm]}& DS& \textbf{0.99}& \textbf{1.56}& \textbf{3.10}& \textbf{1.40}& \textbf{1.37}& \textbf{1.18}& \textbf{9.71}\\
                                  & GLORYS& 7.59& 7.91& 12.91& 8.01& 9.71& 9.09& 18.50\\
                                  & ORAS5& 7.03& 11.47& 9.87& 9.04& 8.76& 8.46& 15.87\\
                                  \midrule
        \multirow{3}{*}{Correlation [-]}& DS& \textbf{0.999}& \textbf{0.996}& \textbf{0.96}& \textbf{0.996}& \textbf{0.996}& \textbf{0.997}& \textbf{0.88}\\
                                        & GLORYS& 0.95& 0.90& 0.89& 0.91& 0.84& 0.93& 0.71\\
                                        & ORAS5& 0.88& 0.74& 0.72& 0.83& 0.80& 0.82& 0.75\\
        \bottomrule
    \end{tabular}
    \label{table:BigBasinMassConservation}
    \end{adjustbox}
\end{table}

\clearpage

\subsection{Comparison with in-situ bottom pressure measurements}
\begin{figure}[!ht]
    \centering
    \begin{adjustbox}{width=17cm,center}
    \includegraphics[width=17cm]{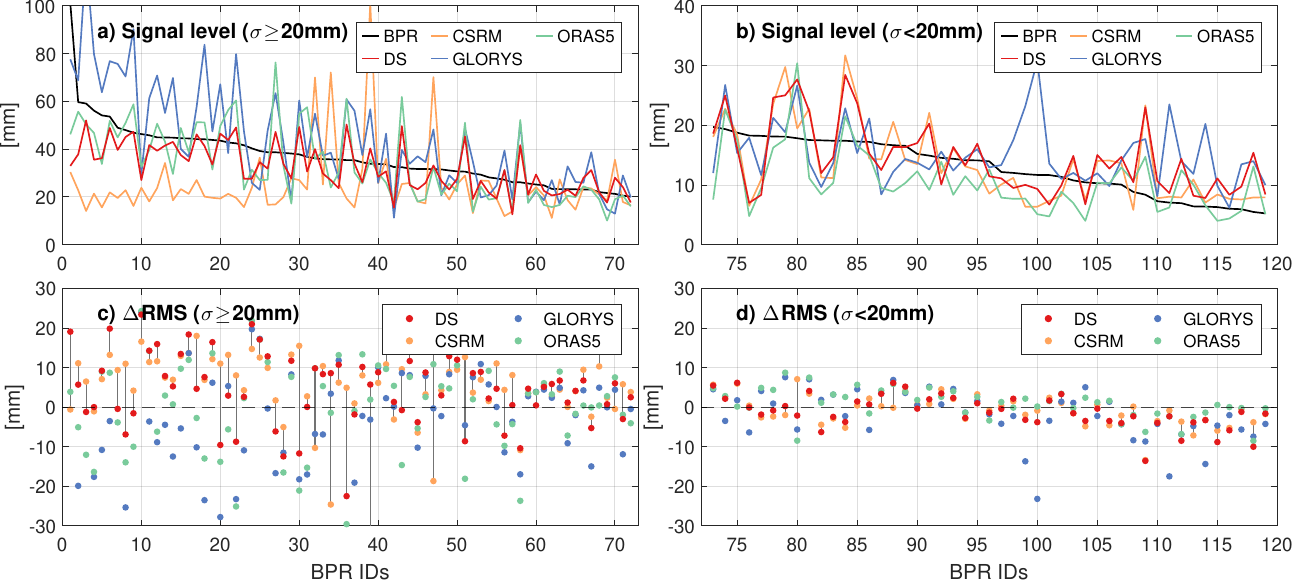}
    \end{adjustbox}
    \caption{Signal levels in terms of standard deviation and RMS reductions in millimeters at the 119 valid BPR stations. The stations are sorted by the descending signal levels of the monthly BPR measurements and split into two groups for better visualization. The first row depicts the signal levels of the five different $p_b$ data for stations that have standard deviations larger than \SI{20}{\milli\meter} (a) and smaller than \SI{20}{\milli\meter} (b). The second row depicts the RMS reduction of the four $p_b$ products at the same stations.}
    \label{fig:Eva_BPR_SignalAndRMSred}
\end{figure}

After examining the large-scale mass variability, we compare the signal levels of the $p_b$ products with in-situ BPR measurements to evaluate their ability to reflect small-scale variations. Upon sorting the data by standard deviations of the BPR series (Fig.~\ref{fig:Eva_BPR_SignalAndRMSred}a,b), we find that the CSRM $p_b$ anomalies fail to follow the change in signal levels across the BPR sites. The CSRM signal levels always hover around \SI{20}{\milli\meter}, interrupted by some unrealistic large peaks (e.g., at stations 34, 39, and 47 in the Kuroshio region). The phenomenon of similar signal levels at many different locations likely bears on the coarse resolution of satellite-based $p_b$ anomalies that smooths out local variability. The situation is improved with all three high-resolution $p_b$ products, the signal levels of which generally follow the ones reflected by the BPR measurements. However, GLORYS tends to overestimate the signal levels at stations with standard deviations larger than \SI{40}{\milli\meter}. ORAS5 and the downscaled $p_b$ anomalies have similar signal content and agree well with the in-situ measurements. We further show the RMS reduction, defined as standard deviations of BPR measurements minus RMS between $p_b$ products and BPR measurements, in Fig.~\ref{fig:Eva_BPR_SignalAndRMSred}(c,d). We find clear RMS reductions for the downscaled and CSRM products, especially at stations with signals larger than \SI{20}{\milli\meter}, whereas subtraction of GLORYS and ORAS5 from the BPR series tends to increase the RMS. Table~\ref{table:BPR} provides average statistics over all BPR sites. The highest correlations and RMS reductions at stations with pronounced $p_b$ signals ($\geq$~\SI{20}{\milli\meter}) are given by the downscaled product, followed by CSRM. The positive RMS reductions for CSRM, along with the relatively stable signals across BPR sites (Fig.~\ref{fig:Eva_BPR_SignalAndRMSred}a,b), suggests that the GRACE(-FO) products cannot resolve the magnitudes of small-scale $p_b$ variability but provide a credible representation of the spatially broad, lower-magnitude $p_b$ background variability.

\begin{table}[ht!]
    \caption{The average signal levels, Pearson correlations, and RMS reductions of the four $p_b$ products compared to monthly BPR measurements. The average signal levels of monthly BPR measurements are shown for comparison. The stations are divided into three groups by considering their signal levels ($\sigma$): all valid stations, greater equal to \SI{20}{\milli\meter}, and smaller than \SI{20}{\milli\meter}.}
    \centering
    \begin{adjustbox}{width=89mm,center}
    \begin{tabular}{ccccc}
        \toprule
        & & All& $\sigma\geq\SI{20}{\milli\meter}$& $\sigma<\SI{20}{\milli\meter}$\\
        & Num. of BPR& 119& 72& 42\\
        \midrule
        \multirow{5}{*}{Signal level [\SI{}{\milli\meter}]}& BPR& 27.3& 36.6& 12.9\\
                                                           & DS& \textbf{25.3}& 32.5& 14.4\\
                                                           & CSRM& 20.7& 25.3& \textbf{13.5}\\
                                                           & GLORYS& 32.3& 43.5& 15.2\\
                                                           & ORAS5& 24.7& \textbf{33.9}& 10.6\\
                                                           \midrule
       \multirow{4}{*}{Correlation [-]} & DS& \textbf{0.49}& \textbf{0.58}& 0.37\\
                                        & CSRM& 0.48& 0.57& 0.35\\
                                        & GLORYS& 0.47& 0.50& 0.41\\
                                        & ORAS5& 0.45& 0.45& \textbf{0.45}\\
                                        \midrule
       \multirow{4}{*}{$\Delta$RMS [\SI{}{\milli\meter}]} & DS& 2.52& \textbf{4.94}& -1.19\\
                                                           & CSRM& \textbf{2.64}& 4.89& -0.81\\
                                                           & GLORYS& -3.03& -3.67& -2.07\\
                                                           & ORAS5& 0.25& -0.13& \textbf{0.84}\\
        \bottomrule
    \end{tabular}
    \label{table:BPR}
    \end{adjustbox}
\end{table}

\begin{figure}[!ht]
    \centering
    \includegraphics[width=100mm]{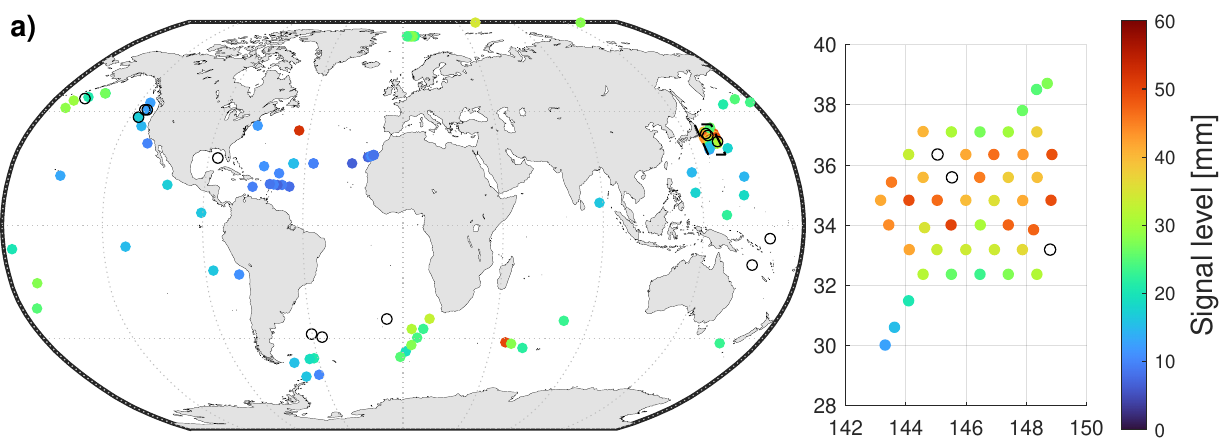}\\
    \includegraphics[width=100mm]{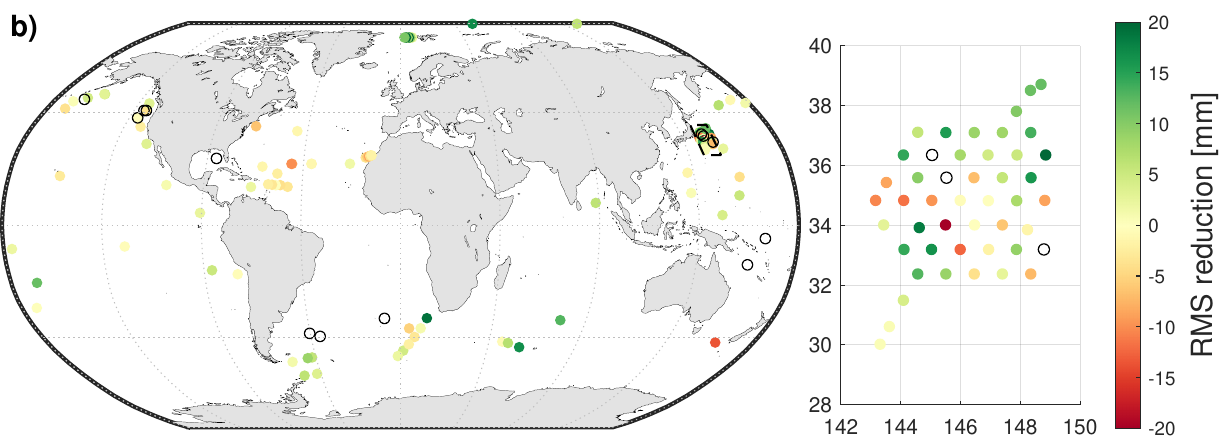}\\
    \includegraphics[width=100mm]{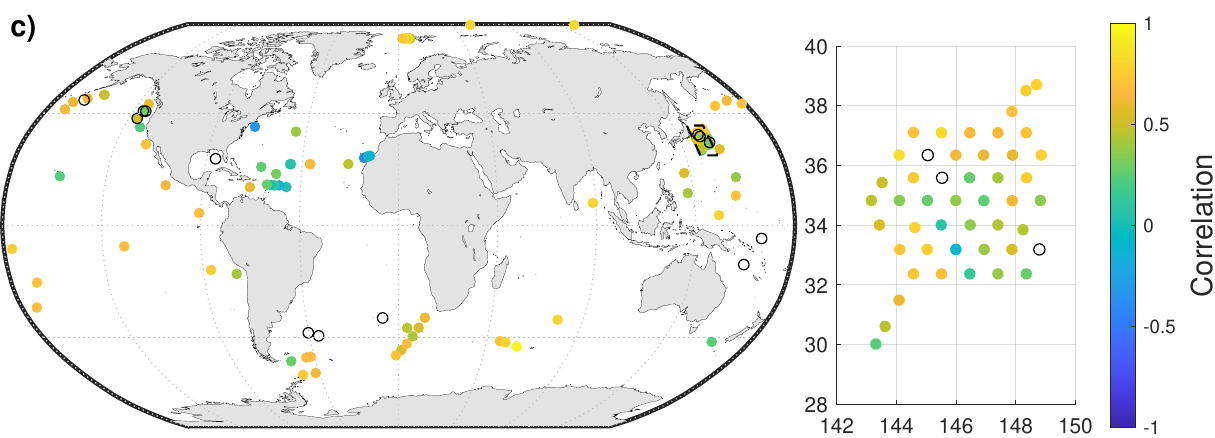}\\
    \caption{The spatial distribution of the 119 BPR stations with the colors depicting a) signal levels (standard deviations) of downscaled $p_b$ anomalies, b) RMS reductions provided by downscaled $p_b$ anomalies, and c) correlations between downscaled $p_b$ and BPR measurements. The Kuroshio region is zoomed in for better visibility.}
    \label{fig:Eva_BPR_Map}
\end{figure}

To further study the performance of our product in different regions, we visualize signal levels, RMS reductions, and correlations in Fig.~\ref{fig:Eva_BPR_Map}. It is clear that most of the low correlations and negative RMS reductions are associated with low signal levels, such as in the North Atlantic and along the coastlines of North America. In particular, variability and phases may be ambiguously defined in areas of weak signals, resulting in mismatches between the in-situ measurements and the downscaled $p_b$ anomalies. Some exceptions from such assumed behavior are found in the Kuroshio region, where the downscaled $p_b$ product yields negative correlations at some stations with strong signals. Although the effective resolution of the downscaled product is significantly improved compared to the original CSRM product, it may not be sufficient to fully resolve the truly local variability measured by the BPR stations, especially in the regions with high dynamics. It remains to be seen if targeting higher spatial resolution than 0.25$^{\circ}$ can improve the agreement with BPR observations.

\subsection{Comparison with in-situ sea-level measurements}
\begin{figure}[!tb]
    \centering
    \begin{adjustbox}{width=160mm,center}
    \includegraphics[width=160mm]{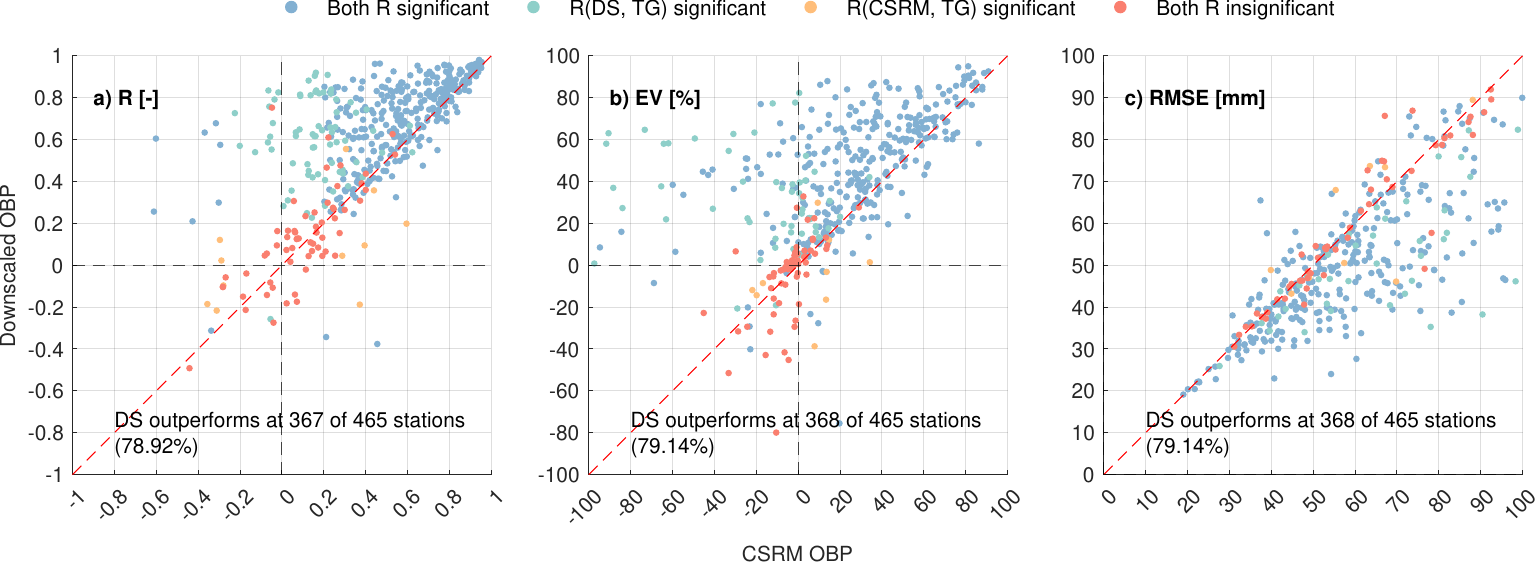}
    \end{adjustbox}
    \caption{Comparison of the full time series of downscaled $p_b$ (DS) and CSRM $p_b$ with in-situ tide gauge (TG) measurements. Three station-wise metrics are reported: a) Pearson correlations (R), b) explained variances (EV), and c) root-mean-square errors (RMSE). The statistical test on correlations is a t-test with a confidence level of 0.95.}
    \label{fig:Eva_TG_Scatter_All}
\end{figure}

Further validation has been performed by comparing the $p_b$ products with coastal sea-level changes measured by tide gauges. The Pearson correlations (R), explained variances (EV), and RMSE between downscaled/CSRM $p_b$ and tide gauge measurements at the 465 selected stations are shown in Fig.~\ref{fig:Eva_TG_Scatter_All}. The downscaled product outperforms CSRM at about \SI{79}{\percent} stations, highlighting its ability to represent near-coastal ocean mass changes due to its refined spatial resolution. The median values of the three matrices are improved relative to CSRM from 0.44 to 0.67 (R), \SI{14}{\percent} to \SI{38}{\percent} (EV), and \SI{62.3}{\milli\meter} to \SI{52.5}{\milli\meter} (RMSE). These results indicate that the downscaled $p_b$ product better captures the overall magnitude and phases of sea level changes at individual sites, but may still underestimate the signal amplitudes. 

\begin{figure}[!ht]
    \centering
    \begin{adjustbox}{width=18.3cm,center}
    \includegraphics[width=18.3cm]{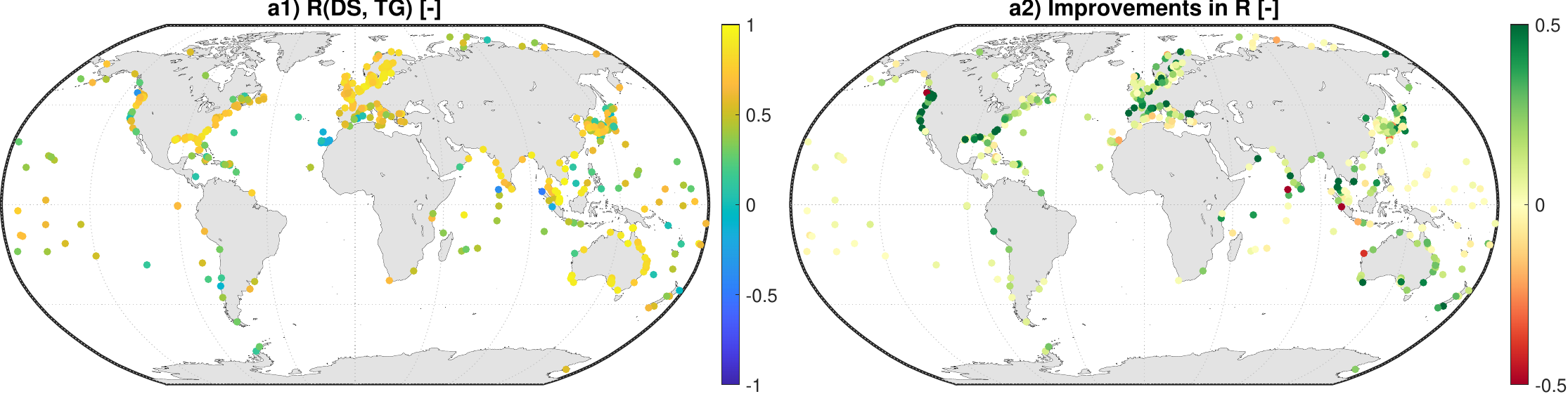}
    \end{adjustbox}
    \begin{adjustbox}{width=18.3cm,center}
    \includegraphics[width=18.3cm]{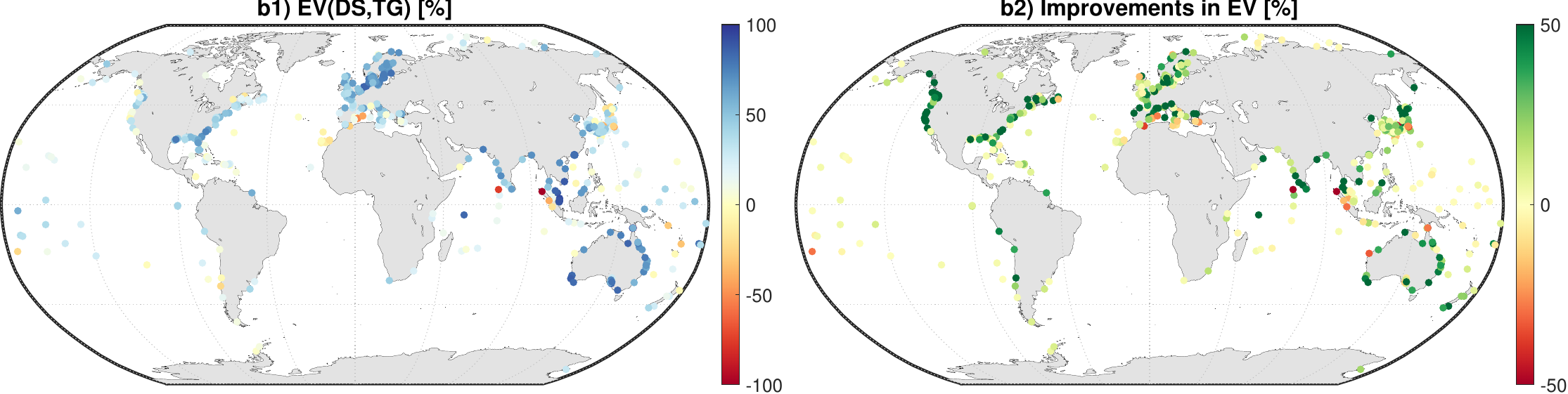}
    \end{adjustbox}
    \begin{adjustbox}{width=18.3cm,center}
    \includegraphics[width=18.3cm]{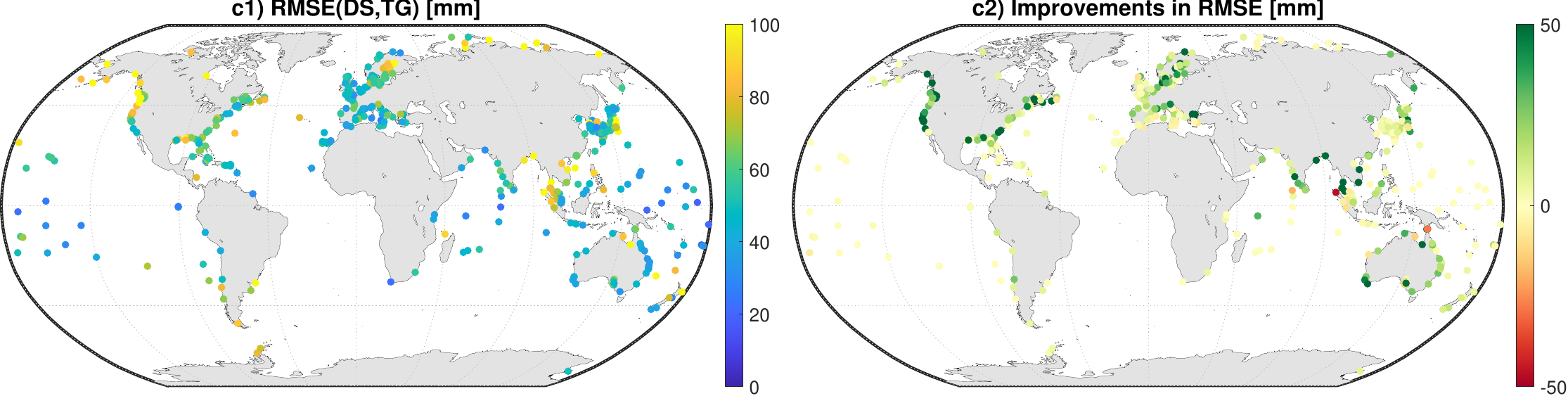}
    \end{adjustbox}
    \caption{The spatial distribution of the three evaluation metrics derived from the full time series of downscaled products (left) and the improvements compared to the ones derived from CSRM (right). The tide gauge (TG) measurements serve as references.}
    \label{fig:Eva_TG_Map_All}
\end{figure}

The spatial distributions of the three metrics derived from downscaled $p_b$ are depicted in Fig.~\ref{fig:Eva_TG_Map_All}, along with improvements compared to CSRM $p_b$. The downscaled product offers distinct improvements along most coastlines, such as North America and Europe. Evidently, our deep learning method allows the downscaled $p_b$ to better separate land and ocean signals and, therefore, improve the fidelity of coastal mass change estimates. At the stations located on the islands in the Pacific Ocean, both downscaled and CSRM $p_b$ variations are very similar since both of them have dominant annual signals and are highly correlated with nearby open-ocean signals~\citep{vinogradov2011lowfrequencySeaLevel,williams2013coherence}. Although the downscaled product provides clear improvements along the coastlines of western North America, its absolute performance in this region is inferior compared to other regions. The issue is perhaps related to large mass changes in nearby glaciated regions, leading to residual leakage errors in the CSRM $p_b$~\citep{chen2019improvedGMOM}. We also acknowledge the influence of large earthquakes on both $p_b$ and tide gauge time series, which contributes to their poor agreement near Sumatra and Japan (see supplementary material).

\begin{figure}[!tb]
    \centering
    \begin{adjustbox}{width=160mm,center}
    \includegraphics[width=160mm]{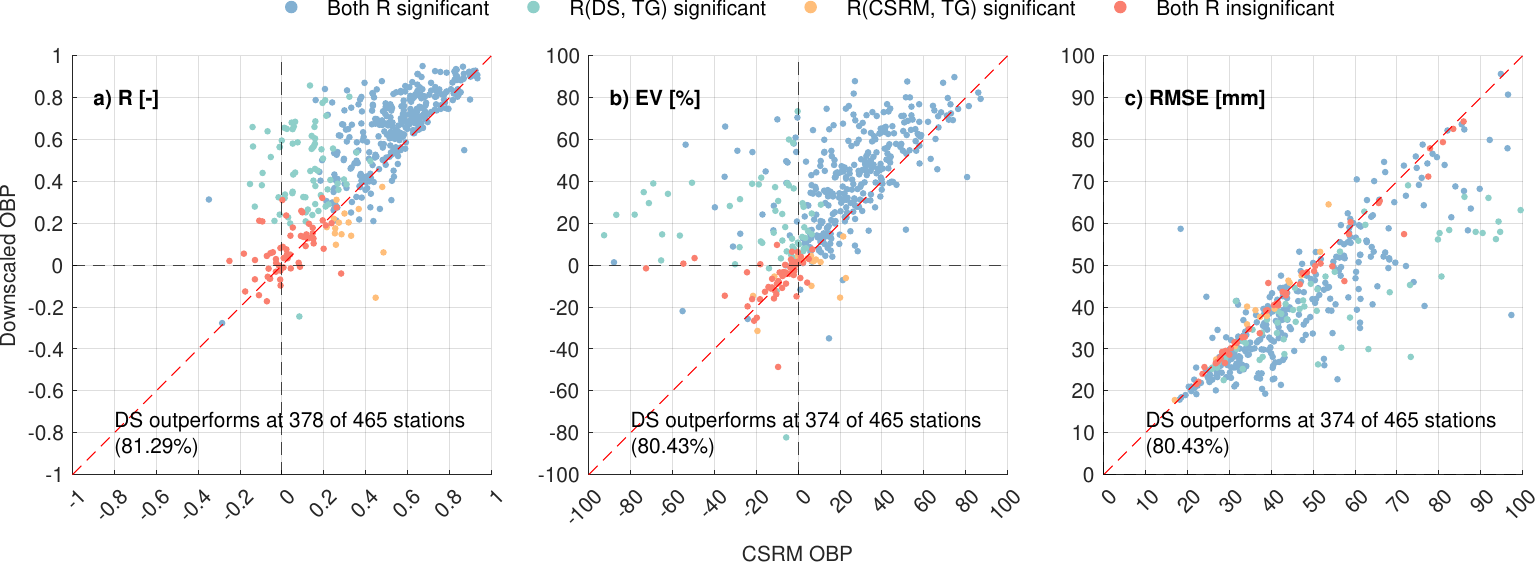}
    \end{adjustbox}
    \caption{Comparison of the intra-annual time series of downscaled $p_b$ (DS) and CSRM $p_b$ with in-situ tide gauge (TG) measurements. Three station-wise metrics are reported: a) Pearson correlations (R), b) explained variances (EV), and c) root-mean-square errors (RMSE). The statistical test on correlations is a t-test with a confidence level of 0.95.}
    \label{fig:Eva_TG_Scatter_rmTrendAnnual}
\end{figure}

\begin{figure}[ht]
    \centering
    \begin{adjustbox}{width=18.3cm,center}
    \includegraphics[width=18.3cm]{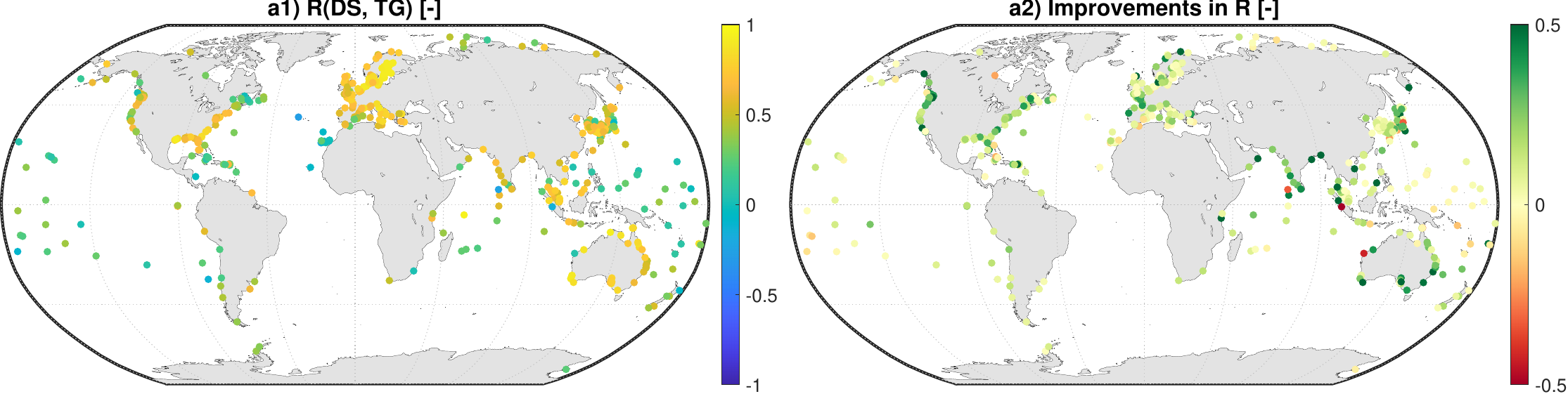}
    \end{adjustbox}
    \begin{adjustbox}{width=18.3cm,center}
    \includegraphics[width=18.3cm]{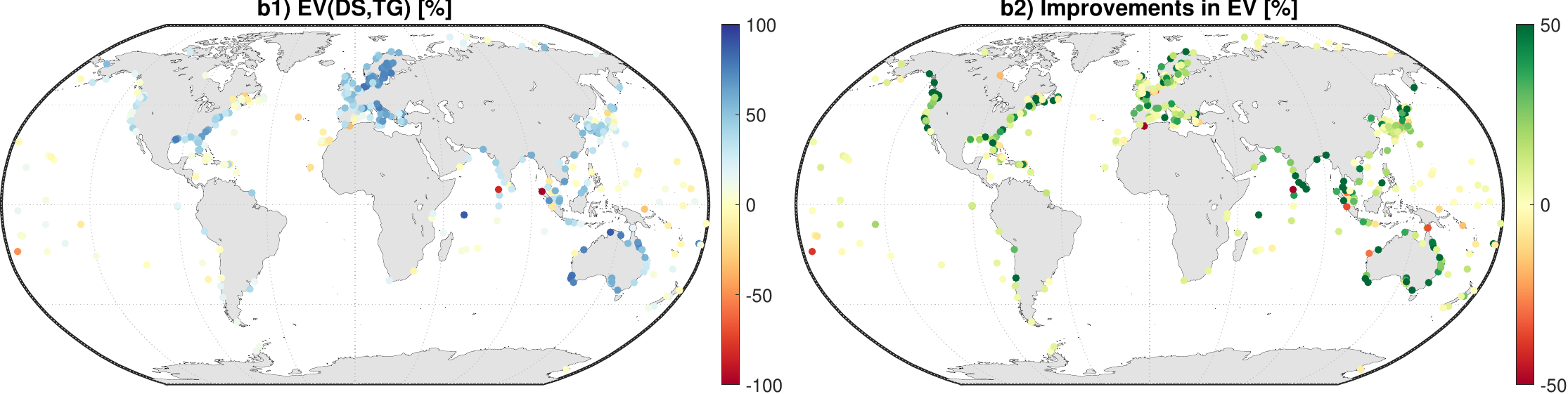}
    \end{adjustbox}
    \begin{adjustbox}{width=18.3cm,center}
    \includegraphics[width=18.3cm]{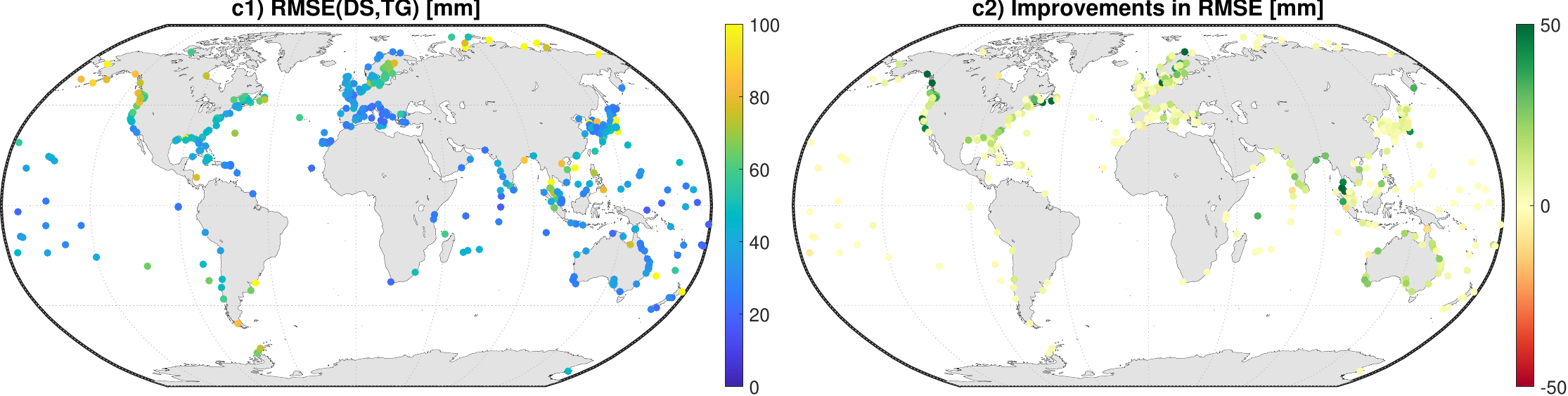}
    \end{adjustbox}
    \caption{The spatial distribution of the three evaluation metrics derived from the intra-annual time series of downscaled products (left) and the improvements compared to the ones derived from CSRM (right). The tide gauge (TG) measurements serve as references.}
    \label{fig:Eva_TG_Map_rmTrendAnnual}
\end{figure}

In order to investigate intra-annual variability, we removed the long-term trends and annual signals from all the time series and generated Fig.~\ref{fig:Eva_TG_Scatter_rmTrendAnnual} and Fig.~\ref{fig:Eva_TG_Map_rmTrendAnnual}. First, the number of stations at which the downscaled product outperforms CSRM increases slightly, from \SI{79}{\percent} to over \SI{80}{\percent}. The median R and EV slightly decrease from 0.67 and \SI{38}{\percent} to 0.61 and \SI{32}{\percent}, while the median RMSE drops from \SI{52.5}{\milli\meter} to \SI{39.3}{\milli\meter}. The results show a clear reduction in signal levels when removing the annual signals. The minor deterioration in R and EV, therefore, hints at the challenge of capturing sea level fluctuations other than the dominant annual oscillation.

The spatial distributions of the intra-annual metrics (Fig.~\ref{fig:Eva_TG_Map_rmTrendAnnual}) are similar to the ones derived from full time series. The most notable differences can be found in R and EV for islands in the Pacific Ocean, where both the downscaled and CSRM $p_b$ have dominant annual signals (Fig.~\ref{fig:OBP_Decompose}b), and local sea level variability correlates strongly with nearby ocean signals~\citep{vinogradov2011lowfrequencySeaLevel}. Therefore, the residual fluctuations after removing the annual term have relatively low signal-to-noise ratios, resulting in diminished values of R and EV. The argument also applies to CSRM $p_b$ anomalies since the downscaled $p_b$ anomalies do not show significant positive or negative changes compared to them. However, the benefits of the downscaled product are evident for regions with strong semi-annual signals, such as the coastlines of North America, Western Europe (especially the Baltic Sea), Australia, and Japan. In these regions, the reduced RMSE relative to Fig.~\ref{fig:Eva_TG_Map_All} owes to lower signal levels after removing the annual component and better agreement with tide gauge measurements. Therefore, R and EV are generally positive, with widespread improvement compared to CSRM (Fig.~\ref{fig:Eva_TG_Map_rmTrendAnnual}a,b).

\clearpage
\section{Conclusions and Outlook}
\label{sec:Conclusions and outlook}
In this study, we successfully downscaled global $p_b$ anomalies derived from monthly GRACE(-FO) products to the eddy-permitting resolution of \SI{0.25}{\degree}. Since high-quality ground truth data with global coverage are not available, we designed a self-supervised pipeline to optimize the network using the supervision signals generated by considering different constraints. Large-scale mass conservation has been demonstrated by examining spatial mean $p_b$ signals over major ocean basins, while the benefit of high-resolution information has been evaluated by comparing with seafloor pressure and coastal tide gauge measurements. The downscaled $p_b$ anomalies closely follow the GRACE(-FO) solutions at the global and basin scale with RMSE at the millimeter level. In terms of high-resolution information, the downscaled $p_b$ anomalies exhibit signal levels consistent with the monthly averaged BPR measurements and provide an average RMS reduction of \SI{2.52}{\milli\meter} over all 119 BPR sites. Along the coastlines, the downscaled $p_b$ anomalies yield better agreements with tide gauge measurements than the GRACE(-FO) $p_b$ anomalies at around \SI{80}{\percent} of 465 globally distributed stations. Our data fusion pipeline based on a self-supervised deep learning model is efficient and amenable to use by other researchers. By training a global model, we try to find the optimum in the loss landscape considering the global inputs. This optimum based on global data is usually not the best solution in a specific region since it compromises generalizability and local capacity. However, with the demonstrated global generalizability, users can easily apply our pipeline to other target regions, by training a model from scratch or considering our model as pre-trained and fine-tune it in the particular region in the context of transfer learning~\citep{weiss2016SurveyTrensferLearning}.
 
The short-scale $p_b$ information fed by the network into the GRACE(-FO) solutions should enable investigations of mass change signals beyond the effective resolution of GRACE(-FO) solutions, similar to analyses by \cite{gou2024Global} for hydrological basins. One key application would be to monitor AMOC transport variability in different latitudes from knowledge of eastern and western boundary pressures on the continental slope \citep{bingham2008determining,roussenov2008BoundaryWaveCommunication}. Further, the downscaled product carries the imprints of mesoscale eddies on $p_b$, which could be examined for trends and interannual variability~\citep{beech2022longtermEddy}. We are also envisioning benefits for Earth rotation studies, particularly efforts to improve estimates of oceanic angular momentum and the resulting excitation of Earth rotation parameters \citep{gross2003wobbles,kiani2022PM,gou2023LOD}. In general, knowledge of $p_b$ variability only constrains the mass term of this excitation, but recovery of the motion term from pressures across topographic gradients is possible under certain assumptions \citep{ponte2022GlobalOceanResponse}.

Despite the progress in methodology, we acknowledge that some of our validations, such as those against BPR measurements (Table~\ref{table:BPR}), are inconclusive. These evaluations demonstrate the ability of our downscaling pipeline to restore signal levels, but the RMS reductions given by the downscaled products do not show clear benefits compared to the original GRACE(-FO) data. Since most of the utilized BPR series are rather short (median length: 22 months), a conclusive validation requires extensions to more recent years and new measurement sites. In coastal regions, the downscaled product represents mass changes clearly better than the original GRACE(-FO) products, demonstrating its ability to resolve small-scale signals and restore signal levels.
\citep{ali2022Downscaling}. Nevertheless, the errors inherent to the GRACE(-FO) solutions~\citep{wiese2016quantifying} necessarily affect our downscaled products. New releases of the atmosphere-ocean de-aliasing product~\citep{dobslaw2017AOD1BRL06,shihora2022AOD1BRL07} or future satellite gravimetry missions~\citep{heller2023MAGIC,daras2024MAGICApplication} will likely reduce noise levels, improve the effective resolution of the derived gravity fields, and therefore enhance the fidelity of our product.

Since our method is based on level-3 products and does not have specific requirements on the input grids, it should be easily applied to other GRACE(-FO) products, ocean models, or reanalysis outputs with different spatio-temporal discretizations than the fields analyzed here. For example, we can get observation-constrained high-frequency products by applying the method to daily or weekly GRACE(-FO) solutions, which resolve $p_b$ signals to wavelengths of around \SI{1000}{\kilo\meter}~\citep{kvas2019itsg}. We may also combine the improved gravimetry products with an eddy-rich ocean model or reanalysis with a spatial resolution of $1/12^\circ$ or higher. Such downscaled products could be particularly beneficial for studying ocean dynamics and energetics on sub-seasonal and sub-monthly time scales~\citep{weijer2010almostfree,yu2018VariationsArgentineGyre,rohith2019basin,ponte2022GlobalOceanResponse}.

\section*{Data Availability}
\noindent
The downscaled products generated by this study are available to the editors and reviewers during the review process and will be released to the public together with core codes upon publication. All the other codes and intermediate datasets are available from the corresponding author upon a reasonable request. The raw data used in this study are available as follows.
CSRM: \url{https://www2.csr.utexas.edu/grace/RL06\_mascons.html} (Accessed: 18.02.2023). PSMSL: \url{https://psmsl.org/data/} (Accessed: 23.11.2023).
Ocean reanalysis products: \url{https://data.marine.copernicus.eu/product/GLOBAL\_MULTIYEAR\_PHY\_ENS\_001\_031/} (Accessed: 12.01.2023). ERA5: \url{https://cds.climate.copernicus.eu/cdsapp\#!/dataset/reanalysis-era5-single-levels-monthly-means} (Accessed: 19.05.2023).

\section*{Author contributions}
\noindent
\textbf{Junyang Gou}: Conceptualization, Methodology, Software, Validation, Formal analysis, Investigation, Data Curation, Visualization, Writing – original draft. \textbf{Lara Börger}: Software, Data curation, Writing – original draft. \textbf{Michael Schindelegger}: Conceptualization, Software, Supervision, Writing – review \& editing. \textbf{Benedikt Soja}: Conceptualization, Supervision, Writing – review \& editing.

\section*{Competing Interest}
\noindent
The authors declare no competing interests.

\section*{Acknowledgment}
\noindent
LB was supported by the German Research Foundation (DFG, Project no. 459392861), as was MS (DFG, Project no. 388296632).

\bibliographystyle{elsarticle-harv} 

\bibliography{mybibfile.bib}

\begin{thebibliography}{109}
\expandafter\ifx\csname natexlab\endcsname\relax\def\natexlab#1{#1}\fi
\providecommand{\url}[1]{\texttt{#1}}
\providecommand{\href}[2]{#2}
\providecommand{\path}[1]{#1}
\providecommand{\DOIprefix}{doi:}
\providecommand{\ArXivprefix}{arXiv:}
\providecommand{\URLprefix}{URL: }
\providecommand{\Pubmedprefix}{pmid:}
\providecommand{\doi}[1]{\href{http://dx.doi.org/#1}{\path{#1}}}
\providecommand{\Pubmed}[1]{\href{pmid:#1}{\path{#1}}}
\providecommand{\bibinfo}[2]{#2}
\ifx\xfnm\relax \def\xfnm[#1]{\unskip,\space#1}\fi
\bibitem[{Abadi et~al.(2015)Abadi, Agarwal, Barham, Brevdo, Chen, Citro, Corrado, Davis, Dean, Devin, Ghemawat, Goodfellow, Harp, Irving, Isard, Jia, Jozefowicz, Kaiser, Kudlur, Levenberg, Man\'{e}, Monga, Moore, Murray, Olah, Schuster, Shlens, Steiner, Sutskever, Talwar, Tucker, Vanhoucke, Vasudevan, Vi\'{e}gas, Vinyals, Warden, Wattenberg, Wicke, Yu and Zheng}]{tensorflow2015whitepaper}
\bibinfo{author}{Abadi, M.}, \bibinfo{author}{Agarwal, A.}, \bibinfo{author}{Barham, P.}, \bibinfo{author}{Brevdo, E.}, \bibinfo{author}{Chen, Z.}, \bibinfo{author}{Citro, C.}, \bibinfo{author}{Corrado, G.S.}, \bibinfo{author}{Davis, A.}, \bibinfo{author}{Dean, J.}, \bibinfo{author}{Devin, M.}, \bibinfo{author}{Ghemawat, S.}, \bibinfo{author}{Goodfellow, I.}, \bibinfo{author}{Harp, A.}, \bibinfo{author}{Irving, G.}, \bibinfo{author}{Isard, M.}, \bibinfo{author}{Jia, Y.}, \bibinfo{author}{Jozefowicz, R.}, \bibinfo{author}{Kaiser, L.}, \bibinfo{author}{Kudlur, M.}, \bibinfo{author}{Levenberg, J.}, \bibinfo{author}{Man\'{e}, D.}, \bibinfo{author}{Monga, R.}, \bibinfo{author}{Moore, S.}, \bibinfo{author}{Murray, D.}, \bibinfo{author}{Olah, C.}, \bibinfo{author}{Schuster, M.}, \bibinfo{author}{Shlens, J.}, \bibinfo{author}{Steiner, B.}, \bibinfo{author}{Sutskever, I.}, \bibinfo{author}{Talwar, K.}, \bibinfo{author}{Tucker, P.}, \bibinfo{author}{Vanhoucke, V.}, \bibinfo{author}{Vasudevan, V.},
  \bibinfo{author}{Vi\'{e}gas, F.}, \bibinfo{author}{Vinyals, O.}, \bibinfo{author}{Warden, P.}, \bibinfo{author}{Wattenberg, M.}, \bibinfo{author}{Wicke, M.}, \bibinfo{author}{Yu, Y.}, \bibinfo{author}{Zheng, X.}, \bibinfo{year}{2015}.
\newblock \bibinfo{title}{{TensorFlow}: Large-scale machine learning on heterogeneous systems}.
\newblock \bibinfo{note}{Software available from https://www.tensorflow.org/}.
\bibitem[{Ali et~al.(2022)Ali, Liu, Fu, Cheema, Pal, Arshad, Pham and Zhang}]{ali2022Downscaling}
\bibinfo{author}{Ali, S.}, \bibinfo{author}{Liu, D.}, \bibinfo{author}{Fu, Q.}, \bibinfo{author}{Cheema, M.J.M.}, \bibinfo{author}{Pal, S.C.}, \bibinfo{author}{Arshad, A.}, \bibinfo{author}{Pham, Q.B.}, \bibinfo{author}{Zhang, L.}, \bibinfo{year}{2022}.
\newblock \bibinfo{title}{{Constructing high-resolution groundwater drought at spatio-temporal scale using GRACE satellite data based on machine learning in the Indus Basin}}.
\newblock \bibinfo{journal}{Journal of Hydrology} \bibinfo{volume}{612}, \bibinfo{pages}{128295}.
\newblock \DOIprefix\doi{https://doi.org/10.1016/j.jhydrol.2022.128295}.
\bibitem[{Amante and Eakins(2009)}]{amanteETOPO1ArcminuteGlobal2009}
\bibinfo{author}{Amante, C.}, \bibinfo{author}{Eakins, B.}, \bibinfo{year}{2009}.
\newblock \bibinfo{title}{{ETOPO1} 1 arc-minute global relief model: Procedures, data sources and analysis. {{NOAA Technical Memorandum NESDIS NGDC-24}}}.
\newblock \DOIprefix\doi{https://doi.org/10.7289/V5C8276M}. \bibinfo{note}{last access: 6 January 2023}.
\bibitem[{Androsov et~al.(2020)Androsov, Boebel, Schr{\"o}ter, Danilov, Macrander and Ivanciu}]{androsov2020OBPmodel}
\bibinfo{author}{Androsov, A.}, \bibinfo{author}{Boebel, O.}, \bibinfo{author}{Schr{\"o}ter, J.}, \bibinfo{author}{Danilov, S.}, \bibinfo{author}{Macrander, A.}, \bibinfo{author}{Ivanciu, I.}, \bibinfo{year}{2020}.
\newblock \bibinfo{title}{{Ocean bottom pressure variability: Can it be reliably modeled?}}
\newblock \bibinfo{journal}{Journal of Geophysical Research: Oceans} \bibinfo{volume}{125}, \bibinfo{pages}{e2019JC015469}.
\newblock \DOIprefix\doi{https://doi.org/10.1029/2019JC015469}.
\bibitem[{Beech et~al.(2022)Beech, Rackow, Semmler, Danilov, Wang and Jung}]{beech2022longtermEddy}
\bibinfo{author}{Beech, N.}, \bibinfo{author}{Rackow, T.}, \bibinfo{author}{Semmler, T.}, \bibinfo{author}{Danilov, S.}, \bibinfo{author}{Wang, Q.}, \bibinfo{author}{Jung, T.}, \bibinfo{year}{2022}.
\newblock \bibinfo{title}{{Long-term evolution of ocean eddy activity in a warming world}}.
\newblock \bibinfo{journal}{Nature Climate Change} \bibinfo{volume}{12}, \bibinfo{pages}{910--917}.
\newblock \DOIprefix\doi{https://doi.org/10.1038/s41558-022-01478-3}.
\bibitem[{Bentel et~al.(2015)Bentel, Landerer and Boening}]{bentel2015monitoring}
\bibinfo{author}{Bentel, K.}, \bibinfo{author}{Landerer, F.W.}, \bibinfo{author}{Boening, C.}, \bibinfo{year}{2015}.
\newblock \bibinfo{title}{{Monitoring Atlantic overturning circulation and transport variability with GRACE-type ocean bottom pressure observations -- a sensitivity study}}.
\newblock \bibinfo{journal}{Ocean Science} \bibinfo{volume}{11}, \bibinfo{pages}{953--963}.
\newblock \DOIprefix\doi{https://doi.org/10.5194/os-11-953-2015}.
\bibitem[{Bingham and Hughes(2008a)}]{bingham2008determining}
\bibinfo{author}{Bingham, R.J.}, \bibinfo{author}{Hughes, C.W.}, \bibinfo{year}{2008}a.
\newblock \bibinfo{title}{{Determining North Atlantic meridional transport variability from pressure on the western boundary: A model investigation}}.
\newblock \bibinfo{journal}{Journal of Geophysical Research: Oceans} \bibinfo{volume}{113}.
\newblock \DOIprefix\doi{https://doi.org/10.1029/2007JC004679}.
\bibitem[{Bingham and Hughes(2008b)}]{bingham2008relationshipSLAandOBP_025}
\bibinfo{author}{Bingham, R.J.}, \bibinfo{author}{Hughes, C.W.}, \bibinfo{year}{2008}b.
\newblock \bibinfo{title}{The relationship between sea-level and bottom pressure variability in an eddy permitting ocean model}.
\newblock \bibinfo{journal}{Geophysical Research Letters} \bibinfo{volume}{35}.
\newblock \DOIprefix\doi{https://doi.org/10.1029/2007GL032662}.
\bibitem[{Bingham and Hughes(2012)}]{bingham2012local}
\bibinfo{author}{Bingham, R.J.}, \bibinfo{author}{Hughes, C.W.}, \bibinfo{year}{2012}.
\newblock \bibinfo{title}{{Local diagnostics to estimate density-induced sea level variations over topography and along coastlines}}.
\newblock \bibinfo{journal}{Journal of Geophysical Research: Oceans} \bibinfo{volume}{117}.
\newblock \DOIprefix\doi{https://doi.org/10.1029/2011JC007276}.
\bibitem[{Bourlard and Kamp(1988)}]{bourlard1988autoencoder1}
\bibinfo{author}{Bourlard, H.}, \bibinfo{author}{Kamp, Y.}, \bibinfo{year}{1988}.
\newblock \bibinfo{title}{{Auto-association by multilayer perceptrons and singular value decomposition}}.
\newblock \bibinfo{journal}{Biological cybernetics} \bibinfo{volume}{59}, \bibinfo{pages}{291--294}.
\newblock \DOIprefix\doi{https://doi.org/10.1007/BF00332918}.
\bibitem[{Börger et~al.(2023)Börger, Schindelegger, Dobslaw and Salstein}]{boergerOceanReanalysesEarthRotation2023}
\bibinfo{author}{Börger, L.}, \bibinfo{author}{Schindelegger, M.}, \bibinfo{author}{Dobslaw, H.}, \bibinfo{author}{Salstein, D.}, \bibinfo{year}{2023}.
\newblock \bibinfo{title}{{Are ocean reanalyses useful for Earth rotation research?}}
\newblock \bibinfo{journal}{Earth and Space Science} \bibinfo{volume}{10}, \bibinfo{pages}{e2022EA002700}.
\newblock \DOIprefix\doi{https://doi.org/10.1029/2022EA002700}.
\bibitem[{Chambers et~al.(2010)Chambers, Wahr, Tamisiea and Nerem}]{chambers2010GRACEOcean}
\bibinfo{author}{Chambers, D.P.}, \bibinfo{author}{Wahr, J.}, \bibinfo{author}{Tamisiea, M.E.}, \bibinfo{author}{Nerem, R.S.}, \bibinfo{year}{2010}.
\newblock \bibinfo{title}{{Ocean mass from GRACE and glacial isostatic adjustment}}.
\newblock \bibinfo{journal}{Journal of Geophysical Research: Solid Earth} \bibinfo{volume}{115}.
\newblock \DOIprefix\doi{https://doi.org/10.1029/2010JB007530}.
\bibitem[{Chen et~al.(2022)Chen, Cazenave, Dahle, Llovel, Panet, Pfeffer and Moreira}]{chen2022GRACEreview}
\bibinfo{author}{Chen, J.}, \bibinfo{author}{Cazenave, A.}, \bibinfo{author}{Dahle, C.}, \bibinfo{author}{Llovel, W.}, \bibinfo{author}{Panet, I.}, \bibinfo{author}{Pfeffer, J.}, \bibinfo{author}{Moreira, L.}, \bibinfo{year}{2022}.
\newblock \bibinfo{title}{{Applications and challenges of GRACE and GRACE Follow-On satellite gravimetry}}.
\newblock \bibinfo{journal}{Surveys in Geophysics} , \bibinfo{pages}{1--41}\DOIprefix\doi{https://doi.org/10.1007/s10712-021-09685-x}.
\bibitem[{Chen et~al.(2019a)Chen, Tapley, Seo, Wilson and Ries}]{chen2019GRACEOcean}
\bibinfo{author}{Chen, J.}, \bibinfo{author}{Tapley, B.}, \bibinfo{author}{Seo, K.W.}, \bibinfo{author}{Wilson, C.}, \bibinfo{author}{Ries, J.}, \bibinfo{year}{2019}a.
\newblock \bibinfo{title}{{Improved quantification of global mean ocean mass change using GRACE satellite gravimetry measurements}}.
\newblock \bibinfo{journal}{Geophysical Research Letters} \bibinfo{volume}{46}, \bibinfo{pages}{13984--13991}.
\newblock \DOIprefix\doi{https://doi.org/10.1029/2019GL085519}.
\bibitem[{Chen et~al.(2019b)Chen, Tapley, Seo, Wilson and Ries}]{chen2019improvedGMOM}
\bibinfo{author}{Chen, J.}, \bibinfo{author}{Tapley, B.}, \bibinfo{author}{Seo, K.W.}, \bibinfo{author}{Wilson, C.}, \bibinfo{author}{Ries, J.}, \bibinfo{year}{2019}b.
\newblock \bibinfo{title}{{Improved quantification of global mean ocean mass change using GRACE satellite gravimetry measurements}}.
\newblock \bibinfo{journal}{Geophysical Research Letters} \bibinfo{volume}{46}, \bibinfo{pages}{13984--13991}.
\newblock \DOIprefix\doi{https://doi.org/10.1029/2019GL085519}.
\bibitem[{Chen et~al.(2007)Chen, Wilson, Tapley and Grand}]{chen2007grace4earthquake}
\bibinfo{author}{Chen, J.L.}, \bibinfo{author}{Wilson, C.R.}, \bibinfo{author}{Tapley, B.D.}, \bibinfo{author}{Grand, S.}, \bibinfo{year}{2007}.
\newblock \bibinfo{title}{{GRACE detects coseismic and postseismic deformation from the Sumatra-Andaman earthquake}}.
\newblock \bibinfo{journal}{Geophysical Research Letters} \bibinfo{volume}{34}.
\newblock \DOIprefix\doi{https://doi.org/10.1029/2007GL030356}.
\bibitem[{Church et~al.(2013)Church, Clark, Cazenave, Gregory, Jevrejeva, Levermann, Merrifield, Milne, Nerem, Nunn et~al.}]{church2013SeaLevelChange}
\bibinfo{author}{Church, J.A.}, \bibinfo{author}{Clark, P.U.}, \bibinfo{author}{Cazenave, A.}, \bibinfo{author}{Gregory, J.M.}, \bibinfo{author}{Jevrejeva, S.}, \bibinfo{author}{Levermann, A.}, \bibinfo{author}{Merrifield, M.A.}, \bibinfo{author}{Milne, G.A.}, \bibinfo{author}{Nerem, R.S.}, \bibinfo{author}{Nunn, P.D.}, et~al., \bibinfo{year}{2013}.
\newblock \bibinfo{title}{Sea level change}.
\newblock \bibinfo{type}{Technical Report}. PM Cambridge University Press.
\bibitem[{{Copernicus Marine Service}(2019)}]{cmemsGlobalOceanEnsemble2019}
\bibinfo{author}{{Copernicus Marine Service}}, \bibinfo{year}{2019}.
\newblock \bibinfo{title}{Global ocean ensemble physics reanalysis}.
\newblock \DOIprefix\doi{https://doi.org/10.48670/MOI-00024}. \bibinfo{note}{last access: 19 August 2023}.
\bibitem[{Coulson et~al.(2022)Coulson, Dangendorf, Mitrovica, Tamisiea, Pan and Sandwell}]{coulson2022IceFingerprints}
\bibinfo{author}{Coulson, S.}, \bibinfo{author}{Dangendorf, S.}, \bibinfo{author}{Mitrovica, J.X.}, \bibinfo{author}{Tamisiea, M.E.}, \bibinfo{author}{Pan, L.}, \bibinfo{author}{Sandwell, D.T.}, \bibinfo{year}{2022}.
\newblock \bibinfo{title}{{A detection of the sea level fingerprint of Greenland Ice Sheet melt}}.
\newblock \bibinfo{journal}{Science} \bibinfo{volume}{377}, \bibinfo{pages}{1550--1554}.
\newblock \DOIprefix\doi{https://doi.org/10.1126/science.abo0926}.
\bibitem[{Daras et~al.(2024)Daras, March, Pail, Hughes, Braitenberg, G{\"u}ntner, Eicker, Wouters, Heller-Kaikov, Pivetta et~al.}]{daras2024MAGICApplication}
\bibinfo{author}{Daras, I.}, \bibinfo{author}{March, G.}, \bibinfo{author}{Pail, R.}, \bibinfo{author}{Hughes, C.}, \bibinfo{author}{Braitenberg, C.}, \bibinfo{author}{G{\"u}ntner, A.}, \bibinfo{author}{Eicker, A.}, \bibinfo{author}{Wouters, B.}, \bibinfo{author}{Heller-Kaikov, B.}, \bibinfo{author}{Pivetta, T.}, et~al., \bibinfo{year}{2024}.
\newblock \bibinfo{title}{{Mass-change And Geosciences International Constellation (MAGIC) expected impact on science and applications}}.
\newblock \bibinfo{journal}{Geophysical Journal International} \bibinfo{volume}{236}, \bibinfo{pages}{1288--1308}.
\newblock \DOIprefix\doi{https://doi.org/10.1093/gji/ggad472}.
\bibitem[{Dee et~al.(2011)Dee, Uppala, Simmons, Berrisford, Poli, Kobayashi, Andrae, Balmaseda, Balsamo, Bauer, Bechtold, Beljaars, {van de Berg}, Bidlot, Bormann, Delsol, Dragani, Fuentes, Geer, Haimberger, Healy, Hersbach, H{\'o}lm, Isaksen, K{\aa}llberg, K{\"o}hler, Matricardi, McNally, {Monge-Sanz}, Morcrette, Park, Peubey, {de Rosnay}, Tavolato, Th{\'e}paut and Vitart}]{deeERAInterimConfiguration2011}
\bibinfo{author}{Dee, D.P.}, \bibinfo{author}{Uppala, S.M.}, \bibinfo{author}{Simmons, A.J.}, \bibinfo{author}{Berrisford, P.}, \bibinfo{author}{Poli, P.}, \bibinfo{author}{Kobayashi, S.}, \bibinfo{author}{Andrae, U.}, \bibinfo{author}{Balmaseda, M.A.}, \bibinfo{author}{Balsamo, G.}, \bibinfo{author}{Bauer, P.}, \bibinfo{author}{Bechtold, P.}, \bibinfo{author}{Beljaars, A.C.M.}, \bibinfo{author}{{van de Berg}, L.}, \bibinfo{author}{Bidlot, J.}, \bibinfo{author}{Bormann, N.}, \bibinfo{author}{Delsol, C.}, \bibinfo{author}{Dragani, R.}, \bibinfo{author}{Fuentes, M.}, \bibinfo{author}{Geer, A.J.}, \bibinfo{author}{Haimberger, L.}, \bibinfo{author}{Healy, S.B.}, \bibinfo{author}{Hersbach, H.}, \bibinfo{author}{H{\'o}lm, E.V.}, \bibinfo{author}{Isaksen, L.}, \bibinfo{author}{K{\aa}llberg, P.}, \bibinfo{author}{K{\"o}hler, M.}, \bibinfo{author}{Matricardi, M.}, \bibinfo{author}{McNally, A.P.}, \bibinfo{author}{{Monge-Sanz}, B.M.}, \bibinfo{author}{Morcrette, J.J.}, \bibinfo{author}{Park, B.K.},
  \bibinfo{author}{Peubey, C.}, \bibinfo{author}{{de Rosnay}, P.}, \bibinfo{author}{Tavolato, C.}, \bibinfo{author}{Th{\'e}paut, J.N.}, \bibinfo{author}{Vitart, F.}, \bibinfo{year}{2011}.
\newblock \bibinfo{title}{{The ERA-Interim reanalysis: configuration and performance of the data assimilation system}}.
\newblock \bibinfo{journal}{Q.J.R. Meteorol. Soc.} \bibinfo{volume}{137}, \bibinfo{pages}{553--597}.
\newblock \DOIprefix\doi{https://doi.org/10.1002/qj.828}.
\bibitem[{Delman and Landerer(2022)}]{delman2022downscalingOBP}
\bibinfo{author}{Delman, A.}, \bibinfo{author}{Landerer, F.}, \bibinfo{year}{2022}.
\newblock \bibinfo{title}{{Downscaling satellite-based estimates of ocean bottom pressure for tracking deep ocean mass transport}}.
\newblock \bibinfo{journal}{Remote Sensing} \bibinfo{volume}{14}, \bibinfo{pages}{1764}.
\newblock \DOIprefix\doi{https://doi.org/10.3390/rs14071764}.
\bibitem[{Desportes et~al.(2019)Desportes, Garric, R{\'e}gnier, Dr{\'e}villon, Parent, Garric, Drillet, Masina, Storto, Mirouze, Cipollone, Zuo, Balmaseda, Peterson, Wood, Jackson, Mulet, Greiner and Gounou}]{desportesQID2019}
\bibinfo{author}{Desportes, C.}, \bibinfo{author}{Garric, G.}, \bibinfo{author}{R{\'e}gnier, C.}, \bibinfo{author}{Dr{\'e}villon, M.}, \bibinfo{author}{Parent, L.}, \bibinfo{author}{Garric, G.}, \bibinfo{author}{Drillet, Y.}, \bibinfo{author}{Masina, S.}, \bibinfo{author}{Storto, A.}, \bibinfo{author}{Mirouze, I.}, \bibinfo{author}{Cipollone, A.}, \bibinfo{author}{Zuo, H.}, \bibinfo{author}{Balmaseda, M.}, \bibinfo{author}{Peterson, D.}, \bibinfo{author}{Wood, R.}, \bibinfo{author}{Jackson, L.}, \bibinfo{author}{Mulet, S.}, \bibinfo{author}{Greiner, E.}, \bibinfo{author}{Gounou, A.}, \bibinfo{year}{2019}.
\newblock \bibinfo{title}{Quality information document for global ocean reanalysis multi-model ensemble products {GREP GLOBAL-REANALYSIS-PHY-001-03}}.
\newblock \bibinfo{publisher}{{Copernicus Marine Environment Monitoring Service}}.
\bibitem[{Dobslaw et~al.(2017)Dobslaw, Bergmann-Wolf, Dill, Poropat, Thomas, Dahle, Esselborn, K{\"o}nig and Flechtner}]{dobslaw2017AOD1BRL06}
\bibinfo{author}{Dobslaw, H.}, \bibinfo{author}{Bergmann-Wolf, I.}, \bibinfo{author}{Dill, R.}, \bibinfo{author}{Poropat, L.}, \bibinfo{author}{Thomas, M.}, \bibinfo{author}{Dahle, C.}, \bibinfo{author}{Esselborn, S.}, \bibinfo{author}{K{\"o}nig, R.}, \bibinfo{author}{Flechtner, F.}, \bibinfo{year}{2017}.
\newblock \bibinfo{title}{{A new high-resolution model of non-tidal atmosphere and ocean mass variability for de-aliasing of satellite gravity observations: AOD1B RL06}}.
\newblock \bibinfo{journal}{Geophysical Journal International} \bibinfo{volume}{211}, \bibinfo{pages}{263--269}.
\newblock \DOIprefix\doi{https://doi.org/10.1093/gji/ggx302}.
\bibitem[{Egbert and Erofeeva(2002)}]{Egbert2002}
\bibinfo{author}{Egbert, G.D.}, \bibinfo{author}{Erofeeva, S.Y.}, \bibinfo{year}{2002}.
\newblock \bibinfo{title}{Efficient inverse modeling of barotropic ocean tides}.
\newblock \bibinfo{journal}{Journal of Atmospheric and Oceanic Technology} \bibinfo{volume}{19}, \bibinfo{pages}{183--204}.
\newblock \DOIprefix\doi{https://doi.org/10.1175/1520-0426(2002)019<0183:EIMOBO>2.0.CO;2}.
\bibitem[{Elipot et~al.(2013)Elipot, Hughes, Olhede and Toole}]{elipot2013coherence}
\bibinfo{author}{Elipot, S.}, \bibinfo{author}{Hughes, C.}, \bibinfo{author}{Olhede, S.}, \bibinfo{author}{Toole, J.}, \bibinfo{year}{2013}.
\newblock \bibinfo{title}{{Coherence of western boundary pressure at the RAPID WAVE array: Boundary wave adjustments or deep western boundary current advection?}}
\newblock \bibinfo{journal}{Journal of Physical oceanography} \bibinfo{volume}{43}, \bibinfo{pages}{744--765}.
\newblock \DOIprefix\doi{https://doi.org/10.1175/JPO-D-12-067.1}.
\bibitem[{{\relax Flanders Marine Institute}(2021)}]{GOAS2021}
\bibinfo{author}{{\relax Flanders Marine Institute}}, \bibinfo{year}{2021}.
\newblock \bibinfo{title}{{MS Windows NT} kernel description}.
\newblock \DOIprefix\doi{https://doi.org/10.14284/542}. \bibinfo{note}{{Accessed: 15.03.2023}}.
\bibitem[{Forget et~al.(2015)Forget, Campin, Heimbach, Hill, Ponte and Wunsch}]{forget2015ecco4}
\bibinfo{author}{Forget, G.}, \bibinfo{author}{Campin, J.M.}, \bibinfo{author}{Heimbach, P.}, \bibinfo{author}{Hill, C.}, \bibinfo{author}{Ponte, R.}, \bibinfo{author}{Wunsch, C.}, \bibinfo{year}{2015}.
\newblock \bibinfo{title}{{ECCO version 4: An integrated framework for non-linear inverse modeling and global ocean state estimation}}.
\newblock \bibinfo{journal}{Geoscientific Model Development} \bibinfo{volume}{8}, \bibinfo{pages}{3071--3104}.
\newblock \DOIprefix\doi{https://doi.org/10.5194/gmd-8-3071-2015}.
\bibitem[{Fox-Kemper et~al.(2019)Fox-Kemper, Adcroft, B{\"o}ning, Chassignet, Curchitser, Danabasoglu, Eden, England, Gerdes, Greatbatch et~al.}]{fox2019challenges}
\bibinfo{author}{Fox-Kemper, B.}, \bibinfo{author}{Adcroft, A.}, \bibinfo{author}{B{\"o}ning, C.W.}, \bibinfo{author}{Chassignet, E.P.}, \bibinfo{author}{Curchitser, E.}, \bibinfo{author}{Danabasoglu, G.}, \bibinfo{author}{Eden, C.}, \bibinfo{author}{England, M.H.}, \bibinfo{author}{Gerdes, R.}, \bibinfo{author}{Greatbatch, R.J.}, et~al., \bibinfo{year}{2019}.
\newblock \bibinfo{title}{Challenges and prospects in ocean circulation models}.
\newblock \bibinfo{journal}{Frontiers in Marine Science} \bibinfo{volume}{6}, \bibinfo{pages}{65}.
\newblock \DOIprefix\doi{https://doi.org/10.3389/fmars.2019.00065}.
\bibitem[{Ghobadi-Far et~al.(2020)Ghobadi-Far, Han, Allgeyer, Tregoning, Sauber, Behzadpour, Mayer-G{\"u}rr, Sneeuw and Okal}]{ghobadi2020grace4earthquakes}
\bibinfo{author}{Ghobadi-Far, K.}, \bibinfo{author}{Han, S.C.}, \bibinfo{author}{Allgeyer, S.}, \bibinfo{author}{Tregoning, P.}, \bibinfo{author}{Sauber, J.}, \bibinfo{author}{Behzadpour, S.}, \bibinfo{author}{Mayer-G{\"u}rr, T.}, \bibinfo{author}{Sneeuw, N.}, \bibinfo{author}{Okal, E.}, \bibinfo{year}{2020}.
\newblock \bibinfo{title}{{GRACE gravitational measurements of tsunamis after the 2004, 2010, and 2011 great earthquakes}}.
\newblock \bibinfo{journal}{Journal of Geodesy} \bibinfo{volume}{94}, \bibinfo{pages}{1--9}.
\newblock \DOIprefix\doi{https://doi.org/10.1007/s00190-020-01395-3}.
\bibitem[{Good et~al.(2013)Good, Martin and Rayner}]{good2013EN4}
\bibinfo{author}{Good, S.A.}, \bibinfo{author}{Martin, M.J.}, \bibinfo{author}{Rayner, N.A.}, \bibinfo{year}{2013}.
\newblock \bibinfo{title}{{EN4: Quality controlled ocean temperature and salinity profiles and monthly objective analyses with uncertainty estimates}}.
\newblock \bibinfo{journal}{Journal of Geophysical Research: Oceans} \bibinfo{volume}{118}, \bibinfo{pages}{6704--6716}.
\newblock \DOIprefix\doi{https://doi.org/10.1002/2013JC009067}.
\bibitem[{Goodfellow et~al.(2016)Goodfellow, Bengio and Courville}]{goodfellow2016deeplearning}
\bibinfo{author}{Goodfellow, I.}, \bibinfo{author}{Bengio, Y.}, \bibinfo{author}{Courville, A.}, \bibinfo{year}{2016}.
\newblock \bibinfo{title}{{Deep learning}}.
\newblock \bibinfo{publisher}{MIT press}.
\bibitem[{Gou et~al.(2023)Gou, Kiani~Shahvandi, Hohensinn and Soja}]{gou2023LOD}
\bibinfo{author}{Gou, J.}, \bibinfo{author}{Kiani~Shahvandi, M.}, \bibinfo{author}{Hohensinn, R.}, \bibinfo{author}{Soja, B.}, \bibinfo{year}{2023}.
\newblock \bibinfo{title}{{Ultra-short-term prediction of LOD using LSTM neural networks}}.
\newblock \bibinfo{journal}{Journal of Geodesy} \bibinfo{volume}{97}, \bibinfo{pages}{52}.
\newblock \DOIprefix\doi{https://doi.org/10.1007/s00190-023-01745-x}.
\bibitem[{Gou and Soja(2024)}]{gou2024Global}
\bibinfo{author}{Gou, J.}, \bibinfo{author}{Soja, B.}, \bibinfo{year}{2024}.
\newblock \bibinfo{title}{Global high-resolution total water storage anomalies from self-supervised data assimilation using deep learning algorithms}.
\newblock \bibinfo{journal}{Nature Water} \bibinfo{volume}{2}, \bibinfo{pages}{139--150}.
\newblock \DOIprefix\doi{https://doi.org/10.1038/s44221-024-00194-w}.
\bibitem[{Gross et~al.(2003)Gross, Fukumori and Menemenlis}]{gross2003wobbles}
\bibinfo{author}{Gross, R.S.}, \bibinfo{author}{Fukumori, I.}, \bibinfo{author}{Menemenlis, D.}, \bibinfo{year}{2003}.
\newblock \bibinfo{title}{{Atmospheric and oceanic excitation of the Earth's wobbles during 1980--2000}}.
\newblock \bibinfo{journal}{Journal of Geophysical Research: Solid Earth} \bibinfo{volume}{108}, \bibinfo{pages}{2370}.
\newblock \DOIprefix\doi{https://doi.org/10.1029/2002JB002143}.
\bibitem[{He et~al.(2016)He, Zhang, Ren and Sun}]{He2016Resnet}
\bibinfo{author}{He, K.}, \bibinfo{author}{Zhang, X.}, \bibinfo{author}{Ren, S.}, \bibinfo{author}{Sun, J.}, \bibinfo{year}{2016}.
\newblock \bibinfo{title}{Deep residual learning for image recognition}, in: \bibinfo{booktitle}{Proceedings of the IEEE Conference on Computer Vision and Pattern Recognition (CVPR)}.
\bibitem[{Heller-Kaikov et~al.(2023)Heller-Kaikov, Pail and Daras}]{heller2023MAGIC}
\bibinfo{author}{Heller-Kaikov, B.}, \bibinfo{author}{Pail, R.}, \bibinfo{author}{Daras, I.}, \bibinfo{year}{2023}.
\newblock \bibinfo{title}{{Mission design aspects for the mass change and geoscience international constellation (MAGIC)}}.
\newblock \bibinfo{journal}{Geophysical Journal International} \bibinfo{volume}{235}, \bibinfo{pages}{718--735}.
\newblock \DOIprefix\doi{https://doi.org/10.1093/gji/ggad266}.
\bibitem[{Hersbach et~al.(2020)Hersbach, Bell, Berrisford, Hirahara, Hor{\'a}nyi, Mu{\~n}oz-Sabater, Nicolas, Peubey, Radu, Schepers et~al.}]{hersbach2020ERA5}
\bibinfo{author}{Hersbach, H.}, \bibinfo{author}{Bell, B.}, \bibinfo{author}{Berrisford, P.}, \bibinfo{author}{Hirahara, S.}, \bibinfo{author}{Hor{\'a}nyi, A.}, \bibinfo{author}{Mu{\~n}oz-Sabater, J.}, \bibinfo{author}{Nicolas, J.}, \bibinfo{author}{Peubey, C.}, \bibinfo{author}{Radu, R.}, \bibinfo{author}{Schepers, D.}, et~al., \bibinfo{year}{2020}.
\newblock \bibinfo{title}{{The ERA5 global reanalysis}}.
\newblock \bibinfo{journal}{Quarterly Journal of the Royal Meteorological Society} \bibinfo{volume}{146}, \bibinfo{pages}{1999--2049}.
\newblock \DOIprefix\doi{https://doi.org/10.1002/qj.3803}.
\bibitem[{Hinton and Salakhutdinov(2006)}]{Hinton2006autoencoder2}
\bibinfo{author}{Hinton, G.E.}, \bibinfo{author}{Salakhutdinov, R.R.}, \bibinfo{year}{2006}.
\newblock \bibinfo{title}{Reducing the dimensionality of data with neural networks}.
\newblock \bibinfo{journal}{Science} \bibinfo{volume}{313}, \bibinfo{pages}{504--507}.
\newblock \DOIprefix\doi{https://doi.org/10.1126/science.1127647}.
\bibitem[{Holgate et~al.(2013)Holgate, Matthews, Woodworth, Rickards, Tamisiea, Bradshaw, Foden, Gordon, Jevrejeva and Pugh}]{holgate2013PSMSL}
\bibinfo{author}{Holgate, S.J.}, \bibinfo{author}{Matthews, A.}, \bibinfo{author}{Woodworth, P.L.}, \bibinfo{author}{Rickards, L.J.}, \bibinfo{author}{Tamisiea, M.E.}, \bibinfo{author}{Bradshaw, E.}, \bibinfo{author}{Foden, P.R.}, \bibinfo{author}{Gordon, K.M.}, \bibinfo{author}{Jevrejeva, S.}, \bibinfo{author}{Pugh, J.}, \bibinfo{year}{2013}.
\newblock \bibinfo{title}{{New data systems and products at the permanent service for mean sea level}}.
\newblock \bibinfo{journal}{Journal of Coastal Research} \bibinfo{volume}{29}, \bibinfo{pages}{493--504}.
\newblock \DOIprefix\doi{https://doi.org/10.2112/JCOASTRES-D-12-00175.1}.
\bibitem[{Hsu and Velicogna(2017)}]{hsu2017IceFingerprints}
\bibinfo{author}{Hsu, C.W.}, \bibinfo{author}{Velicogna, I.}, \bibinfo{year}{2017}.
\newblock \bibinfo{title}{{Detection of sea level fingerprints derived from GRACE gravity data}}.
\newblock \bibinfo{journal}{Geophysical Research Letters} \bibinfo{volume}{44}, \bibinfo{pages}{8953--8961}.
\newblock \DOIprefix\doi{https://doi.org/10.1002/2017GL074070}.
\bibitem[{Hughes et~al.(2018)Hughes, Williams, Blaker, Coward and Stepanov}]{hughes2018window2deepocean}
\bibinfo{author}{Hughes, C.W.}, \bibinfo{author}{Williams, J.}, \bibinfo{author}{Blaker, A.}, \bibinfo{author}{Coward, A.}, \bibinfo{author}{Stepanov, V.}, \bibinfo{year}{2018}.
\newblock \bibinfo{title}{{A window on the deep ocean: the special value of ocean bottom pressure for monitoring the large-scale, deep-ocean circulation}}.
\newblock \bibinfo{journal}{Progress in Oceanography} \bibinfo{volume}{161}, \bibinfo{pages}{19--46}.
\newblock \DOIprefix\doi{https://doi.org/10.1016/j.pocean.2018.01.011}.
\bibitem[{Ioffe and Szegedy(2015)}]{ioffe2015BatchNorm}
\bibinfo{author}{Ioffe, S.}, \bibinfo{author}{Szegedy, C.}, \bibinfo{year}{2015}.
\newblock \bibinfo{title}{{Batch normalization: Accelerating deep network training by reducing internal covariate shift}}, in: \bibinfo{editor}{Bach, F.}, \bibinfo{editor}{Blei, D.} (Eds.), \bibinfo{booktitle}{Proceedings of the 32nd International Conference on Machine Learning}, \bibinfo{publisher}{PMLR}. pp. \bibinfo{pages}{448--456}.
\bibitem[{Irrgang et~al.(2021)Irrgang, Boers, Sonnewald, Barnes, Kadow, Staneva and Saynisch-Wagner}]{irrgang2021towards}
\bibinfo{author}{Irrgang, C.}, \bibinfo{author}{Boers, N.}, \bibinfo{author}{Sonnewald, M.}, \bibinfo{author}{Barnes, E.A.}, \bibinfo{author}{Kadow, C.}, \bibinfo{author}{Staneva, J.}, \bibinfo{author}{Saynisch-Wagner, J.}, \bibinfo{year}{2021}.
\newblock \bibinfo{title}{{Towards neural Earth system modelling by integrating artificial intelligence in Earth system science}}.
\newblock \bibinfo{journal}{Nature Machine Intelligence} \bibinfo{volume}{3}, \bibinfo{pages}{667--674}.
\newblock \DOIprefix\doi{https://doi.org/10.1038/s42256-021-00374-3}.
\bibitem[{Irrgang et~al.(2020)Irrgang, Saynisch-Wagner, Dill, Boergens and Thomas}]{irrgang2020self}
\bibinfo{author}{Irrgang, C.}, \bibinfo{author}{Saynisch-Wagner, J.}, \bibinfo{author}{Dill, R.}, \bibinfo{author}{Boergens, E.}, \bibinfo{author}{Thomas, M.}, \bibinfo{year}{2020}.
\newblock \bibinfo{title}{{Self-validating deep learning for recovering terrestrial water storage from gravity and altimetry measurements}}.
\newblock \bibinfo{journal}{Geophysical Research Letters} \bibinfo{volume}{47}, \bibinfo{pages}{e2020GL089258}.
\newblock \DOIprefix\doi{https://doi.org/10.1029/2020GL089258}.
\bibitem[{Johnson and Chambers(2013)}]{johnson13seasonal}
\bibinfo{author}{Johnson, G.C.}, \bibinfo{author}{Chambers, D.P.}, \bibinfo{year}{2013}.
\newblock \bibinfo{title}{Ocean bottom pressure seasonal cycles and decadal trends from {GRACE} {R}elease-05: {O}cean circulation implications}.
\newblock \bibinfo{journal}{Journal of Geophysical Research: Oceans} \bibinfo{volume}{118}, \bibinfo{pages}{4228--4240}.
\newblock \DOIprefix\doi{https://doi.org/10.1002/jgrc.20307}.
\bibitem[{Jord{\`a} and Gomis(2013)}]{jorda2013StericvsMass}
\bibinfo{author}{Jord{\`a}, G.}, \bibinfo{author}{Gomis, D.}, \bibinfo{year}{2013}.
\newblock \bibinfo{title}{{On the interpretation of the steric and mass components of sea level variability: The case of the Mediterranean basin}}.
\newblock \bibinfo{journal}{Journal of Geophysical Research: Oceans} \bibinfo{volume}{118}, \bibinfo{pages}{953--963}.
\newblock \DOIprefix\doi{https://doi.org/10.1002/jgrc.20060}.
\bibitem[{Kiani~Shahvandi et~al.(2022)Kiani~Shahvandi, Schartner and Soja}]{kiani2022PM}
\bibinfo{author}{Kiani~Shahvandi, M.}, \bibinfo{author}{Schartner, M.}, \bibinfo{author}{Soja, B.}, \bibinfo{year}{2022}.
\newblock \bibinfo{title}{{Neural ODE differential learning and its application in polar motion prediction}}.
\newblock \bibinfo{journal}{Journal of Geophysical Research: Solid Earth} \bibinfo{volume}{127}, \bibinfo{pages}{e2022JB024775}.
\newblock \DOIprefix\doi{https://doi.org/10.1029/2022JB024775}.
\bibitem[{Kingma and Ba(2014)}]{kingma2014Adam}
\bibinfo{author}{Kingma, D.P.}, \bibinfo{author}{Ba, J.}, \bibinfo{year}{2014}.
\newblock \bibinfo{title}{{Adam: A method for stochastic optimization}}.
\newblock \bibinfo{journal}{arXiv preprint arXiv:1412.6980} \DOIprefix\doi{https://doi.org/10.48550/arXiv.1412.6980}.
\bibitem[{K{\"o}hl et~al.(2012)K{\"o}hl, Siegismund and Stammer}]{kohl2012impactOBPassimilation}
\bibinfo{author}{K{\"o}hl, A.}, \bibinfo{author}{Siegismund, F.}, \bibinfo{author}{Stammer, D.}, \bibinfo{year}{2012}.
\newblock \bibinfo{title}{{Impact of assimilating bottom pressure anomalies from GRACE on ocean circulation estimates}}.
\newblock \bibinfo{journal}{Journal of Geophysical Research: Oceans} \bibinfo{volume}{117}.
\newblock \DOIprefix\doi{https://doi.org/10.1029/2011JC007623}.
\bibitem[{Kvas et~al.(2019)Kvas, Behzadpour, Ellmer, Klinger, Strasser, Zehentner and Mayer-G{\"u}rr}]{kvas2019itsg}
\bibinfo{author}{Kvas, A.}, \bibinfo{author}{Behzadpour, S.}, \bibinfo{author}{Ellmer, M.}, \bibinfo{author}{Klinger, B.}, \bibinfo{author}{Strasser, S.}, \bibinfo{author}{Zehentner, N.}, \bibinfo{author}{Mayer-G{\"u}rr, T.}, \bibinfo{year}{2019}.
\newblock \bibinfo{title}{{ITSG-Grace2018: Overview and evaluation of a new GRACE-only gravity field time series}}.
\newblock \bibinfo{journal}{Journal of Geophysical Research: Solid Earth} \bibinfo{volume}{124}, \bibinfo{pages}{9332--9344}.
\newblock \DOIprefix\doi{https://doi.org/10.1029/2019JB017415}.
\bibitem[{Landerer and Cooley(2021)}]{JPL2021GRACEL3handbook}
\bibinfo{author}{Landerer, F.W.}, \bibinfo{author}{Cooley, S.S.}, \bibinfo{year}{2021}.
\newblock \bibinfo{title}{{Gravity Recovery and Climate Experiment Follow-On (GRACE-FO) Level-3 data product user handbook}}.
\newblock \URLprefix \url{https://podaac-tools.jpl.nasa.gov/drive/files/allData/gracefo/docs/GRACE-FO_L3_Handbook_JPL.pdf}.
\bibitem[{Landerer et~al.(2020)Landerer, Flechtner, Save, Webb, Bandikova, Bertiger, Bettadpur, Byun, Dahle, Dobslaw et~al.}]{landerer2020GRACE-FO}
\bibinfo{author}{Landerer, F.W.}, \bibinfo{author}{Flechtner, F.M.}, \bibinfo{author}{Save, H.}, \bibinfo{author}{Webb, F.H.}, \bibinfo{author}{Bandikova, T.}, \bibinfo{author}{Bertiger, W.I.}, \bibinfo{author}{Bettadpur, S.V.}, \bibinfo{author}{Byun, S.H.}, \bibinfo{author}{Dahle, C.}, \bibinfo{author}{Dobslaw, H.}, et~al., \bibinfo{year}{2020}.
\newblock \bibinfo{title}{{Extending the global mass change data record: GRACE Follow-On instrument and science data performance}}.
\newblock \bibinfo{journal}{Geophysical Research Letters} \bibinfo{volume}{47}, \bibinfo{pages}{e2020GL088306}.
\newblock \DOIprefix\doi{https://doi.org/10.1029/2020GL088306}.
\bibitem[{Landerer et~al.(2015)Landerer, Wiese, Bentel, Boening and Watkins}]{landerer2015GRACE4AMOC}
\bibinfo{author}{Landerer, F.W.}, \bibinfo{author}{Wiese, D.N.}, \bibinfo{author}{Bentel, K.}, \bibinfo{author}{Boening, C.}, \bibinfo{author}{Watkins, M.M.}, \bibinfo{year}{2015}.
\newblock \bibinfo{title}{{North Atlantic meridional overturning circulation variations from GRACE ocean bottom pressure anomalies}}.
\newblock \bibinfo{journal}{Geophysical Research Letters} \bibinfo{volume}{42}, \bibinfo{pages}{8114--8121}.
\newblock \DOIprefix\doi{https://doi.org/10.1002/2015GL065730}.
\bibitem[{LeCun et~al.(1998)LeCun, Bottou, Bengio and Haffner}]{lecun1998CNN}
\bibinfo{author}{LeCun, Y.}, \bibinfo{author}{Bottou, L.}, \bibinfo{author}{Bengio, Y.}, \bibinfo{author}{Haffner, P.}, \bibinfo{year}{1998}.
\newblock \bibinfo{title}{{Gradient-based learning applied to document recognition}}.
\newblock \bibinfo{journal}{Proceedings of the IEEE} \bibinfo{volume}{86}, \bibinfo{pages}{2278--2324}.
\newblock \DOIprefix\doi{https://doi.org/10.1109/5.726791}.
\bibitem[{Lellouche et~al.(2013)Lellouche, Le~Galloudec, Dr{\'e}villon, R{\'e}gnier, Greiner, Garric, Ferry, Desportes, Testut, Bricaud et~al.}]{lellouche2013GLORYS}
\bibinfo{author}{Lellouche, J.M.}, \bibinfo{author}{Le~Galloudec, O.}, \bibinfo{author}{Dr{\'e}villon, M.}, \bibinfo{author}{R{\'e}gnier, C.}, \bibinfo{author}{Greiner, E.}, \bibinfo{author}{Garric, G.}, \bibinfo{author}{Ferry, N.}, \bibinfo{author}{Desportes, C.}, \bibinfo{author}{Testut, C.E.}, \bibinfo{author}{Bricaud, C.}, et~al., \bibinfo{year}{2013}.
\newblock \bibinfo{title}{{Evaluation of global monitoring and forecasting systems at Mercator Oc{\'e}an}}.
\newblock \bibinfo{journal}{Ocean Science} \bibinfo{volume}{9}, \bibinfo{pages}{57--81}.
\newblock \DOIprefix\doi{https://doi.org/10.5194/os-9-57-2013}.
\bibitem[{Loomis et~al.(2019)Loomis, Luthcke and Sabaka}]{loomis2019GSFCMascon}
\bibinfo{author}{Loomis, B.}, \bibinfo{author}{Luthcke, S.}, \bibinfo{author}{Sabaka, T.}, \bibinfo{year}{2019}.
\newblock \bibinfo{title}{{Regularization and error characterization of GRACE mascons}}.
\newblock \bibinfo{journal}{Journal of Geodesy} \bibinfo{volume}{93}, \bibinfo{pages}{1381--1398}.
\newblock \DOIprefix\doi{https://doi.org/10.1007/s00190-019-01252-y}.
\bibitem[{Macrander et~al.(2010)Macrander, B{\"o}ning, Boebel and Schr{\"o}ter}]{Macrander2010BPR}
\bibinfo{author}{Macrander, A.}, \bibinfo{author}{B{\"o}ning, C.}, \bibinfo{author}{Boebel, O.}, \bibinfo{author}{Schr{\"o}ter, J.}, \bibinfo{year}{2010}.
\newblock \bibinfo{title}{Validation of GRACE gravity fields by in-situ data of ocean bottom pressure}. \bibinfo{publisher}{Springer Berlin Heidelberg}, \bibinfo{address}{Berlin, Heidelberg}.
\newblock pp. \bibinfo{pages}{169--185}.
\newblock \DOIprefix\doi{10.1007/978-3-642-10228-8_14}.
\bibitem[{McCarthy et~al.(2020)McCarthy, Brown, Flagg, Goni, Houpert, Hughes, Hummels, Inall, Jochumsen, Larsen et~al.}]{mccarthy2020ReviewAMOC}
\bibinfo{author}{McCarthy, G.D.}, \bibinfo{author}{Brown, P.J.}, \bibinfo{author}{Flagg, C.N.}, \bibinfo{author}{Goni, G.}, \bibinfo{author}{Houpert, L.}, \bibinfo{author}{Hughes, C.W.}, \bibinfo{author}{Hummels, R.}, \bibinfo{author}{Inall, M.}, \bibinfo{author}{Jochumsen, K.}, \bibinfo{author}{Larsen, K.}, et~al., \bibinfo{year}{2020}.
\newblock \bibinfo{title}{{Sustainable observations of the AMOC: Methodology and technology}}.
\newblock \bibinfo{journal}{Reviews of Geophysics} \bibinfo{volume}{58}, \bibinfo{pages}{e2019RG000654}.
\newblock \DOIprefix\doi{https://doi.org/10.1029/2019RG000654}.
\bibitem[{McDougall and Barker(2011)}]{mcdougall2011GSW}
\bibinfo{author}{McDougall, T.J.}, \bibinfo{author}{Barker, P.M.}, \bibinfo{year}{2011}.
\newblock \bibinfo{title}{{Getting started with TEOS-10 and the Gibbs Seawater (GSW) oceanographic toolbox}}.
\newblock \bibinfo{journal}{Scor/Iapso WG} \bibinfo{volume}{127}, \bibinfo{pages}{1--28}.
\bibitem[{Menemenlis et~al.(2008)Menemenlis, Campin, Heimbach, Hill, Lee, Nguyen, Schodlok and Zhang}]{menemenlis2008ecco2}
\bibinfo{author}{Menemenlis, D.}, \bibinfo{author}{Campin, J.M.}, \bibinfo{author}{Heimbach, P.}, \bibinfo{author}{Hill, C.}, \bibinfo{author}{Lee, T.}, \bibinfo{author}{Nguyen, A.}, \bibinfo{author}{Schodlok, M.}, \bibinfo{author}{Zhang, H.}, \bibinfo{year}{2008}.
\newblock \bibinfo{title}{{ECCO2: High resolution global ocean and sea ice data synthesis}}.
\newblock \bibinfo{journal}{Mercator Ocean Quarterly Newsletter} \bibinfo{volume}{31}, \bibinfo{pages}{13--21}.
\bibitem[{Miro and Famiglietti(2018)}]{miro2018downscaling}
\bibinfo{author}{Miro, M.E.}, \bibinfo{author}{Famiglietti, J.S.}, \bibinfo{year}{2018}.
\newblock \bibinfo{title}{{Downscaling GRACE remote sensing datasets to high-resolution groundwater storage change maps of California’s Central Valley}}.
\newblock \bibinfo{journal}{Remote Sensing} \bibinfo{volume}{10}, \bibinfo{pages}{143}.
\newblock \DOIprefix\doi{https://doi.org/10.3390/rs10010143}.
\bibitem[{Mu et~al.(2020)Mu, Xu and Xu}]{mu2020investigation}
\bibinfo{author}{Mu, D.}, \bibinfo{author}{Xu, T.}, \bibinfo{author}{Xu, G.}, \bibinfo{year}{2020}.
\newblock \bibinfo{title}{{An investigation of mass changes in the Bohai Sea observed by GRACE}}.
\newblock \bibinfo{journal}{Journal of Geodesy} \bibinfo{volume}{94}, \bibinfo{pages}{79}.
\newblock \DOIprefix\doi{https://doi.org/10.1007/s00190-020-01408-1}.
\bibitem[{Niu et~al.(2022)Niu, Cheng, Qin, Ou, Yang and Huang}]{niu2022mechanisms}
\bibinfo{author}{Niu, Y.}, \bibinfo{author}{Cheng, X.}, \bibinfo{author}{Qin, J.}, \bibinfo{author}{Ou, N.}, \bibinfo{author}{Yang, C.}, \bibinfo{author}{Huang, D.}, \bibinfo{year}{2022}.
\newblock \bibinfo{title}{{Mechanisms of interannual variability of ocean bottom pressure in the southern Indian ocean}}.
\newblock \bibinfo{journal}{Frontiers in Marine Science} \bibinfo{volume}{9}, \bibinfo{pages}{916592}.
\newblock \DOIprefix\doi{https://doi.org/10.3389/fmars.2022.916592}.
\bibitem[{Olbers et~al.(2012)Olbers, Willebrand and Eden}]{olbers2012oceandynamics}
\bibinfo{author}{Olbers, D.}, \bibinfo{author}{Willebrand, J.}, \bibinfo{author}{Eden, C.}, \bibinfo{year}{2012}.
\newblock \bibinfo{title}{{Ocean dynamics}}.
\newblock \bibinfo{publisher}{Springer Science \& Business Media}.
\bibitem[{Oldenburg et~al.(2022)Oldenburg, Wills, Armour and Thompson}]{oldenburg2022resolution}
\bibinfo{author}{Oldenburg, D.}, \bibinfo{author}{Wills, R.C.}, \bibinfo{author}{Armour, K.C.}, \bibinfo{author}{Thompson, L.}, \bibinfo{year}{2022}.
\newblock \bibinfo{title}{{Resolution dependence of atmosphere--ocean interactions and water mass transformation in the North Atlantic}}.
\newblock \bibinfo{journal}{Journal of Geophysical Research: Oceans} \bibinfo{volume}{127}, \bibinfo{pages}{e2021JC018102}.
\newblock \DOIprefix\doi{https://doi.org/10.1029/2021JC018102}.
\bibitem[{Peralta-Ferriz et~al.(2014)Peralta-Ferriz, Morison, Wallace, Bonin and Zhang}]{peralta2014arctic}
\bibinfo{author}{Peralta-Ferriz, C.}, \bibinfo{author}{Morison, J.H.}, \bibinfo{author}{Wallace, J.M.}, \bibinfo{author}{Bonin, J.A.}, \bibinfo{author}{Zhang, J.}, \bibinfo{year}{2014}.
\newblock \bibinfo{title}{{Arctic ocean circulation patterns revealed by GRACE}}.
\newblock \bibinfo{journal}{Journal of Climate} \bibinfo{volume}{27}, \bibinfo{pages}{1445--1468}.
\newblock \DOIprefix\doi{https://doi.org/10.1175/JCLI-D-13-00013.1}.
\bibitem[{Piecuch et~al.(2015)Piecuch, Fukumori, Ponte and Wang}]{piecuch2015vertical}
\bibinfo{author}{Piecuch, C.G.}, \bibinfo{author}{Fukumori, I.}, \bibinfo{author}{Ponte, R.M.}, \bibinfo{author}{Wang, O.}, \bibinfo{year}{2015}.
\newblock \bibinfo{title}{Vertical structure of ocean pressure variations with application to satellite-gravimetric observations}.
\newblock \bibinfo{journal}{Journal of Atmospheric and Oceanic Technology} \bibinfo{volume}{32}, \bibinfo{pages}{603--613}.
\newblock \DOIprefix\doi{https://doi.org/10.1175/JTECH-D-14-00156.1}.
\bibitem[{Piecuch et~al.(2018)Piecuch, Landerer and Ponte}]{piecuch2018tide}
\bibinfo{author}{Piecuch, C.G.}, \bibinfo{author}{Landerer, F.W.}, \bibinfo{author}{Ponte, R.M.}, \bibinfo{year}{2018}.
\newblock \bibinfo{title}{{Tide gauge records reveal improved processing of gravity recovery and climate experiment time-variable mass solutions over the coastal ocean}}.
\newblock \bibinfo{journal}{Geophysical Journal International} \bibinfo{volume}{214}, \bibinfo{pages}{1401--1412}.
\newblock \DOIprefix\doi{https://doi.org/10.1093/gji/ggy207}.
\bibitem[{Ponte(1994)}]{ponte1994IB}
\bibinfo{author}{Ponte, R.M.}, \bibinfo{year}{1994}.
\newblock \bibinfo{title}{Understanding the relation between wind-and pressure-driven sea level variability}.
\newblock \bibinfo{journal}{Journal of Geophysical Research: Oceans} \bibinfo{volume}{99}, \bibinfo{pages}{8033--8039}.
\newblock \DOIprefix\doi{https://doi.org/10.1029/94JC00217}.
\bibitem[{Ponte(1999)}]{ponteGlobalOBP1999}
\bibinfo{author}{Ponte, R.M.}, \bibinfo{year}{1999}.
\newblock \bibinfo{title}{A preliminary model study of the large-scale seasonal cycle in bottom pressure over the global ocean}.
\newblock \bibinfo{journal}{Journal of Geophysical Research: Oceans} \bibinfo{volume}{104}, \bibinfo{pages}{1289--1300}.
\newblock \DOIprefix\doi{https://doi.org/10.1029/1998JC900028}.
\bibitem[{Ponte and Schindelegger(2022)}]{ponte2022GlobalOceanResponse}
\bibinfo{author}{Ponte, R.M.}, \bibinfo{author}{Schindelegger, M.}, \bibinfo{year}{2022}.
\newblock \bibinfo{title}{{Global ocean response to the 5-day Rossby-Haurwitz atmospheric mode seen by GRACE}}.
\newblock \bibinfo{journal}{Journal of Geophysical Research: Oceans} \bibinfo{volume}{127}.
\newblock \DOIprefix\doi{https://doi.org/10.1029/2021JC018302}.
\bibitem[{Poropat et~al.(2018)Poropat, Dobslaw, Zhang, Macrander, Boebel and Thomas}]{poropat2018BPRpostcessing}
\bibinfo{author}{Poropat, L.}, \bibinfo{author}{Dobslaw, H.}, \bibinfo{author}{Zhang, L.}, \bibinfo{author}{Macrander, A.}, \bibinfo{author}{Boebel, O.}, \bibinfo{author}{Thomas, M.}, \bibinfo{year}{2018}.
\newblock \bibinfo{title}{{Time variations in ocean bottom pressure from a few hours to many years: In situ data, numerical models, and GRACE satellite gravimetry}}.
\newblock \bibinfo{journal}{Journal of Geophysical Research: Oceans} \bibinfo{volume}{123}, \bibinfo{pages}{5612--5623}.
\newblock \DOIprefix\doi{https://doi.org/10.1029/2018JC014108}.
\bibitem[{PSMSL(2023)}]{PSMSLData}
\bibinfo{author}{PSMSL}, \bibinfo{year}{2023}.
\newblock \bibinfo{title}{{Permanent Service for Mean Sea Level (PSMSL), tide gauge data}}.
\newblock \bibinfo{note}{Retrieved 09 Oct 2023 from \url{http://www.psmsl.org/data/obtaining/}}.
\bibitem[{Qin et~al.(2022)Qin, Cheng, Yang, Ou and Xiong}]{qin2022mechanism}
\bibinfo{author}{Qin, J.}, \bibinfo{author}{Cheng, X.}, \bibinfo{author}{Yang, C.}, \bibinfo{author}{Ou, N.}, \bibinfo{author}{Xiong, X.}, \bibinfo{year}{2022}.
\newblock \bibinfo{title}{{Mechanism of interannual variability of ocean bottom pressure in the South Pacific}}.
\newblock \bibinfo{journal}{Climate Dynamics} \bibinfo{volume}{59}, \bibinfo{pages}{2103--2116}.
\newblock \DOIprefix\doi{https://doi.org/10.1007/s00382-022-06198-0}.
\bibitem[{Reichstein et~al.(2019)Reichstein, Camps-Valls, Stevens, Jung, Denzler, Carvalhais et~al.}]{reichstein2019deeplearning4geoscience}
\bibinfo{author}{Reichstein, M.}, \bibinfo{author}{Camps-Valls, G.}, \bibinfo{author}{Stevens, B.}, \bibinfo{author}{Jung, M.}, \bibinfo{author}{Denzler, J.}, \bibinfo{author}{Carvalhais, N.}, et~al., \bibinfo{year}{2019}.
\newblock \bibinfo{title}{{Deep learning and process understanding for data-driven Earth system science}}.
\newblock \bibinfo{journal}{Nature} \bibinfo{volume}{566}, \bibinfo{pages}{195--204}.
\newblock \DOIprefix\doi{https://doi.org/10.1038/s41586-019-0912-1}.
\bibitem[{Richard~Peltier et~al.(2018)Richard~Peltier, Argus and Drummond}]{richard2018GIAcorrection}
\bibinfo{author}{Richard~Peltier, W.}, \bibinfo{author}{Argus, D.F.}, \bibinfo{author}{Drummond, R.}, \bibinfo{year}{2018}.
\newblock \bibinfo{title}{{Comment on “An assessment of the ICE-6G\_C (VM5a) glacial isostatic adjustment model” by Purcell et al.}}
\newblock \bibinfo{journal}{Journal of Geophysical Research: Solid Earth} \bibinfo{volume}{123}, \bibinfo{pages}{2019--2028}.
\newblock \DOIprefix\doi{https://doi.org/10.1002/2016JB013844}.
\bibitem[{Rohith et~al.(2019)Rohith, Paul, Durand, Testut, Prerna, Afroosa, Ramakrishna and Shenoi}]{rohith2019basin}
\bibinfo{author}{Rohith, B.}, \bibinfo{author}{Paul, A.}, \bibinfo{author}{Durand, F.}, \bibinfo{author}{Testut, L.}, \bibinfo{author}{Prerna, S.}, \bibinfo{author}{Afroosa, M.}, \bibinfo{author}{Ramakrishna, S.}, \bibinfo{author}{Shenoi, S.}, \bibinfo{year}{2019}.
\newblock \bibinfo{title}{{Basin-wide sea level coherency in the tropical Indian Ocean driven by Madden--Julian Oscillation}}.
\newblock \bibinfo{journal}{Nature Communications} \bibinfo{volume}{10}, \bibinfo{pages}{1257}.
\newblock \DOIprefix\doi{https://doi.org/10.1038/s41467-019-09243-5}.
\bibitem[{Ronneberger et~al.(2015)Ronneberger, Fischer and Brox}]{ronneberger2015U-net}
\bibinfo{author}{Ronneberger, O.}, \bibinfo{author}{Fischer, P.}, \bibinfo{author}{Brox, T.}, \bibinfo{year}{2015}.
\newblock \bibinfo{title}{U-net: Convolutional networks for biomedical image segmentation}, in: \bibinfo{booktitle}{Medical Image Computing and Computer-Assisted Intervention--MICCAI 2015: 18th International Conference, Munich, Germany, October 5-9, 2015, Proceedings, Part III 18}, \bibinfo{organization}{Springer}. pp. \bibinfo{pages}{234--241}.
\bibitem[{Roussenov et~al.(2008)Roussenov, Williams, Hughes and Bingham}]{roussenov2008BoundaryWaveCommunication}
\bibinfo{author}{Roussenov, V.M.}, \bibinfo{author}{Williams, R.G.}, \bibinfo{author}{Hughes, C.W.}, \bibinfo{author}{Bingham, R.J.}, \bibinfo{year}{2008}.
\newblock \bibinfo{title}{{Boundary wave communication of bottom pressure and overturning changes for the North Atlantic}}.
\newblock \bibinfo{journal}{Journal of Geophysical Research: Oceans} \bibinfo{volume}{113}.
\newblock \DOIprefix\doi{https://doi.org/10.1029/2007JC004501}.
\bibitem[{Sakumura et~al.(2014)Sakumura, Bettadpur and Bruinsma}]{sakumura2014ensemble}
\bibinfo{author}{Sakumura, C.}, \bibinfo{author}{Bettadpur, S.}, \bibinfo{author}{Bruinsma, S.}, \bibinfo{year}{2014}.
\newblock \bibinfo{title}{{Ensemble prediction and intercomparison analysis of GRACE time-variable gravity field models}}.
\newblock \bibinfo{journal}{Geophysical Research Letters} \bibinfo{volume}{41}, \bibinfo{pages}{1389--1397}.
\newblock \DOIprefix\doi{https://doi.org/10.1002/2013GL058632}.
\bibitem[{Save et~al.(2016)Save, Bettadpur and Tapley}]{save2016CSRMascon}
\bibinfo{author}{Save, H.}, \bibinfo{author}{Bettadpur, S.}, \bibinfo{author}{Tapley, B.D.}, \bibinfo{year}{2016}.
\newblock \bibinfo{title}{{High-resolution CSR GRACE RL05 mascons}}.
\newblock \bibinfo{journal}{Journal of Geophysical Research: Solid Earth} \bibinfo{volume}{121}, \bibinfo{pages}{7547--7569}.
\newblock \DOIprefix\doi{https://doi.org/10.1002/2016JB013007}.
\bibitem[{Schindelegger et~al.(2021)Schindelegger, Harker, Ponte, Dobslaw and Salstein}]{schindelegger2021DailyGRACE}
\bibinfo{author}{Schindelegger, M.}, \bibinfo{author}{Harker, A.A.}, \bibinfo{author}{Ponte, R.M.}, \bibinfo{author}{Dobslaw, H.}, \bibinfo{author}{Salstein, D.A.}, \bibinfo{year}{2021}.
\newblock \bibinfo{title}{{Convergence of daily GRACE solutions and models of submonthly ocean bottom pressure variability}}.
\newblock \bibinfo{journal}{Journal of Geophysical Research: Oceans} \bibinfo{volume}{126}, \bibinfo{pages}{e2020JC017031}.
\newblock \DOIprefix\doi{https://doi.org/10.1029/2020JC017031}.
\bibitem[{Schneider et~al.(2023)Schneider, Behera, Boccaletti, Deser, Emanuel, Ferrari, Leung, Lin, M{\"u}ller, Navarra et~al.}]{schneider2023AI4ClimateModel}
\bibinfo{author}{Schneider, T.}, \bibinfo{author}{Behera, S.}, \bibinfo{author}{Boccaletti, G.}, \bibinfo{author}{Deser, C.}, \bibinfo{author}{Emanuel, K.}, \bibinfo{author}{Ferrari, R.}, \bibinfo{author}{Leung, L.R.}, \bibinfo{author}{Lin, N.}, \bibinfo{author}{M{\"u}ller, T.}, \bibinfo{author}{Navarra, A.}, et~al., \bibinfo{year}{2023}.
\newblock \bibinfo{title}{{Harnessing AI and computing to advance climate modelling and prediction}}.
\newblock \bibinfo{journal}{Nature Climate Change} \bibinfo{volume}{13}, \bibinfo{pages}{887--889}.
\newblock \DOIprefix\doi{https://doi.org/10.1038/s41558-023-01769-3}.
\bibitem[{Schneider et~al.(2017)Schneider, Lan, Stuart and Teixeira}]{schneider2017EarthSystemModeling}
\bibinfo{author}{Schneider, T.}, \bibinfo{author}{Lan, S.}, \bibinfo{author}{Stuart, A.}, \bibinfo{author}{Teixeira, J.}, \bibinfo{year}{2017}.
\newblock \bibinfo{title}{Earth system modeling 2.0: A blueprint for models that learn from observations and targeted high-resolution simulations}.
\newblock \bibinfo{journal}{Geophysical Research Letters} \bibinfo{volume}{44}, \bibinfo{pages}{12--396}.
\newblock \DOIprefix\doi{https://doi.org/10.1002/2017GL076101}.
\bibitem[{Seyoum et~al.(2019)Seyoum, Kwon and Milewski}]{seyoum2019downscaling}
\bibinfo{author}{Seyoum, W.M.}, \bibinfo{author}{Kwon, D.}, \bibinfo{author}{Milewski, A.M.}, \bibinfo{year}{2019}.
\newblock \bibinfo{title}{{Downscaling GRACE TWSA data into high-resolution groundwater level anomaly using machine learning-based models in a glacial aquifer system}}.
\newblock \bibinfo{journal}{Remote Sensing} \bibinfo{volume}{11}, \bibinfo{pages}{824}.
\newblock \DOIprefix\doi{https://doi.org/10.3390/rs11070824}.
\bibitem[{Shihora et~al.(2022)Shihora, Balidakis, Dill, Dahle, Ghobadi-Far, Bonin and Dobslaw}]{shihora2022AOD1BRL07}
\bibinfo{author}{Shihora, L.}, \bibinfo{author}{Balidakis, K.}, \bibinfo{author}{Dill, R.}, \bibinfo{author}{Dahle, C.}, \bibinfo{author}{Ghobadi-Far, K.}, \bibinfo{author}{Bonin, J.}, \bibinfo{author}{Dobslaw, H.}, \bibinfo{year}{2022}.
\newblock \bibinfo{title}{{Non-tidal background modeling for satellite gravimetry based on operational ECWMF and ERA5 reanalysis data: AOD1B RL07}}.
\newblock \bibinfo{journal}{Journal of Geophysical Research: Solid Earth} \bibinfo{volume}{127}, \bibinfo{pages}{e2022JB024360}.
\newblock \DOIprefix\doi{https://doi.org/10.1029/2022JB024360}.
\bibitem[{Tapley et~al.(2004)Tapley, Bettadpur, Watkins and Reigber}]{tapley2004GracePrinciple}
\bibinfo{author}{Tapley, B.D.}, \bibinfo{author}{Bettadpur, S.}, \bibinfo{author}{Watkins, M.}, \bibinfo{author}{Reigber, C.}, \bibinfo{year}{2004}.
\newblock \bibinfo{title}{{The gravity recovery and climate experiment: Mission overview and early results}}.
\newblock \bibinfo{journal}{Geophysical Research Letters} \bibinfo{volume}{31}.
\newblock \DOIprefix\doi{https://doi.org/10.1029/2004GL019920}.
\bibitem[{{The IMBIE Team}(2020)}]{imbie2020MassLossGreenland}
\bibinfo{author}{{The IMBIE Team}}, \bibinfo{year}{2020}.
\newblock \bibinfo{title}{{Mass balance of the Greenland Ice Sheet from 1992 to 2018}}.
\newblock \bibinfo{journal}{Nature} \bibinfo{volume}{579}, \bibinfo{pages}{233--239}.
\newblock \DOIprefix\doi{https://doi.org/10.1038/s41586-019-1855-2}.
\bibitem[{Velicogna et~al.(2020)Velicogna, Mohajerani, Landerer, Mouginot, Noel, Rignot, Sutterley, van~den Broeke, van Wessem and Wiese}]{velicogna2020MassLossIceSheet}
\bibinfo{author}{Velicogna, I.}, \bibinfo{author}{Mohajerani, Y.}, \bibinfo{author}{Landerer, F.}, \bibinfo{author}{Mouginot, J.}, \bibinfo{author}{Noel, B.}, \bibinfo{author}{Rignot, E.}, \bibinfo{author}{Sutterley, T.}, \bibinfo{author}{van~den Broeke, M.}, \bibinfo{author}{van Wessem, M.}, \bibinfo{author}{Wiese, D.}, \bibinfo{year}{2020}.
\newblock \bibinfo{title}{{Continuity of ice sheet mass loss in Greenland and Antarctica from the GRACE and GRACE Follow-On missions}}.
\newblock \bibinfo{journal}{Geophysical Research Letters} \bibinfo{volume}{47}, \bibinfo{pages}{e2020GL087291}.
\newblock \DOIprefix\doi{https://doi.org/10.1029/2020GL087291}.
\bibitem[{Vinogradov and Ponte(2011)}]{vinogradov2011lowfrequencySeaLevel}
\bibinfo{author}{Vinogradov, S.V.}, \bibinfo{author}{Ponte, R.M.}, \bibinfo{year}{2011}.
\newblock \bibinfo{title}{{Low-frequency variability in coastal sea level from tide gauges and altimetry}}.
\newblock \bibinfo{journal}{Journal of Geophysical Research: Oceans} \bibinfo{volume}{116}.
\newblock \DOIprefix\doi{https://doi.org/10.1029/2011JC007034}.
\bibitem[{Wahr et~al.(2006)Wahr, Swenson and Velicogna}]{wahr2006accuracyGRACE}
\bibinfo{author}{Wahr, J.}, \bibinfo{author}{Swenson, S.}, \bibinfo{author}{Velicogna, I.}, \bibinfo{year}{2006}.
\newblock \bibinfo{title}{{Accuracy of GRACE mass estimates}}.
\newblock \bibinfo{journal}{Geophysical Research Letters} \bibinfo{volume}{33}.
\newblock \DOIprefix\doi{https://doi.org/10.1029/2005GL025305}.
\bibitem[{Wahr et~al.(2004)Wahr, Swenson, Zlotnicki and Velicogna}]{wahr2004GracePrinciple}
\bibinfo{author}{Wahr, J.}, \bibinfo{author}{Swenson, S.}, \bibinfo{author}{Zlotnicki, V.}, \bibinfo{author}{Velicogna, I.}, \bibinfo{year}{2004}.
\newblock \bibinfo{title}{{Time-variable gravity from GRACE: First results}}.
\newblock \bibinfo{journal}{Geophysical Research Letters} \bibinfo{volume}{31}.
\newblock \DOIprefix\doi{https://doi.org/10.1029/2004GL019779}.
\bibitem[{Wang et~al.(2012)Wang, Shum, Simons, Tapley and Dai}]{wang2012grace4earthquake}
\bibinfo{author}{Wang, L.}, \bibinfo{author}{Shum, C.}, \bibinfo{author}{Simons, F.J.}, \bibinfo{author}{Tapley, B.}, \bibinfo{author}{Dai, C.}, \bibinfo{year}{2012}.
\newblock \bibinfo{title}{{Coseismic and postseismic deformation of the 2011 Tohoku-Oki earthquake constrained by GRACE gravimetry}}.
\newblock \bibinfo{journal}{Geophysical Research Letters} \bibinfo{volume}{39}.
\newblock \DOIprefix\doi{https://doi.org/10.1029/2012GL051104}.
\bibitem[{Wang et~al.(2022)Wang, Albrecht, Braham, Mou and Zhu}]{wang2022SSLinRS}
\bibinfo{author}{Wang, Y.}, \bibinfo{author}{Albrecht, C.M.}, \bibinfo{author}{Braham, N.A.A.}, \bibinfo{author}{Mou, L.}, \bibinfo{author}{Zhu, X.X.}, \bibinfo{year}{2022}.
\newblock \bibinfo{title}{{Self-supervised learning in remote sensing: A review}}.
\newblock \bibinfo{journal}{IEEE Geoscience and Remote Sensing Magazine} \bibinfo{volume}{10}, \bibinfo{pages}{213--247}.
\newblock \DOIprefix\doi{https://doi.org/10.1109/MGRS.2022.3198244}.
\bibitem[{Watkins et~al.(2015)Watkins, Wiese, Yuan, Boening and Landerer}]{watkins2015JPLMascon}
\bibinfo{author}{Watkins, M.M.}, \bibinfo{author}{Wiese, D.N.}, \bibinfo{author}{Yuan, D.N.}, \bibinfo{author}{Boening, C.}, \bibinfo{author}{Landerer, F.W.}, \bibinfo{year}{2015}.
\newblock \bibinfo{title}{{Improved methods for observing Earth's time variable mass distribution with GRACE using spherical cap mascons}}.
\newblock \bibinfo{journal}{Journal of Geophysical Research: Solid Earth} \bibinfo{volume}{120}, \bibinfo{pages}{2648--2671}.
\newblock \DOIprefix\doi{https://doi.org/10.1002/2014JB011547}.
\bibitem[{Watts and Kontoyiannis(1990)}]{watts1990BPR}
\bibinfo{author}{Watts, D.R.}, \bibinfo{author}{Kontoyiannis, H.}, \bibinfo{year}{1990}.
\newblock \bibinfo{title}{{Deep-ocean bottom pressure measurement: Drift removal and performance}}.
\newblock \bibinfo{journal}{Journal of Atmospheric and Oceanic Technology} \bibinfo{volume}{7}, \bibinfo{pages}{296--306}.
\newblock \DOIprefix\doi{https://doi.org/10.1175/1520-0426(1990)007<0296:DOBPMD>2.0.CO;2}.
\bibitem[{Weijer(2010)}]{weijer2010almostfree}
\bibinfo{author}{Weijer, W.}, \bibinfo{year}{2010}.
\newblock \bibinfo{title}{{An almost-free barotropic mode in the Australian-Antarctic Basin}}.
\newblock \bibinfo{journal}{Geophysical Research Letters} \bibinfo{volume}{37}.
\newblock \DOIprefix\doi{https://doi.org/10.1029/2010GL042657}.
\bibitem[{Weiss et~al.(2016)Weiss, Khoshgoftaar and Wang}]{weiss2016SurveyTrensferLearning}
\bibinfo{author}{Weiss, K.}, \bibinfo{author}{Khoshgoftaar, T.M.}, \bibinfo{author}{Wang, D.}, \bibinfo{year}{2016}.
\newblock \bibinfo{title}{A survey of transfer learning}.
\newblock \bibinfo{journal}{Journal of Big data} \bibinfo{volume}{3}, \bibinfo{pages}{1--40}.
\newblock \DOIprefix\doi{https://doi.org/10.1186/s40537-016-0043-6}.
\bibitem[{Wiese et~al.(2016)Wiese, Landerer and Watkins}]{wiese2016quantifying}
\bibinfo{author}{Wiese, D.N.}, \bibinfo{author}{Landerer, F.W.}, \bibinfo{author}{Watkins, M.M.}, \bibinfo{year}{2016}.
\newblock \bibinfo{title}{{Quantifying and reducing leakage errors in the JPL RL05M GRACE mascon solution}}.
\newblock \bibinfo{journal}{Water Resources Research} \bibinfo{volume}{52}, \bibinfo{pages}{7490--7502}.
\newblock \DOIprefix\doi{https://doi.org/10.1002/2016WR019344}.
\bibitem[{Williams and Hughes(2013)}]{williams2013coherence}
\bibinfo{author}{Williams, J.}, \bibinfo{author}{Hughes, C.W.}, \bibinfo{year}{2013}.
\newblock \bibinfo{title}{The coherence of small island sea level with the wider ocean: a model study}.
\newblock \bibinfo{journal}{Ocean Science} \bibinfo{volume}{9}, \bibinfo{pages}{111--119}.
\newblock \DOIprefix\doi{https://doi.org/10.5194/os-9-111-2013}.
\bibitem[{Williams and Penna(2011)}]{williams2011non-tidalOceanLoading}
\bibinfo{author}{Williams, S.}, \bibinfo{author}{Penna, N.}, \bibinfo{year}{2011}.
\newblock \bibinfo{title}{{Non-tidal ocean loading effects on geodetic GPS heights}}.
\newblock \bibinfo{journal}{Geophysical Research Letters} \bibinfo{volume}{38}.
\newblock \DOIprefix\doi{https://doi.org/10.1029/2011GL046940}.
\bibitem[{Woodworth et~al.(2019)Woodworth, Melet, Marcos, Ray, W{\"o}ppelmann, Sasaki, Cirano, Hibbert, Huthnance, Monserrat et~al.}]{woodworth2019forcingCoastSeaLevel}
\bibinfo{author}{Woodworth, P.L.}, \bibinfo{author}{Melet, A.}, \bibinfo{author}{Marcos, M.}, \bibinfo{author}{Ray, R.D.}, \bibinfo{author}{W{\"o}ppelmann, G.}, \bibinfo{author}{Sasaki, Y.N.}, \bibinfo{author}{Cirano, M.}, \bibinfo{author}{Hibbert, A.}, \bibinfo{author}{Huthnance, J.M.}, \bibinfo{author}{Monserrat, S.}, et~al., \bibinfo{year}{2019}.
\newblock \bibinfo{title}{{Forcing factors affecting sea level changes at the coast}}.
\newblock \bibinfo{journal}{Surveys in Geophysics} \bibinfo{volume}{40}, \bibinfo{pages}{1351--1397}.
\newblock \DOIprefix\doi{https://doi.org/10.1007/s10712-019-09531-1}.
\bibitem[{Worthington et~al.(2019)Worthington, Frajka-Williams and McCarthy}]{worthington2019AMOCfromBPR}
\bibinfo{author}{Worthington, E.}, \bibinfo{author}{Frajka-Williams, E.}, \bibinfo{author}{McCarthy, G.D.}, \bibinfo{year}{2019}.
\newblock \bibinfo{title}{{Estimating the deep overturning transport variability at 26${}^\circ$N using bottom pressure recorders}}.
\newblock \bibinfo{journal}{Journal of Geophysical Research: Oceans} \bibinfo{volume}{124}, \bibinfo{pages}{335--348}.
\newblock \DOIprefix\doi{https://doi.org/10.1029/2018JC014221}.
\bibitem[{Yin et~al.(2022)Yin, Zhang, Liu, Zhang, Zhang and Chen}]{yin2022Downscaling}
\bibinfo{author}{Yin, W.}, \bibinfo{author}{Zhang, G.}, \bibinfo{author}{Liu, F.}, \bibinfo{author}{Zhang, D.}, \bibinfo{author}{Zhang, X.}, \bibinfo{author}{Chen, S.}, \bibinfo{year}{2022}.
\newblock \bibinfo{title}{{Improving the spatial resolution of GRACE-based groundwater storage estimates using a machine learning algorithm and hydrological model}}.
\newblock \bibinfo{journal}{Hydrogeology Journal} \bibinfo{volume}{30}, \bibinfo{pages}{947--963}.
\newblock \DOIprefix\doi{https://doi.org/10.1007/s10040-021-02447-4}.
\bibitem[{Yu et~al.(2018)Yu, Chao, Garc{\'\i}a-Garc{\'\i}a and Luo}]{yu2018VariationsArgentineGyre}
\bibinfo{author}{Yu, Y.}, \bibinfo{author}{Chao, B.F.}, \bibinfo{author}{Garc{\'\i}a-Garc{\'\i}a, D.}, \bibinfo{author}{Luo, Z.}, \bibinfo{year}{2018}.
\newblock \bibinfo{title}{{Variations of the Argentine Gyre observed in the GRACE time-variable gravity and ocean altimetry measurements}}.
\newblock \bibinfo{journal}{Journal of Geophysical Research: Oceans} \bibinfo{volume}{123}, \bibinfo{pages}{5375--5387}.
\newblock \DOIprefix\doi{https://doi.org/10.1029/2018JC014189}.
\bibitem[{Zhao et~al.(2023)Zhao, Ponte and Penduff}]{zhao2023IntrinsicVariability}
\bibinfo{author}{Zhao, M.}, \bibinfo{author}{Ponte, R.M.}, \bibinfo{author}{Penduff, T.}, \bibinfo{year}{2023}.
\newblock \bibinfo{title}{Global-scale random bottom pressure fluctuations from oceanic intrinsic variability}.
\newblock \bibinfo{journal}{Science Advances} \bibinfo{volume}{9}, \bibinfo{pages}{eadg0278}.
\newblock \DOIprefix\doi{https://doi.org/10.1126/sciadv.adg0278}.
\bibitem[{Zhao et~al.(2021)Zhao, Ponte, Penduff, Close, Llovel and Molines}]{zhao2021IntrinsicVariability}
\bibinfo{author}{Zhao, M.}, \bibinfo{author}{Ponte, R.M.}, \bibinfo{author}{Penduff, T.}, \bibinfo{author}{Close, S.}, \bibinfo{author}{Llovel, W.}, \bibinfo{author}{Molines, J.M.}, \bibinfo{year}{2021}.
\newblock \bibinfo{title}{{Imprints of ocean chaotic intrinsic variability on bottom pressure and implications for data and model analyses}}.
\newblock \bibinfo{journal}{Geophysical Research Letters} \bibinfo{volume}{48}, \bibinfo{pages}{e2021GL096341}.
\newblock \DOIprefix\doi{https://doi.org/10.1029/2021GL096341}.
\bibitem[{Zuo et~al.(2017)Zuo, Balmaseda and Mogensen}]{zuo2017ORAS5}
\bibinfo{author}{Zuo, H.}, \bibinfo{author}{Balmaseda, M.A.}, \bibinfo{author}{Mogensen, K.}, \bibinfo{year}{2017}.
\newblock \bibinfo{title}{{The new eddy-permitting ORAP5 ocean reanalysis: description, evaluation and uncertainties in climate signals}}.
\newblock \bibinfo{journal}{Climate Dynamics} \bibinfo{volume}{49}, \bibinfo{pages}{791--811}.
\newblock \DOIprefix\doi{https://doi.org/10.1007/s00382-015-2675-1}.

\end{thebibliography}
\clearpage



\section*{Supplementary material}

# Supplementary material for Downscaling GRACE-derived ocean bottom pressure anomalies using self-supervised data fusion

**Junyang Gou**[1,*], **Lara Börger**[2], **Michael Schindelegger**[2], **and Benedikt Soja**[1]

[1]Institute of Geodesy and Photogrammetry, ETH Zurich, Zurich, Switzerland
[2]Institute of Geodesy and Geoinformation, University of Bonn, Bonn, Germany
[*]jungou@ethz.ch

## S1 Downscaled products with different mascon products

Differences in the processing strategies among the three mascon providers (Center for Space Research, CSRM; Jet Propulsion Laboratory, JPLM; and Goddard Space Flight Center, GSFC) may lead to differences in the satellite-based monthly ocean bottom pressure ($p_b$) solutions (Sakumura et al., 2014; Landerer & Cooley, 2021). In this study, we applied our downscaling algorithm using the three mascon products to explore possible sensitivities to these input data. The three downscaled products forced by CSRM, JPLM, and GSFC are denoted as *DS-CSRM*, *DS-JPLM*, and *DS-GSFC*, respectively.

Fig. S1 shows the temporal decomposition of the three downscaled $p_b$ fields, along with a similar decomposition for the three corresponding mascon solutions. It is worth noting that the long-term trend component of the three downscaled solutions features the most notable differences, particularly in the Arctic Ocean. The CSRM solution reveals decreasing trends surrounding Greenland, whereas the JPLM and GSFC solutions indicate decreasing trends in Baffin Bay but increasing trends in other Arctic regions. Since the downscaled products are forced to agree with utilized GRACE(-FO) solutions over a large area, they are expected to inherit the long-term information from the corresponding GRACE(-FO) solutions. A similar case can be found in the Southern Ocean: JPLM and GSFC trends show questionable signals in the Bellingshausen Sea and Amundsen Sea, which might be caused by leakage of mass change signals from West Antarctica. The downscaled $p_b$ solutions also show similar signals in these regions. The three mascon solutions agree well in terms of annual signal amplitudes, resulting in consistent annual signals across the three downscaled products. The semi-annual signals and post-fit residuals estimated from the three downscaled products are very similar since they are learned from the same reanalysis products. It is noteworthy that the nominal mascon sizes do not compromise the effectiveness of our method in resolving small-scale signals. Specifically, the downscaled solution based on JPLM with a mascon size of 3° has very similar small-scale patterns as the downscaled solution based on CSRM with a mascon size of 1°. Thus, the high-resolution information seems less affected by the chosen GRACE(-FO) products than the long-term trends.

We further examine the large-scale mass conservation of the different solutions and report evaluations against the mascon ensemble (ME) in Table S1. We can observe that the choice of a particular mascon product dominates the large-scale signals of downscaled products. The statistics for the primary temporal components (trend, annual amplitude, semi-annual amplitude) and other numerical metrics (RMSE and correlation) of the downscaled products are very similar to the ones obtained from their forcing mascons. For example, the largest disparity in long-term trends between the DS-CSRM solution and the ME solution is in the Arctic Ocean (0.05 mm yr$^{-1}$ vs 1.33 mm yr$^{-1}$), which is mainly caused by the differences between the CSRM and ME solutions (0.01 mm yr$^{-1}$ vs 1.33 mm yr$^{-1}$). Similar issues can be found in other regions, such as the Indian Ocean, where JPLM tends to overestimate the trend (2.08 mm yr$^{-1}$), whereas GSFC tends to underestimate it (1.76 mm yr$^{-1}$). Consequently, the downscaled products forced by JPLM and GSFC have similar trends (2.15 mm yr$^{-1}$ vs 1.76 mm yr$^{-1}$). Nevertheless, all the mascon products and downscaled products have similar RMSEs and correlations, indicating similar large-scale accuracy. All of them are of better quality than the reanalysis solutions in terms of large-scale accuracy.

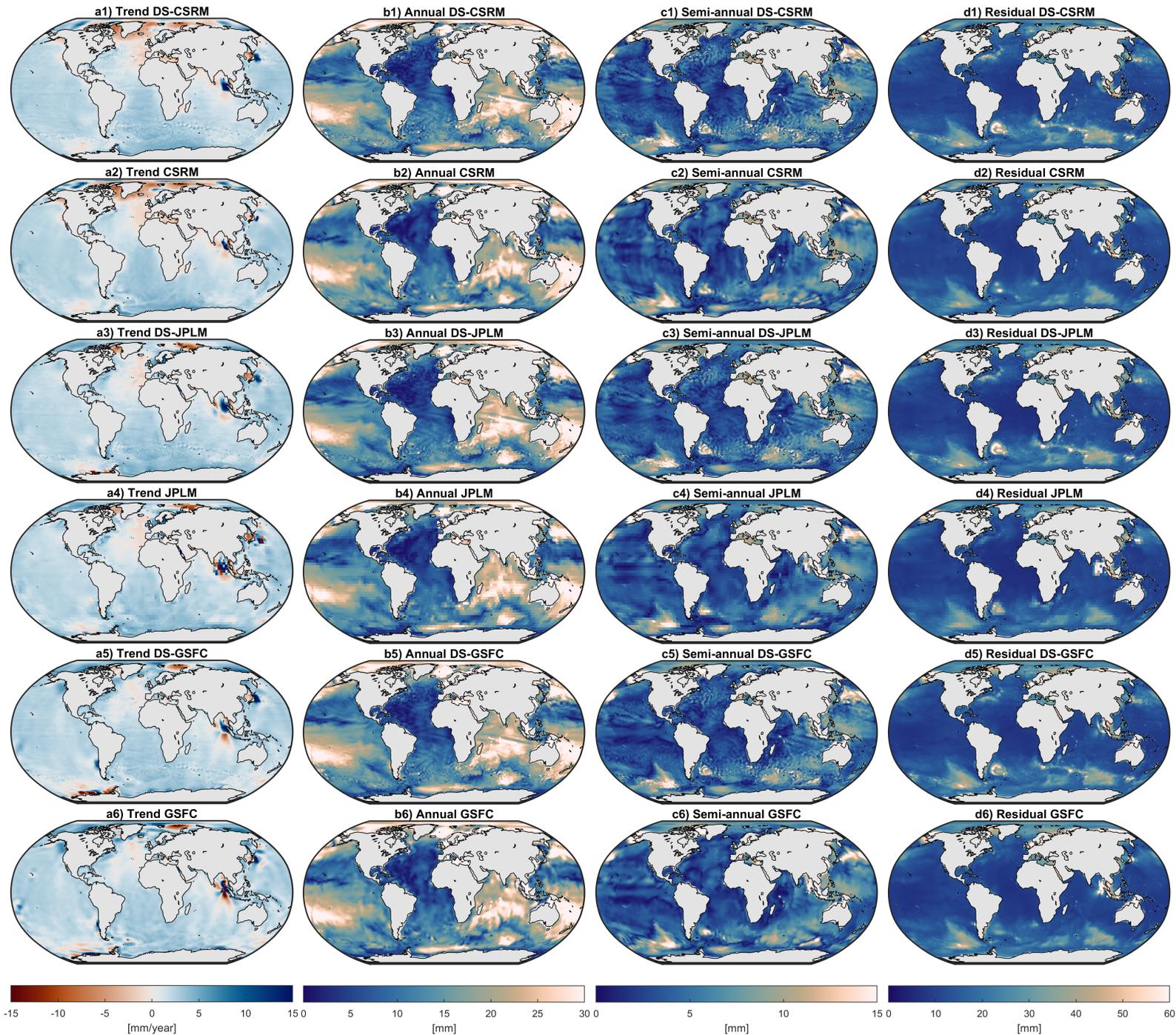

**Figure S1.** Temporal decomposition of six $p_b$ products from April 2002 to December 2020. The four columns show the long-term trends [mm/year], annual amplitudes [mm], semi-annual amplitudes [mm], and standard deviations of residuals [mm]. The six rows depict the signals estimated from CSRM, JPLM, GSFC mascons, and the corresponding downscaled products.

**Table S1.** Comparison of the three downscaled $p_b$ products, three mascon products, and two reanalysis products with the mascon ensemble (ME) solution in the six large ocean basins regarding long-term trends, RMSEs, and Pearson correlations. SA: South Atlantic, NA: North Atlantic, SP: South Pacific, NP: North Pacific, IND: Indian, ARC: Arctic.

|  |  | Global | SA | NA | SP | NP | IND | ARC |
|---|---|---|---|---|---|---|---|---|
| Trends [mm yr$^{-1}$] | ME | 1.99 | 2.72 | 1.22 | 2.06 | 2.23 | 1.90 | 1.33 |
|  | DS-CSRM | 1.87 | 2.67 | 0.88 | 2.14 | 2.10 | 1.80 | 0.05 |
|  | CSRM | 1.89 | 2.71 | 0.90 | 2.18 | 2.12 | 1.84 | 0.01 |
|  | DS-JPLM | 2.10 | 2.79 | 1.20 | 2.13 | 2.16 | 2.15 | 1.49 |
|  | JPLM | 2.04 | 2.73 | 1.18 | 2.08 | 2.11 | 2.08 | 1.47 |
|  | DS-GSFC | 2.04 | 2.71 | 1.59 | 1.89 | 2.46 | 1.76 | 2.47 |
|  | GSFC | 2.04 | 2.71 | 1.59 | 1.91 | 2.46 | 1.76 | 2.52 |
|  | GLORYS | 2.72 | 2.66 | 2.91 | 2.52 | 2.43 | 2.88 | 2.55 |
|  | ORAS5 | 1.97 | 1.72 | 1.58 | 2.02 | 1.99 | 1.95 | 1.35 |
| Annual Amplitudes [mm] | ME | 9.27 | 10.53 | 6.33 | 10.08 | 7.93 | 8.54 | 16.45 |
|  | DS-CSRM | 9.26 | 11.08 | 5.76 | 10.60 | 8.06 | 8.22 | 14.42 |
|  | CSRM | 9.40 | 11.27 | 5.86 | 10.82 | 8.11 | 8.28 | 14.31 |
|  | DS-JPLM | 8.84 | 10.29 | 6.04 | 9.42 | 7.33 | 8.33 | 16.30 |
|  | JPLM | 8.61 | 10.05 | 6.01 | 9.14 | 7.20 | 8.04 | 15.80 |
|  | DS-GSFC | 9.81 | 10.28 | 7.19 | 10.24 | 8.52 | 9.28 | 19.25 |
|  | GSFC | 9.83 | 10.28 | 7.15 | 10.32 | 8.56 | 9.30 | 19.56 |
|  | GLORYS | 11.3 | 12.06 | 11.80 | 11.70 | 8.44 | 10.56 | 21.15 |
|  | ORAS5 | 8.10 | 7.85 | 8.09 | 11.24 | 4.50 | 8.39 | 22.12 |
| Semi-annual Amplitudes [mm] | ME | 0.76 | 1.13 | 1.49 | 0.67 | 1.02 | 1.70 | 3.62 |
|  | DS-CSRM | 0.82 | 1.24 | 1.34 | 1.07 | 1.00 | 1.66 | 3.91 |
|  | CSRM | 0.85 | 1.32 | 1.34 | 1.15 | 1.02 | 1.68 | 3.94 |
|  | DS-JPLM | 0.76 | 1.18 | 1.60 | 0.54 | 1.05 | 1.86 | 3.51 |
|  | JPLM | 0.75 | 1.56 | 1.56 | 0.53 | 1.05 | 1.89 | 3.52 |
|  | DS-GSFC | 0.68 | 0.87 | 1.77 | 0.40 | 0.98 | 1.48 | 3.59 |
|  | GSFC | 0.69 | 0.92 | 1.75 | 0.39 | 1.02 | 1.52 | 3.64 |
|  | GLORYS | 1.73 | 0.14 | 1.00 | 2.87 | 2.00 | 2.89 | 5.43 |
|  | ORAS5 | 0.18 | 1.96 | 1.76 | 1.55 | 1.39 | 2.40 | 4.31 |
| RMSE [mm] | DS-CSRM | 0.99 | 1.56 | 3.10 | 1.40 | 1.37 | 1.18 | 9.71 |
|  | CSRM | 0.85 | 1.64 | 3.06 | 1.62 | 1.27 | 1.01 | 10.03 |
|  | DS-JPLM | 1.03 | 1.53 | 1.29 | 1.30 | 1.07 | 1.97 | 2.99 |
|  | JPLM | 0.80 | 1.36 | 1.31 | 1.30 | 1.27 | 1.56 | 2.83 |
|  | DS-GSFC | 0.74 | 1.84 | 3.31 | 1.46 | 1.78 | 1.42 | 8.69 |
|  | GSFC | 0.74 | 1.86 | 3.33 | 1.41 | 1.80 | 1.38 | 9.01 |
|  | GLORYS | 7.59 | 7.91 | 12.91 | 8.01 | 9.71 | 9.09 | 18.50 |
|  | ORAS5 | 7.03 | 11.47 | 9.87 | 9.04 | 8.76 | 8.46 | 15.87 |
| Correlation [-] | DS-CSRM | 0.999 | 0.996 | 0.96 | 0.996 | 0.996 | 0.997 | 0.88 |
|  | CSRM | 0.999 | 0.996 | 0.96 | 0.996 | 0.996 | 0.998 | 0.87 |
|  | DS-JPLM | 0.998 | 0.997 | 0.99 | 0.996 | 0.997 | 0.996 | 0.99 |
|  | JPLM | 0.998 | 0.997 | 0.99 | 0.996 | 0.997 | 0.996 | 0.99 |
|  | DS-GSFC | 0.999 | 0.996 | 0.988 | 0.996 | 0.998 | 0.994 | 0.97 |
|  | GSFC | 0.999 | 0.996 | 0.99 | 0.996 | 0.998 | 0.994 | 0.96 |
|  | GLORYS | 0.95 | 0.90 | 0.89 | 0.91 | 0.84 | 0.93 | 0.71 |
|  | ORAS5 | 0.88 | 0.74 | 0.72 | 0.83 | 0.80 | 0.82 | 0.75 |

## S2 Extended evaluation against in-situ bottom pressure recorders measurements

We extend the evaluation against in-situ bottom pressure recorders (BPR) to five downscaled $p_b$ products to investigate the impacts of different mascon and reanalysis products on the final downscaled solutions. The first three downscaled products are based on two reanalysis products and one of the three mascon solutions, as in Section S1. The other two downscaled products are generated based on the CSRM solution and only one of the two reanalysis products, denoted as *DS-CSRM (GLORYS only)* and *DS-CSRM (ORAS5 only)*.

Table S2 reports the average correlation and RMS reductions. From these bulk statistics, we can find that the impact of different mascon solutions is minor, with the solutions forced by JPLM or GSFC products slightly outperforming the one forced by CSRM. On the contrary, considering two reanalyses noticeably increases the agreement between the downscaled products and BPR measurements. The benefits of combining two reanalysis products are more obvious by analyzing the signal levels and RMS reductions at individual BPR stations (Fig. S2 and S3). The downscaled product forced by only one reanalysis tends to overfit this particular $p_b$ solution and also inherit its deficiencies. The signal levels of these two downscaled products follow the corresponding reanalysis solutions (Fig. S2c-f). Therefore, they yield negative RMS reductions (Fig. S3c-f). The situation is improved considerably by combining the two reanalysis products, enabling the downscaled solutions to have more realistic signal levels and positive RMS reductions. We find no clear evidence that the disparities in the different GRACE(-FO) $p_b$ estimates influence the downscaling quality (Fig. S2a-b, g-j).

**Table S2.** The average signal levels, Pearson correlations, and RMS reductions of the four $p_b$ products compared to monthly BPR measurements. The average signal levels of monthly BPR measurements are shown for comparison. The stations are divided into three groups by considering their signal levels ($\sigma$): all valid stations, greater equal to 20 mm, and smaller than 20 mm.

|   |   | All | $\sigma \geq 20\,\text{mm}$ | $\sigma < 20\,\text{mm}$ |
|---|---|---|---|---|
|   | Num. of BPR | 119 | 72 | 42 |
| Signal level [mm] | BPR | 27.3 | 36.6 | 12.9 |
|   | DS-CSRM | **25.3** | 32.5 | 14.4 |
|   | DS-CSRM (GLORYS only) | 44.8 | 25.3 | 18.2 |
|   | DS-CSRM (ORAS5 only) | 37.2 | 25.3 | 16.0 |
|   | DS-JPLM | 24.6 | 31.2 | **13.6** |
|   | DS-GSFC | 24.3 | **37.2** | 16.0 |
| Correlation [-] | DS-CSRM | 0.49 | 0.58 | 0.37 |
|   | DS-CSRM (GLORYS only) | 0.46 | 0.53 | 0.34 |
|   | DS-CSRM (ORAS5 only) | 0.44 | 0.49 | 0.37 |
|   | DS-JPLM | **0.53** | **0.61** | 0.41 |
|   | DS-GSFC | **0.53** | 0.60 | **0.42** |
| ΔRMS [mm] | DS-CSRM | 2.52 | 4.94 | -1.19 |
|   | DS-CSRM (GLORYS only) | -3.53 | -3.03 | -4.31 |
|   | DS-CSRM (ORAS5 only) | -1.47 | -0.88 | -2.37 |
|   | DS-JPLM | **3.87** | 6.64 | **-0.38** |
|   | DS-GSFC | **3.87** | 6.78 | -0.60 |

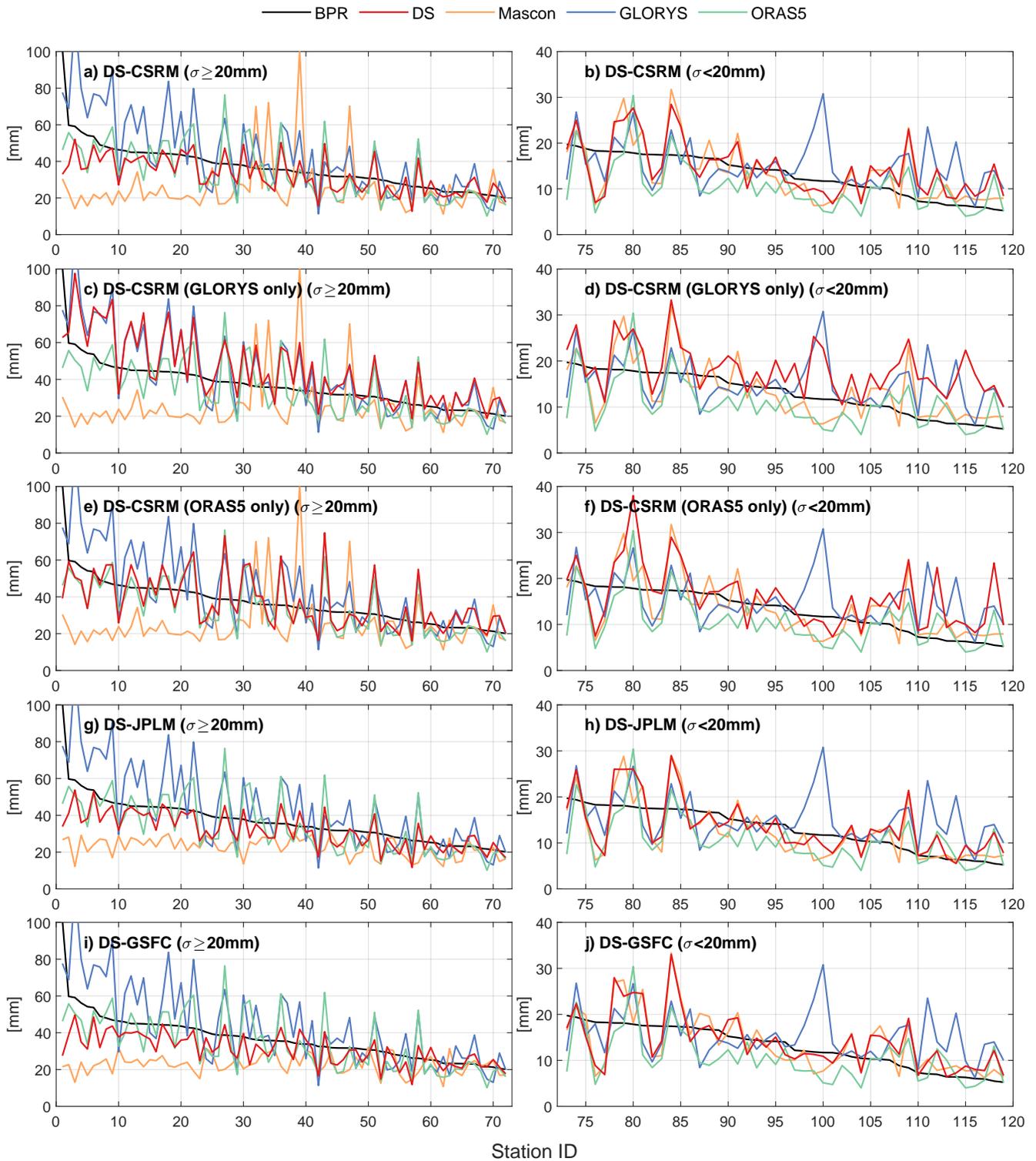

**Figure S2.** The signal levels at the 119 valid in-situ BPR stations. The stations are sorted by the descending signal levels of the monthly BPR measurements and split into two groups for better visualization. Each row shows an individual downscaled product with different inputs. The signal levels of the corresponding mascon product and reanalysis products are shown in all panels for comparison.

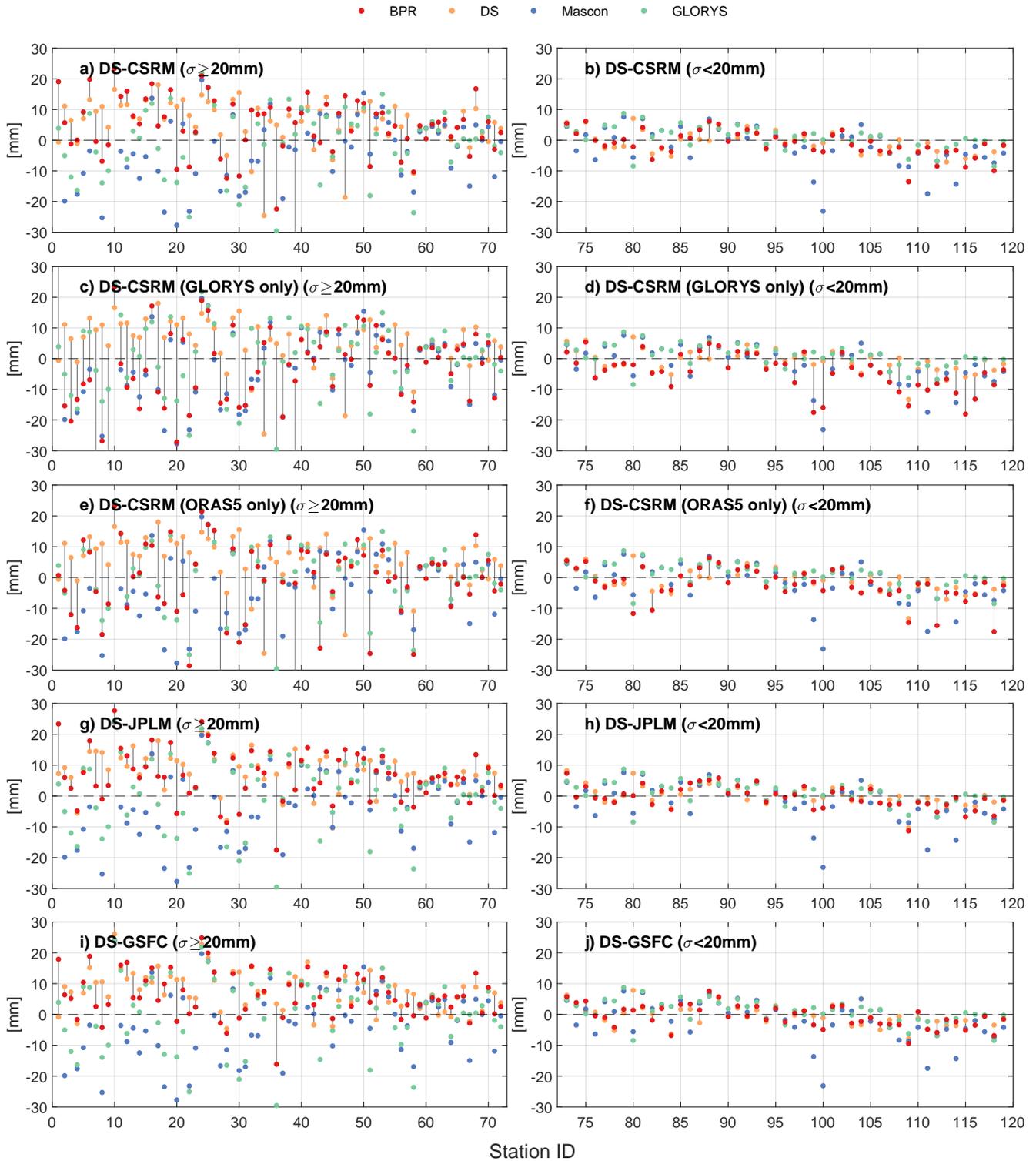

**Figure S3.** The RMS at the 119 valid in-situ BPR stations. The stations are sorted by the descending signal levels of the monthly BPR measurements and split into two groups for better visualization. Each row shows an individual downscaled product with different inputs. The signal levels of the corresponding mascon product and reanalysis products are shown in all panels for comparison.

## S3 Impacts of earthquakes on the evaluation

Large earthquakes and tsunamis may affect our evaluation in two ways. First, the GRACE(-FO) $p_b$ fields are derived from gravity measurements, which carry the imprint of seismological events. Second, the displacements caused by earthquakes may create discontinuities in tide gauge (TG) measurements. We pick four TG stations that may have been impacted by the Sumatra–Andaman earthquake (26 December 2004) and the Tōhoku-Oki earthquake (11 March 2011) and show the time series in Fig. S4. These seismological imprints may cause biased statistics for the evaluations against tide gauge measurements, such as low correlations and high RMSE. However, compared to the GRACE(-FO) solutions, our downscaled solution has more similar variability to the tide gauge measurements, which demonstrates the ability of the downscaling method to restore signal levels.

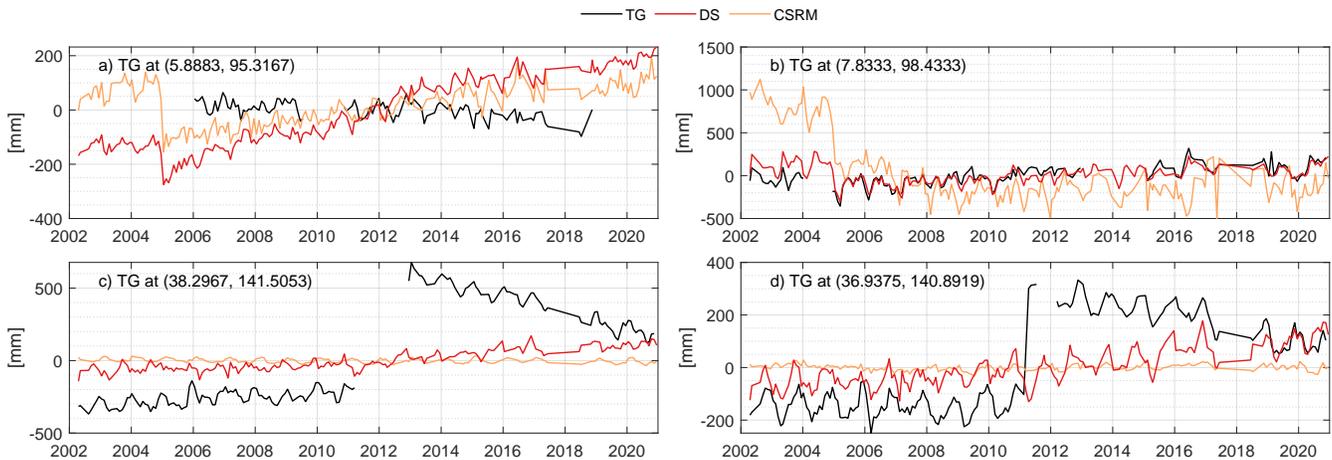

**Figure S4.** Examples of two tide gauge (TG) stations in the Tōhoku region (a, b) and two tide gauge stations in the Indian Ocean (c, d).

## References

Landerer, F. W., & Cooley, S. S. (2021). *Gravity Recovery and Climate Experiment Follow-on (GRACE-FO) Level-3 Data Product User Handbook*. URL: https://podaac-tools.jpl.nasa.gov/drive/files/allData/gracefo/docs/GRACE-FO_L3_Handbook_JPL.pdf.

Sakumura, C., Bettadpur, S., & Bruinsma, S. (2014). Ensemble prediction and intercomparison analysis of GRACE time-variable gravity field models. *Geophysical Research Letters*, *41*, 1389–1397. doi:https://doi.org/10.1002/2013GL058632.



\end{document}